\documentclass[amsmath,showpacs,aps,prb,twocolumn,superscriptaddress,floatfix]{revtex4-1}
\usepackage{amsmath}
\usepackage{dcolumn}
\usepackage{amsfonts}
\usepackage{amssymb}
\usepackage{graphicx}
\usepackage{natbib}
\usepackage{color}
\usepackage[caption=false]{subfig}
\usepackage[colorlinks=false,citecolor=blue,linkcolor=blue]{hyperref}

\begin{document}
\title{Topological phases in layered pyrochlore oxide thin films along the [111] direction}
\author{Xiang Hu} 
\email{phyxiang@gmail.com}
\affiliation{Department of Physics, The University of Texas at Austin, Austin, Texas 78712, USA}
\author{Andreas R\"uegg}
\affiliation{Department of Physics, University of California, Berkeley, CA 94720, USA} 
\author{Gregory A. Fiete}
\affiliation{Department of Physics, The University of Texas at Austin, Austin, Texas 78712, USA}
\date{\today}
\begin{abstract}
We theoretically study a multi-band Hubbard model of pyrochlore oxides of the form A$_2$B$_2$O$_7$, where B is a heavy transition metal ion with strong spin-orbit coupling, in a thin film geometry orientated along the [111] direction. Along this direction, the pyrochlore lattice consists of alternating kagome and triangular lattice planes of B ions. We consider a single kagome layer, a bilayer, and the two different trilayers. As a function of the strength of the spin-orbit coupling, the direct and indirect $d$-orbital hopping, and the band filling, we identify a number of scenarios where a non-interacting time-reversal invariant Z$_2$ topological phase is expected and we suggest some candidate materials. We study the interactions in the half-filled $d$-shell within Hatree-Fock theory and identify parameter regimes where a zero magnetic field Chern insulator with Chern number $\pm1$ can be found. The most promising geometries for topological phases appear to be the bilayer which supports both a Z$_2$ topological insulator and a Chern insulator, and the triangular-kagome-triangular trilayer which supports a relatively robust Chern insulator phase.
\end{abstract}

\pacs{71.10.Fd,71.10.Pm,73.20.-r,73.43.-f}


\maketitle

\section{Introduction}

With the experimental discovery of two\cite{Konig:sci07,Roth:sci09} and three\cite{Hsieh:nat08,Hsieh:sci09,Xia:np09} dimensional topological insulators, the study of topological phases has rapidly intensified over the past few years.\cite{Hasan:rmp10,Moore:nat10,Qi:rmp11} Significant effort has been directed at the identification of new materials\cite{Zhang:np09,Chadov:nm10,JWang:prl11,Feng:prl11,Zhang:prl11,Xiao:prl10,Lin:prl10,Chen:prl10,Yan:epl10,Yan:prb10}
and the role that interactions may play in driving topological phases.\cite{Young:prb08,Ran:prl08,Raghu:prl08,Zhang:prb09,Wen:prb10,Ri:np10,Swingle:prb11,Maciejko:prl10,Levin:prb12,Levin:prl09,Moore:prl08,Neupert:prb11,Witczak-Krempa:prb10,Liu:prb10} In addition, the interplay between magnetism and non-trivial topological bands has also been addressed.\cite{He:prb11, Hohenadler:prb12, Yoshida:prb12, He12, Yoshida12, Hohenadler12}

A number of the studies focused on interaction effects have been directed at topological phases in transition metal oxides.\cite{Shitade:prl09,Pesin:np10,Yang_Kim:prb10,Kargarian:prb11,Wan:prb11,Ruegg:prl12,Kargarian:prb12,Go:prl12,Yang:prb11b,Witczak:prb12}  A new direction that has emerged within this area is the search for topological phases at the interfaces of correlated oxides.\cite{Xiao:nc11,Ruegg11_2,Ruegg:prb12,Yang:prb11a,Wang:prb11}

To date, the prediction of topological phases in transition metal oxide interfaces has focused on the perovskite structure\cite{Xiao:nc11,Ruegg11_2,Ruegg:prb12,Yang:prb11a,Wang:prb11} ABO$_3$, where A is usually a rare earth element, B is a transition metal, and O is oxygen.  An undistorted perovskite has a relatively simple cubic structure with natural cleave planes along the [001] and equivalent directions.  For this reason, most of the interface and superlattice structures of these materials have been grown along this direction.\cite{Mannhart:sci10,Mannhart:mrb08,Zubko:arcmp11,Boris:sci11,Benckiser:nm11,Chakhalian:prl11,Lui:prb11,Son:apl10,Son_2:apl10}  However, theoretical models suggest that thin films, particularly bilayers and trilayers, grown along the [111] direction are more promising for realizing topological phases.\cite{Xiao:nc11,Ruegg11_2,Ruegg:prb12,Yang:prb11a,Wang:prb11}  The underlying reason for the increased favorability of topological phases in thin films grown along the [111] direction is the nature of the energy bands (derived primarily from the transition metal $d$-orbitals and the oxygen $p$-orbitals):  They are both relatively ``flat" ({\it i.e.} dispersionless) along some directions for certain filling fractions, and they possess flat band touching points in relevant tight-binding models. In two dimensions, flat band touchings are perturbatively unstable to interactions and topological phases often emerge as the leading instability.\cite{Sun:prl09,Wen:prb10,FZhang:prl11,Ruegg11_2,Yang:prb11a} Flat bands themselves harbor topological phases if their Chern number is non-zero, including the most interesting case of partial band filling.\cite{Sheng11,Sun:prl11,Neupert:prl11,Tang:prl11,Hu:prb11,Trescher12,Liu:prl12}

\begin{figure}[ht]
 \begin{minipage}{0.3\columnwidth}
\subfloat[Sandwich structure]{
\centering
    \includegraphics[width=\linewidth]{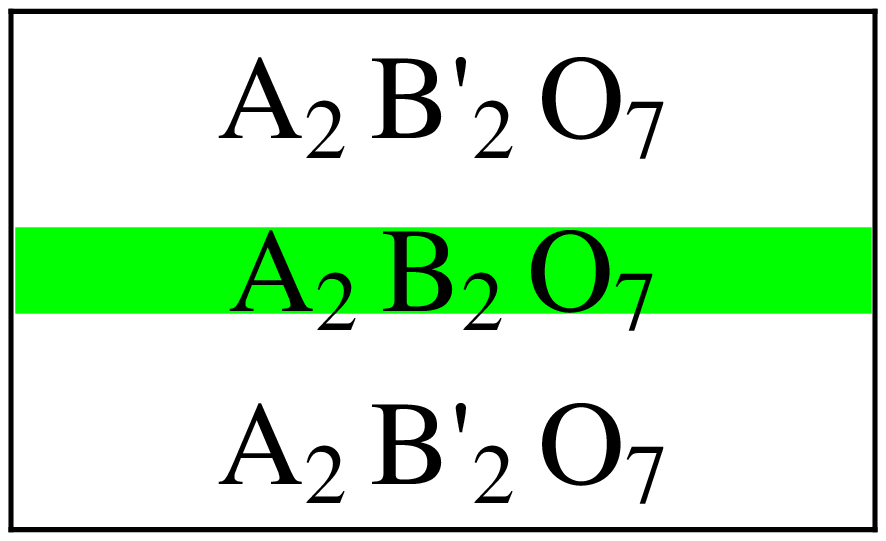}
    \label{fig1a}
    }
    \vspace{0.4cm}
\subfloat[Brillouin zone]{
\centering
    \includegraphics[width=\linewidth]{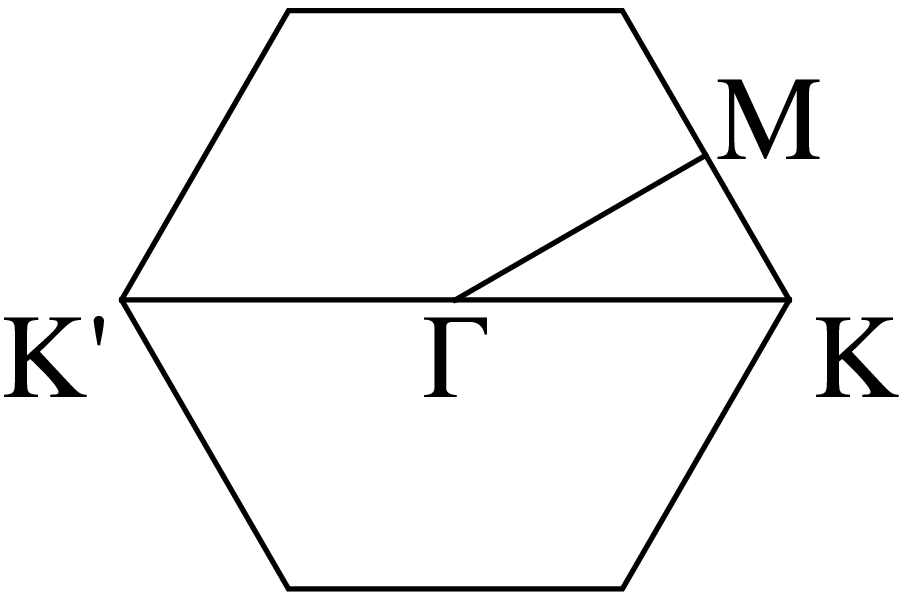}
    \label{fig1b}
}

\end{minipage}
\hfill
\begin{minipage}{0.65\columnwidth}
\subfloat[Pyrochlore lattice structure]{
\centering
    \includegraphics[width=\linewidth]{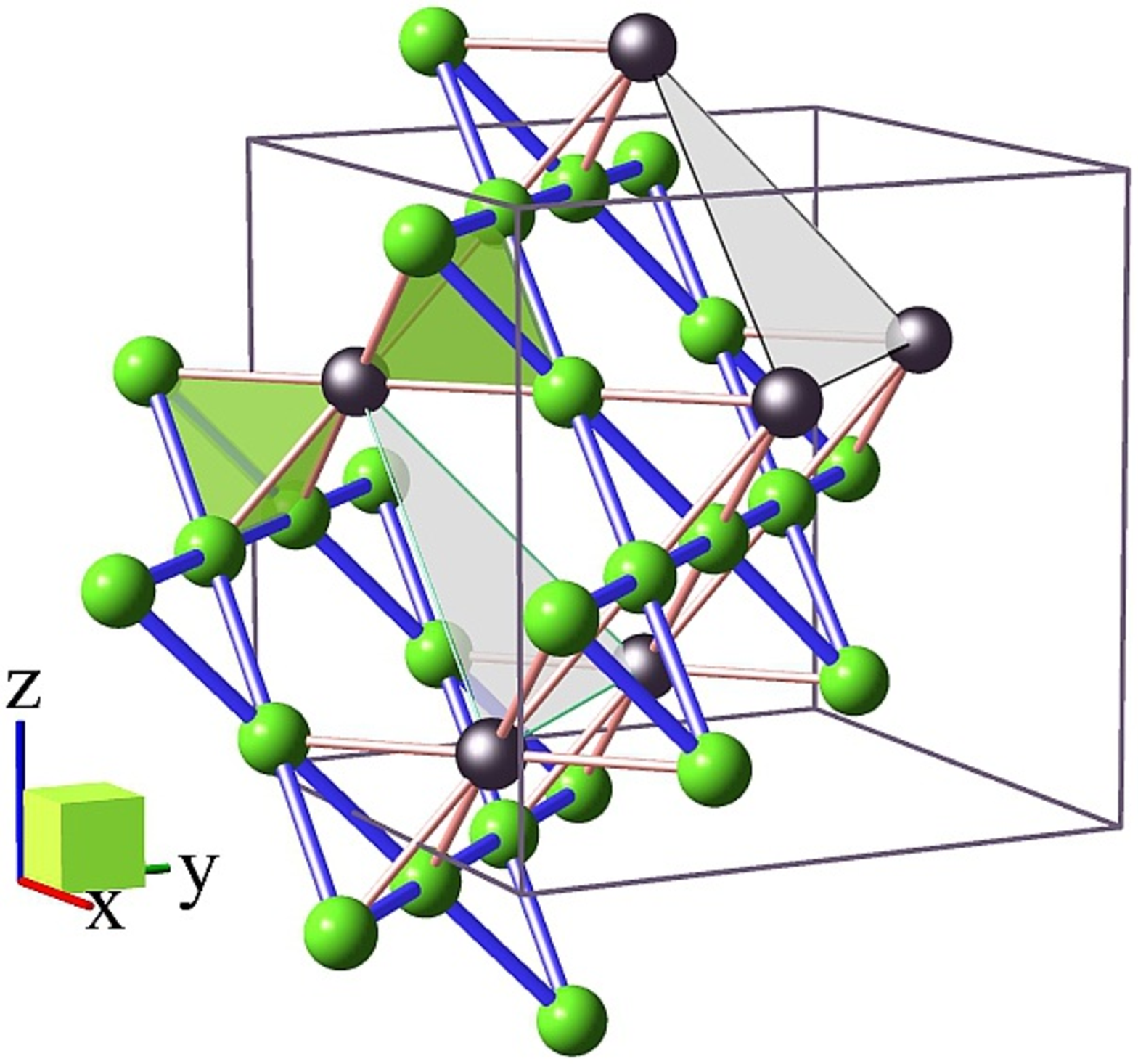}
    \label{fig1c}
    }
    \end{minipage}
 \caption{(color online) (a) Schematic of the experimental geometry. The middle A$_2$B$_2$O$_7$ thin-film layer is the region of physical interest. The capping regions A$_2$B'$_2$O$_7$ would ideally be non-magnetic large gap band insulators of the same lattice constant as A$_2$B$_2$O$_7$ and would only serve to structurally stabilize the thin-film region.  The growth direction of the structure is along [111]. (b) The Brillouin zone of the underlying triangular Bravais lattice of the thin-films. (c) Pyrochlore lattice structure showing alternating kagome and triangular lattice planes along the [111] direction. A kagome-triangular-kagome-triangular tetralayer is shown with green atoms in the kagome planes and gray atoms in the triangular planes.}
\label{fig1} 
\end{figure}

A potential experimental challenge in the implementation of the theoretical proposals\cite{Xiao:nc11,Ruegg11_2,Ruegg:prb12,Yang:prb11a,Wang:prb11} for topological phases in the perovskite ABO$_3$ heterostructures grown along [111] is the difficulty of crystal growth in this direction. As it is not a natural cleavage plane, it is also not a natural growth direction. It is unclear at present whether this issue will become a significant impediment to the exploration of topological phases in oxide thin films and heterostructures, although successful growth of (111)-oriented ABO$_3$ heterostructures was recently reported.\cite{Gibert:nm2012,Herranz:sp2012,Middey:12} In the meantime, it is prudent to suggest other material systems where strong correlation effects and topological phases are likely to coexist.  

In this work we study the potential for realizing topological phases in thin films [see Fig.~\ref{fig1a}] of transition metal oxides of the form A$_2$B$_2$O$_7$, where A is a rare earth element, B is a heavy transition metal ion with strong spin-orbit coupling, and O is oxygen. In this class of materials, the energy bands are expected to be derived primarily from the transition metal ion $d$-orbitals and the oxygen $p$-orbitals,\cite{Pesin:np10,Kargarian:prb11,Wan:prb11,Witczak:prb12} similar to the ABO$_3$ perovskites discussed just above. However, in the case of A$_2$B$_2$O$_7$ the transition metal ions B reside on a pyrochlore lattice--a corner sharing network of tetrahedra shown in Fig.~\ref{fig1c}.  Along the [111] and equivalent directions there are natural cleavage planes that consists of alternating layers of a triangular lattice of B ions and a kagome lattice of B ions.  [See Fig.~\ref{fig1c}.] Therefore, in contrast to the ABO$_3$ materials, the [111] direction is expected to be a good growth direction.  

\begin{figure}[ht]
\centering
\subfloat[Kagome layer]{
\label{fig2a} 
\includegraphics[width=0.45\linewidth]{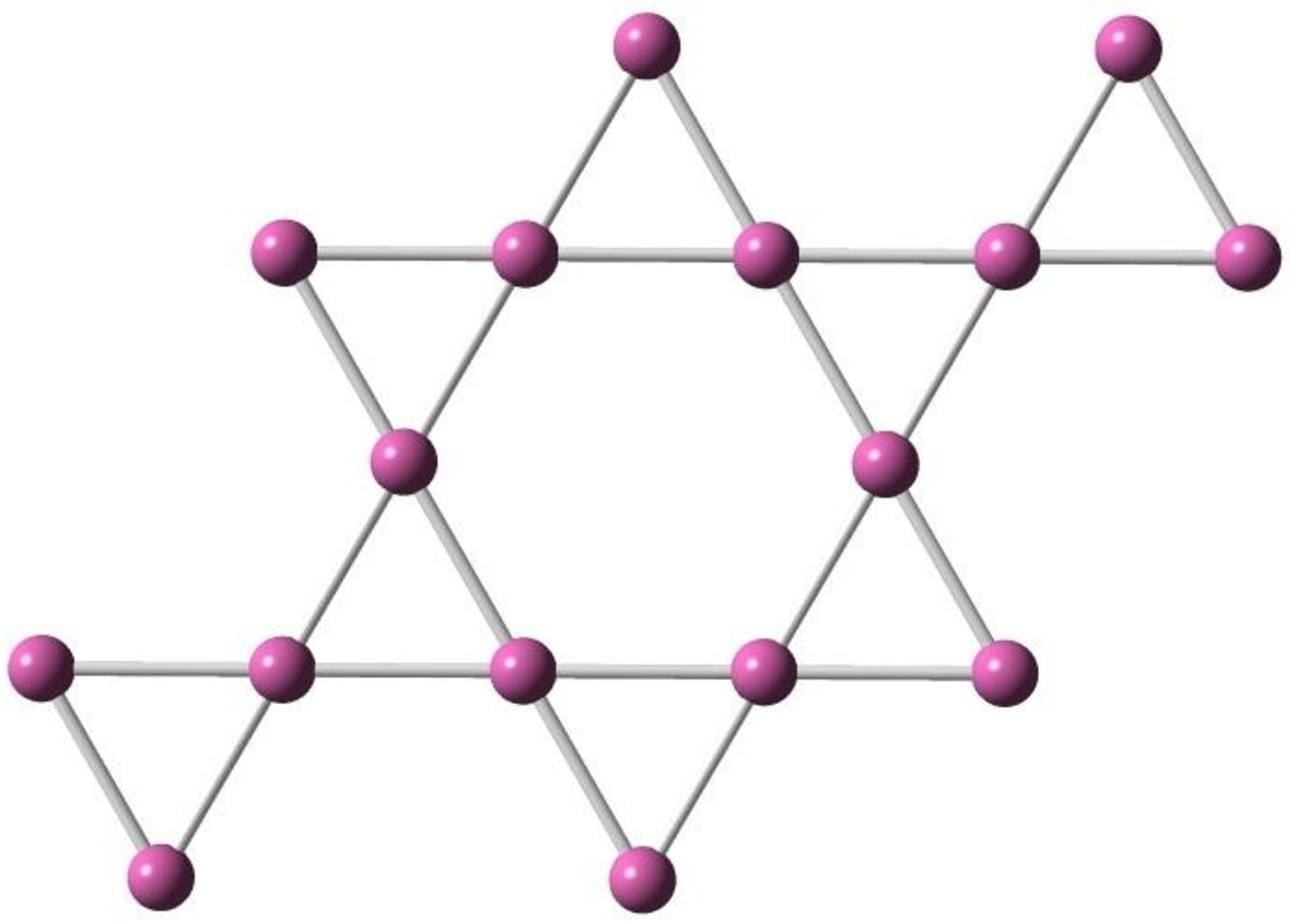}
}
\hfill
\subfloat[Bilayer]{
\label{fig2b} 
\includegraphics[width=0.45\linewidth]{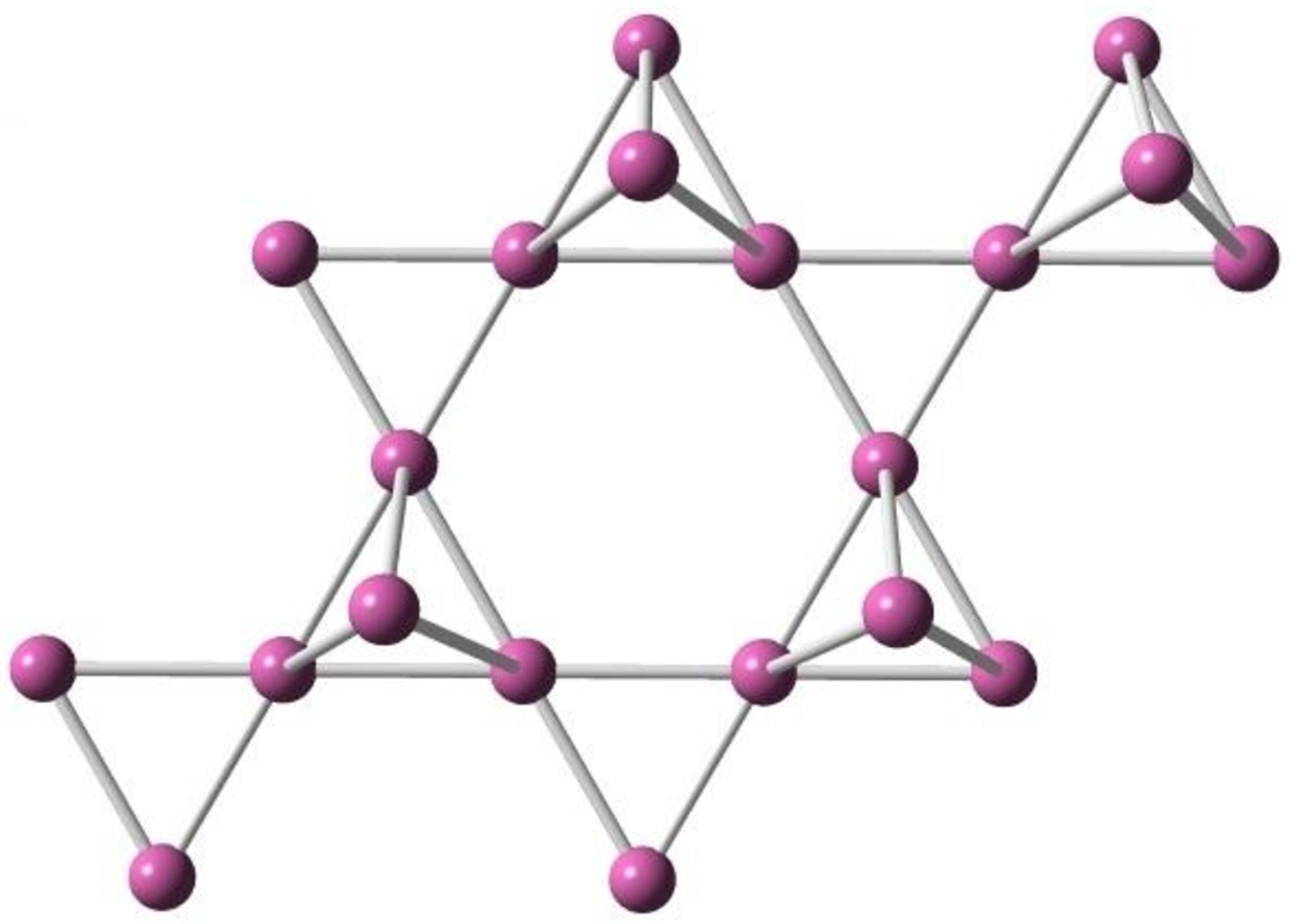}
}
\vspace{0.5cm}
\subfloat[\ Trilayer (TKT)]{
\label{fig2c} 
\includegraphics[width=0.45\linewidth]{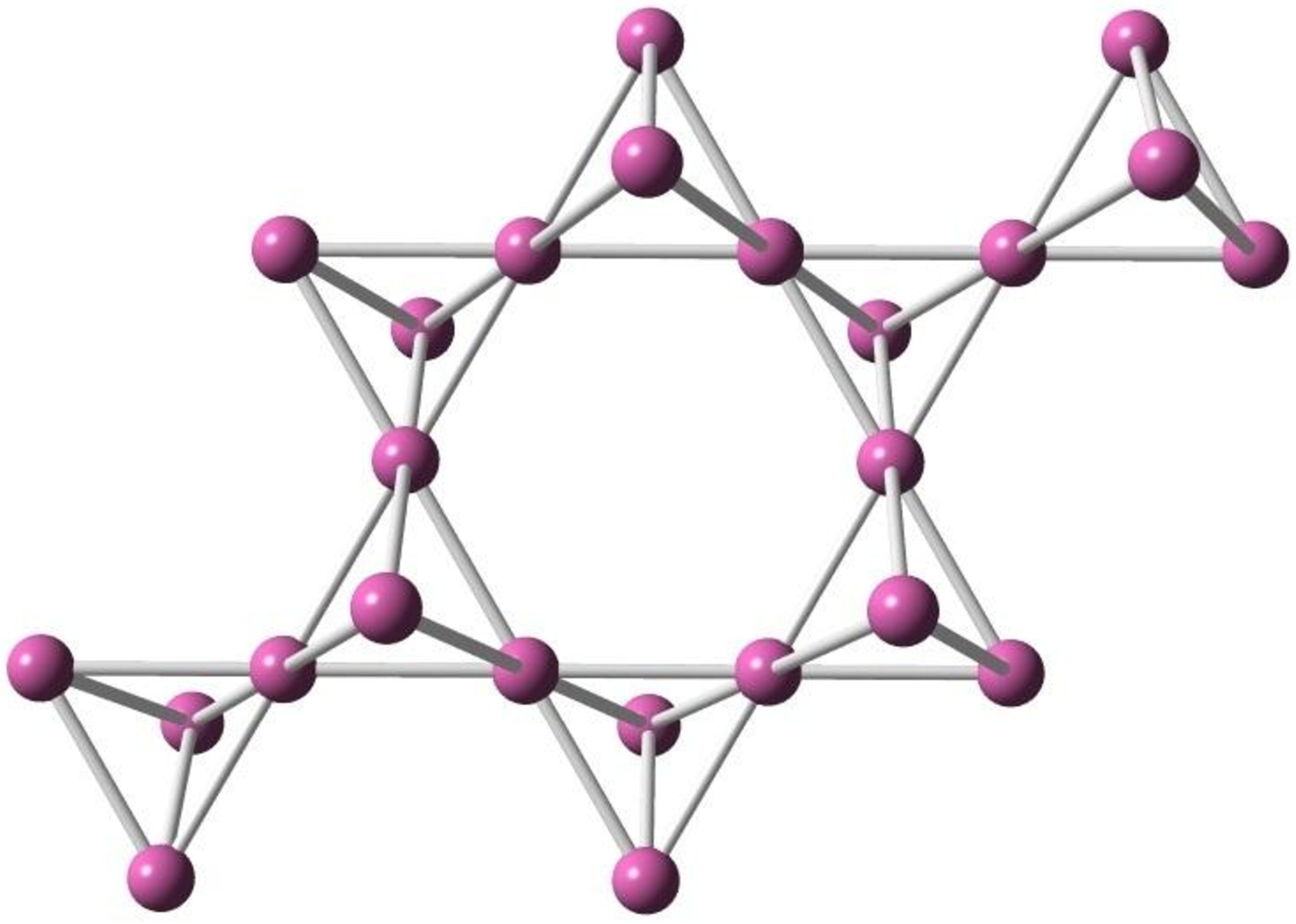}
}
\hfill
\subfloat[\ Trilayer (KTK)]{
\label{fig2d} 
\includegraphics[width=0.45\linewidth]{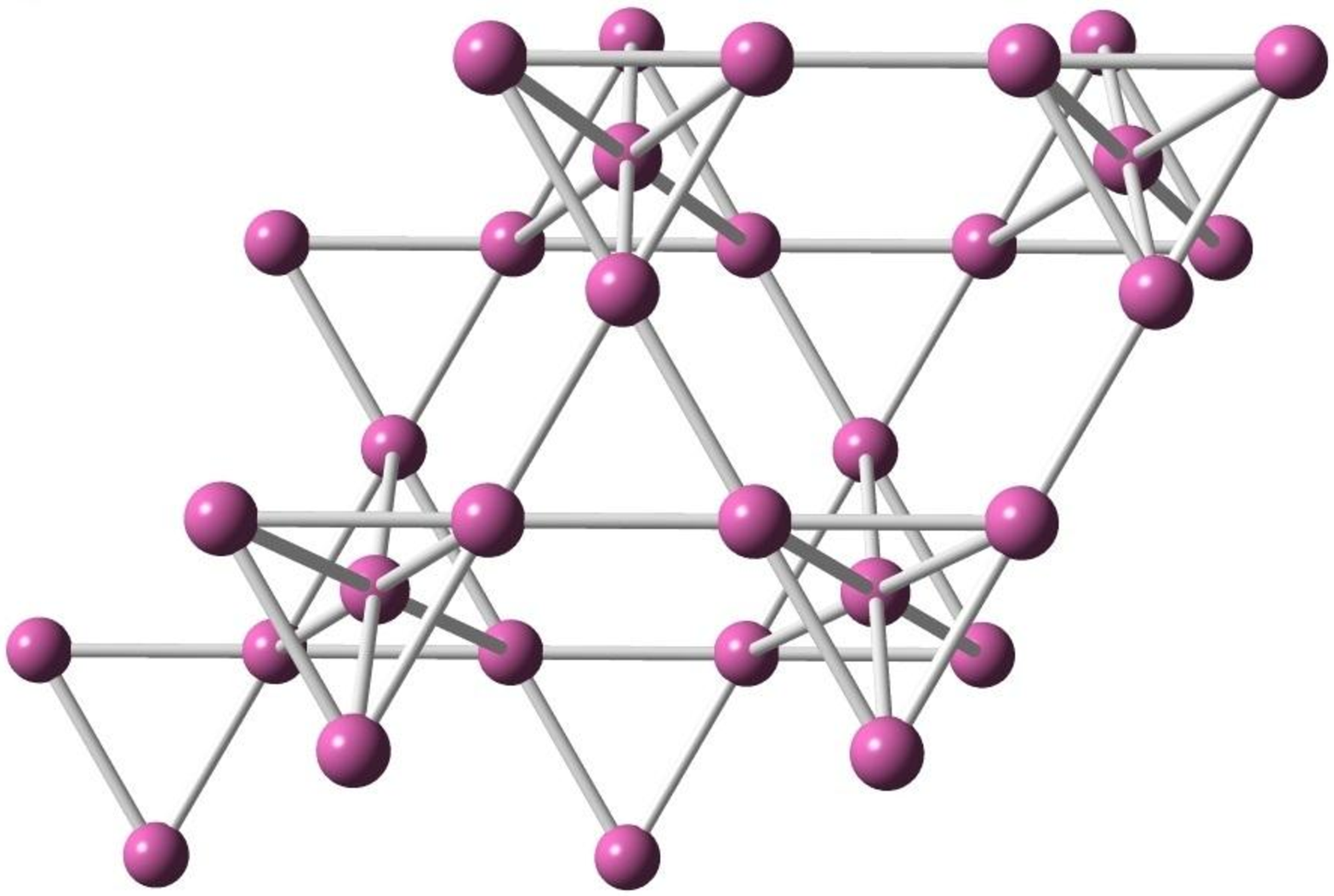}
}
\caption{(color online) Candidate thin films to be used in the A$_2$B$_2$O$_7$ region shown in Fig.\ref{fig1a}.  As shown in Fig.\ref{fig1c}, along the [111] direction the pyrochlore lattice is an alternating series of kagome and triangular lattice planes.  Single, double, and triple layer stackings are shown in (a)-(d).  Among these stackings, the bilayer and TKT trilayer are found to be the most promising for realizing robust (to model Hamiltonian parameters) $Z_2$ topological insulator and Chern insulator (zero magnetic field quantum Hall) phases.}
\label{fig2} 
\end{figure}

In this paper we explore the favorability of topological phases in heterostructures grown along the [111] direction with a geometry such as that shown in Fig.~\ref{fig1a}. The ``bread slices" of the ``sandwich" composed of the material A$_2$B'$_2$O$_7$, would ideally be a non-magnetic band insulator which would differ from the thin-film layer only in the ``B" element to minimize lattice strain effects.\cite{Note_Strain} We consider a central thin-film layer that could consist of: (i) a single kagome layer, (ii) a triangular-kagome bilayer, (iii) a triangular-kagome-triangular trilayer (TKT), and (iv) a kagome-triangular-kagome trilayer (KTK).  These thin film structures are shown in Fig.~\ref{fig2}. We study a multi-band Hubbard model with an on-site atomic spin-orbit coupling of the $\vec L \cdot \vec S$ form, Eq.~\eqref{eq:H0}. The interactions are included with the standard on-site Hubbard term, Eq.~\eqref{eq:H_U}, and we treat them within the unrestricted Hatree-Fock approximation.  

Our main results are that we find $d$-shell filling fractions in {\em all} systems studied that favor $Z_2$ time-reversal invariant topological insulators (TI) at small and vanishing interactions. The specific filling fractions in each system that support a TI phase are shown in Figs.~\ref{fig:singlelayer}-\ref{fig:KTK}, along with representative band dispersions along high-symmetry directions for two values of spin-orbit coupling and two values of hopping parameters. In order to compare our results for the case of arbitrary interactions with those obtained in bulk A$_2$B$_2$O$_7$ pyrochlores\cite{Go:prl12,Witczak:prb12} we focus on the case of strong spin-orbit coupling and a half-filled $d$-shell. Our main results for the non-interacting case are summarized in the phase diagrams in Fig.~\ref{fig:phase_U_0} and for the interacting case are summarized in the phase diagrams in Fig.~\ref{fig:U_phase}. Two results are worth highlighting: (i) The bilayer possesses a TI at small interactions {\em and} a narrow region of a zero magnetic field Chern insulator with Chern number $\pm 1$ at intermediate interaction strength. (ii) The TKT system possesses a relative wide range interactions of intermediate strength where a Chern insulator phase with Chern number $\pm 1$ is stabilized.  

In our study, we have swept out a wide portion of the parameter space accessible within the tight-binding model and Hartree-Fock calculations employed (especially in the case of a half-filled $d$-shell).  It is important to list some materials that might serve as a useful starting point in an experimental search for the topological phases we find here. For half-filled $d$-shell materials, materials such as A$_2$Ir$_2$O$_7$ with A=La, Y might be good candidates for the thin films considered here.  For the ``capping" layers A$_2$B'$_2$O$_7$, possible candidates are A$_2$Hf$_2$O$_7$ for A=La, Y.\cite{Li:jap07,Seguini:apl06,Pruneda:prb07}  Undoubtedly there are others, and we hope material growth experts will pick up the thread and apply their own expertise to suggest the most promising candidates.\cite{Gardner:rmp10,Rabe:arcmp10}  

Our paper is organized as follows. In Sec.~\ref{sec:non-interacting} we describe the effective tight-binding model we consider, which includes an atomic spin-orbit coupling term that drives topological insulator phases and both direct $d$-orbital hopping between transition metal ions and the more commonly considered indirect $d$-orbital hopping via the oxygen $p$-orbitals.  We present phase diagrams and band structures for a variety of representative thin films. In Sec.~\ref{sec:interacting} we include an additional on-site Hubbard interaction that  we study within the Hartree-Fock approximation. We present phase diagrams for the case of half-filled $d$ shells for a variety different thin films and highlight the conditions that favor interaction-driven topological phases with a non-zero Chern number. In Sec.~\ref{sec:conclusions} we present the main conclusions of our work, and in the appendicies we provide some technical details related to computing the direct $d$-orbital hopping parameters on the pyrochlore lattice (including the thin film geometries we consider) and the mean-field calculation.

\section{Effective Tight-binding Model}
\label{sec:non-interacting}

As we mentioned in the introduction, we are focused on the physics of the thin film A$_2$B$_2$O$_7$ layer shown in Fig.~\ref{fig1a}. We consider the four varieties of films shown in Fig.~\ref{fig2}. The ``capping" layers A$_2$B'$_2$O$_7$ are assumed to be electronically inert, and this assumption has been substantiated in closely related density functional theory (DFT) studies in the ABO$_3$ perovskite films.\cite{Xiao:nc11,Ruegg:prb12}  

The transition metal ions B in A$_2$B$_2$O$_7$ reside on a three-dimensional pyrochlore lattice [see Fig.~\ref{fig1c}] in which each B atom is centered in an octahedral cage of oxygen atoms.  The transition metal ion B is thus subjected to a cubic crystalline field which splits the $d$-orbitals into a lower-lying $t_{2g}$ manifold and a higher lying (typically on the order of a few electron volts) $e_g$ manifold.\cite{TMO_book,Wan:prb11}  If spin-orbit coupling is also present as we assume here, it further splits the $t_{2g}$ manifold into a lower-lying $j=3/2$ manifold and higher-lying $j=1/2$ manifold, with the size of the splitting given by the size of the spin-orbit coupling.\cite{TMO_book}  To the lowest order, the spin-orbit coupling does not affect the $e_g$ manifold, though at second-order virtual transitions to the $t_{2g}$ manifold do lead to an effective spin-orbit coupling within the $e_g$ manifold.\cite{Xiao:nc11}  
\begin{figure}[htb]
\includegraphics[width=0.49\linewidth]{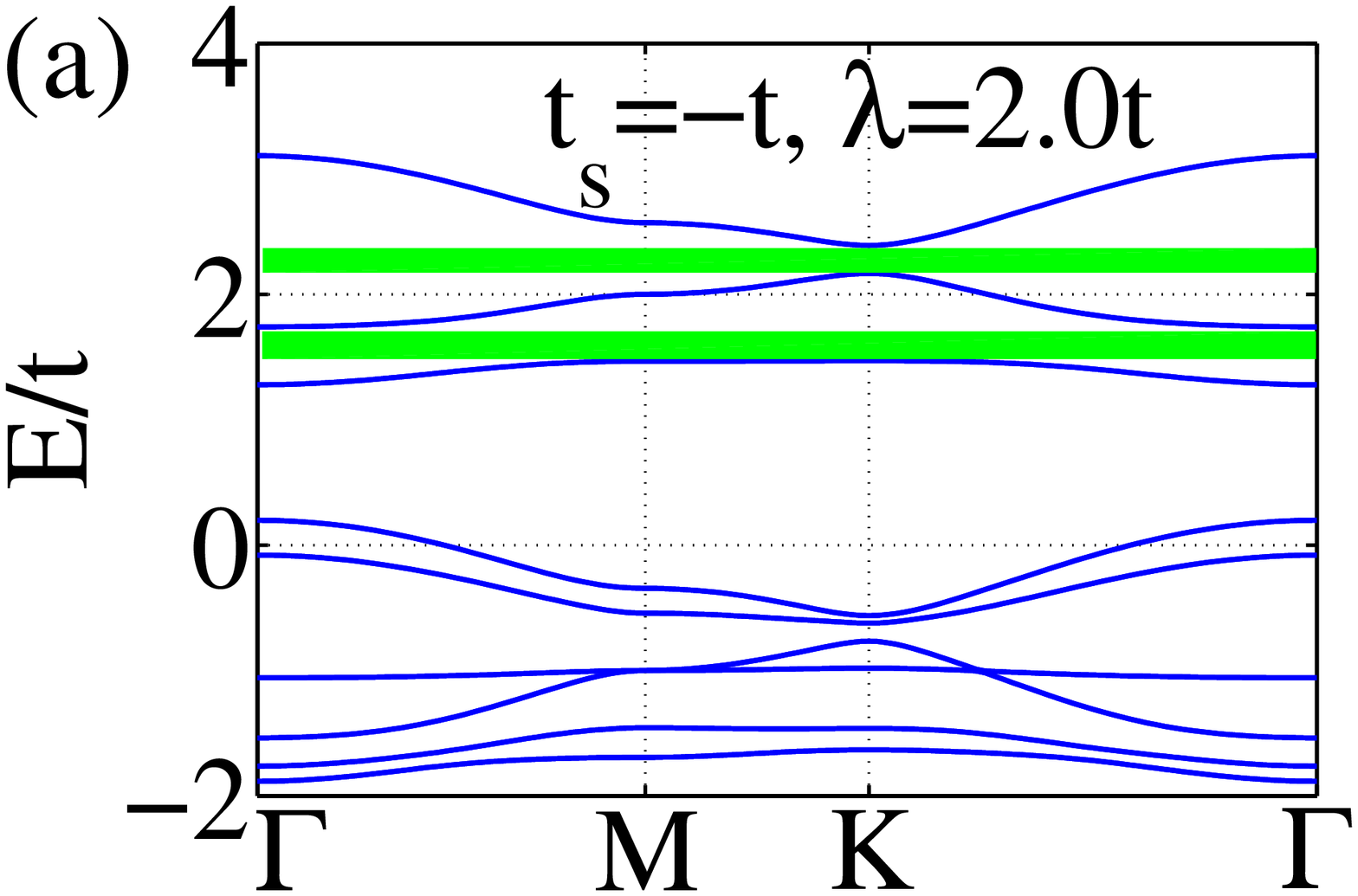}
\includegraphics[width=0.49\linewidth]{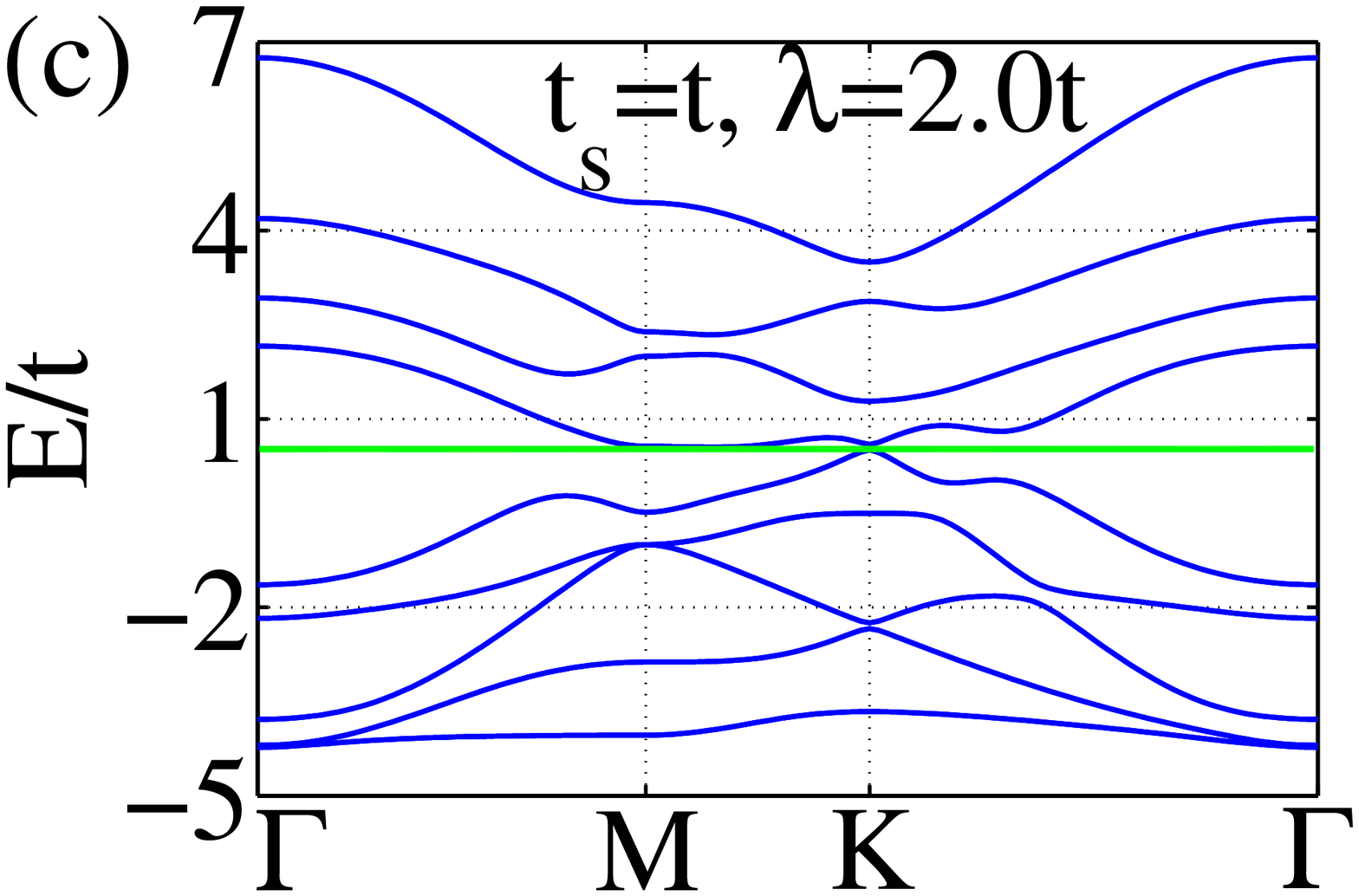}
\includegraphics[width=0.49\linewidth]{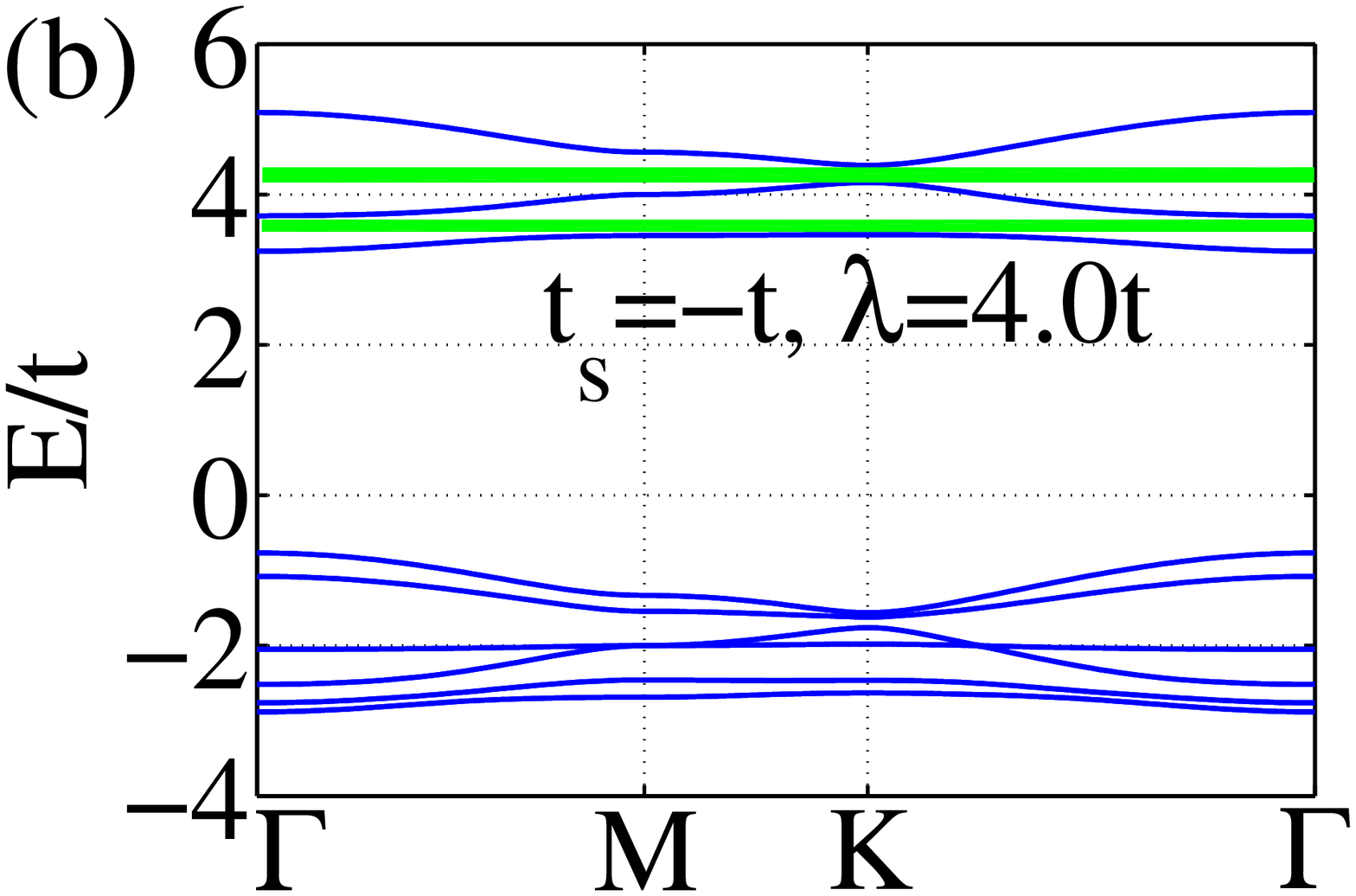}
\includegraphics[width=0.49\linewidth]{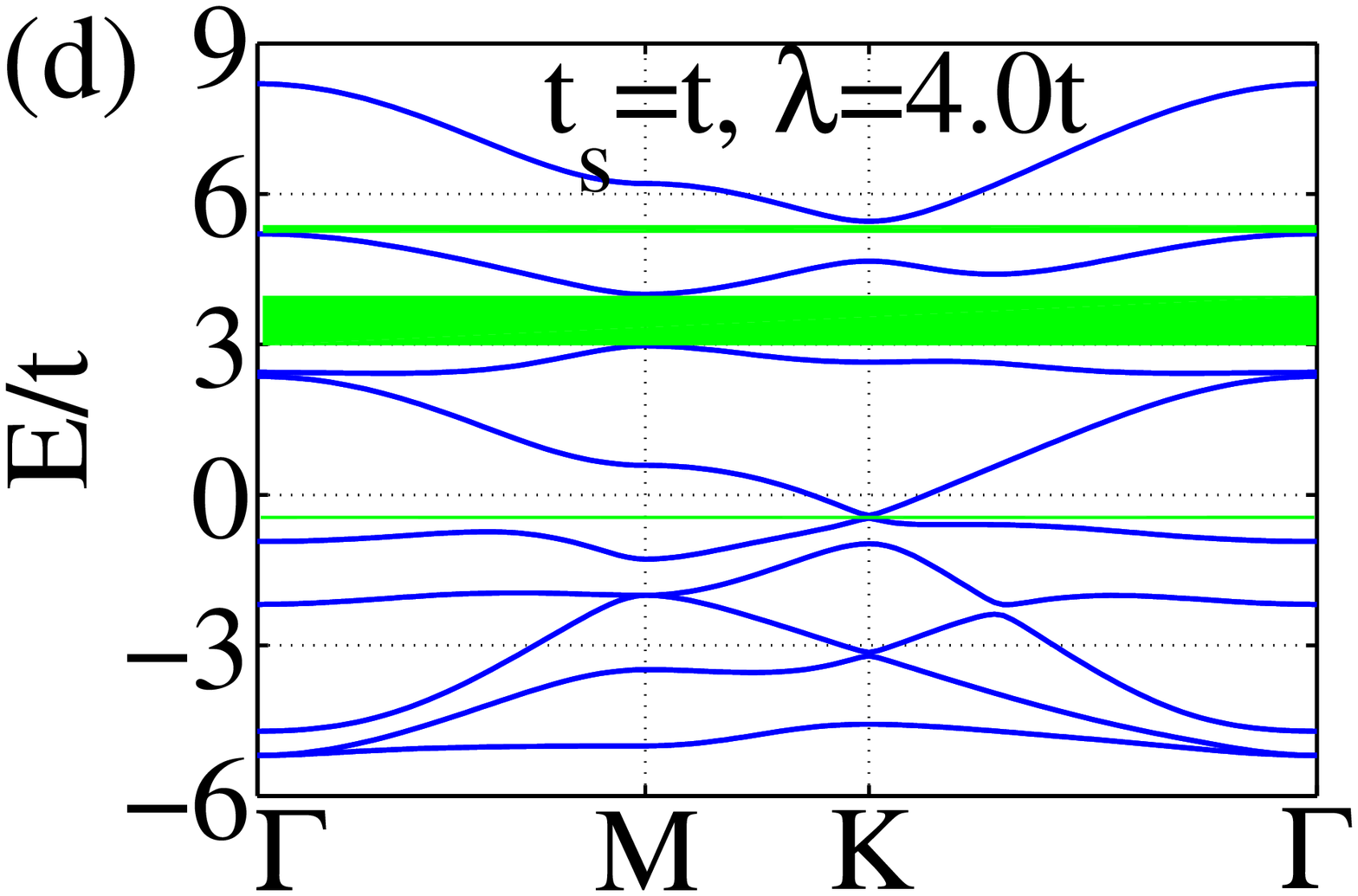}
\caption{(color online) The band structure of Eq.~\eqref{eq:H0} along the high symmetry directions [see Fig.~\ref{fig1b}] of a single kagome layer with $t_p=-2t_s/3$, $t_s=-t,t$ and $\lambda=2t, 4t$ as shown within the figure. Green (light gray) lines indicate filling fractions of of a $Z_2$ TI.  The thickness of the line indicates the size of the gap. In the cases shown $Z_2$ TI occur at $t_{2g}$ filling fractions:
(a) $\frac 7 9$, $\frac 8 9$; (b) $\frac 7 9$, $\frac 8 9$;
(c) $\frac 5 9$; (d) $\frac 5 9$, $\frac 7 9$, $\frac 8 9$.
}
\label{fig:singlelayer}
\end{figure}

\begin{figure}[htb]
\includegraphics[width=0.49\linewidth]{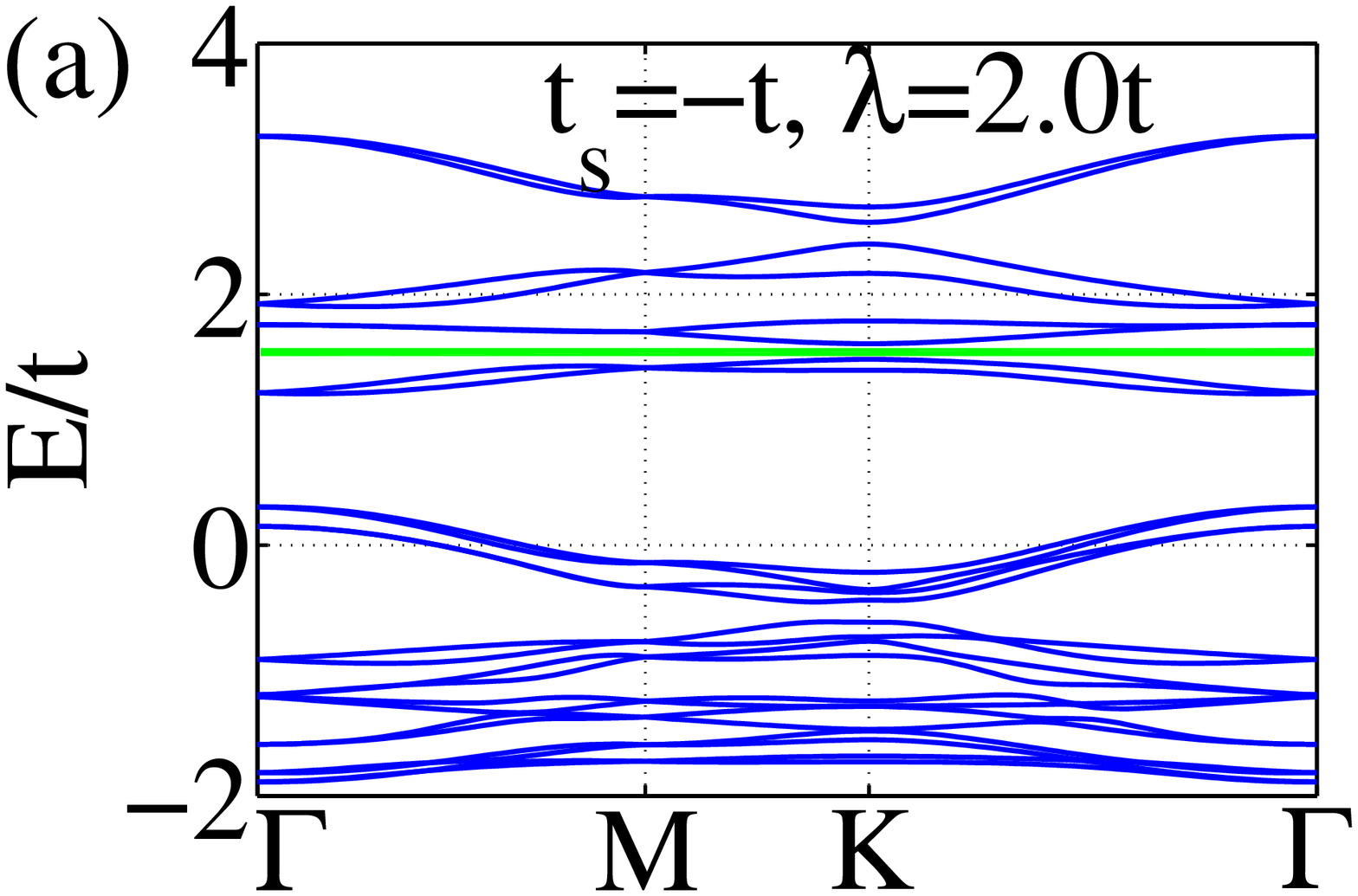}
\includegraphics[width=0.49\linewidth]{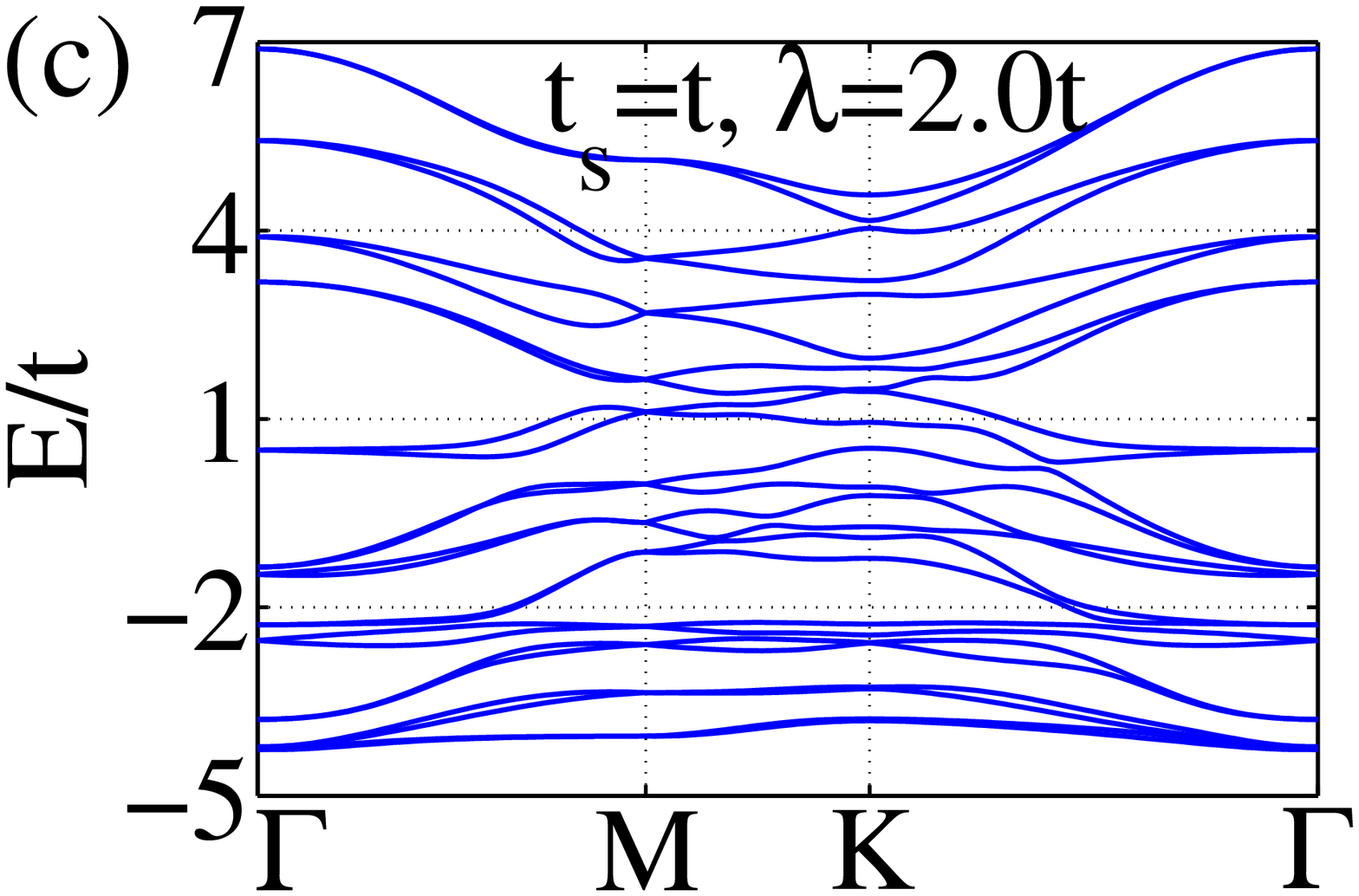}
\includegraphics[width=0.49\linewidth]{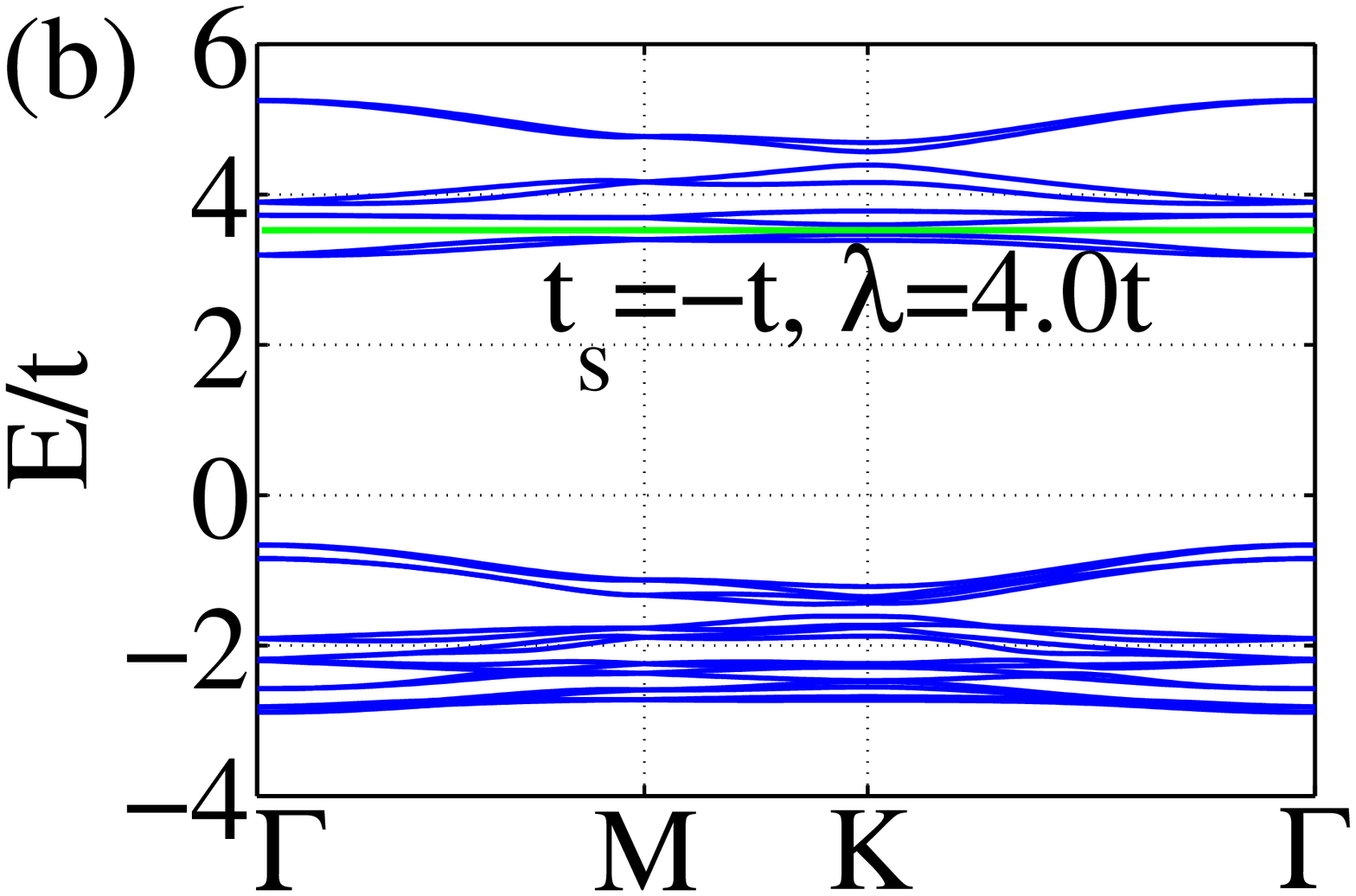}
\includegraphics[width=0.49\linewidth]{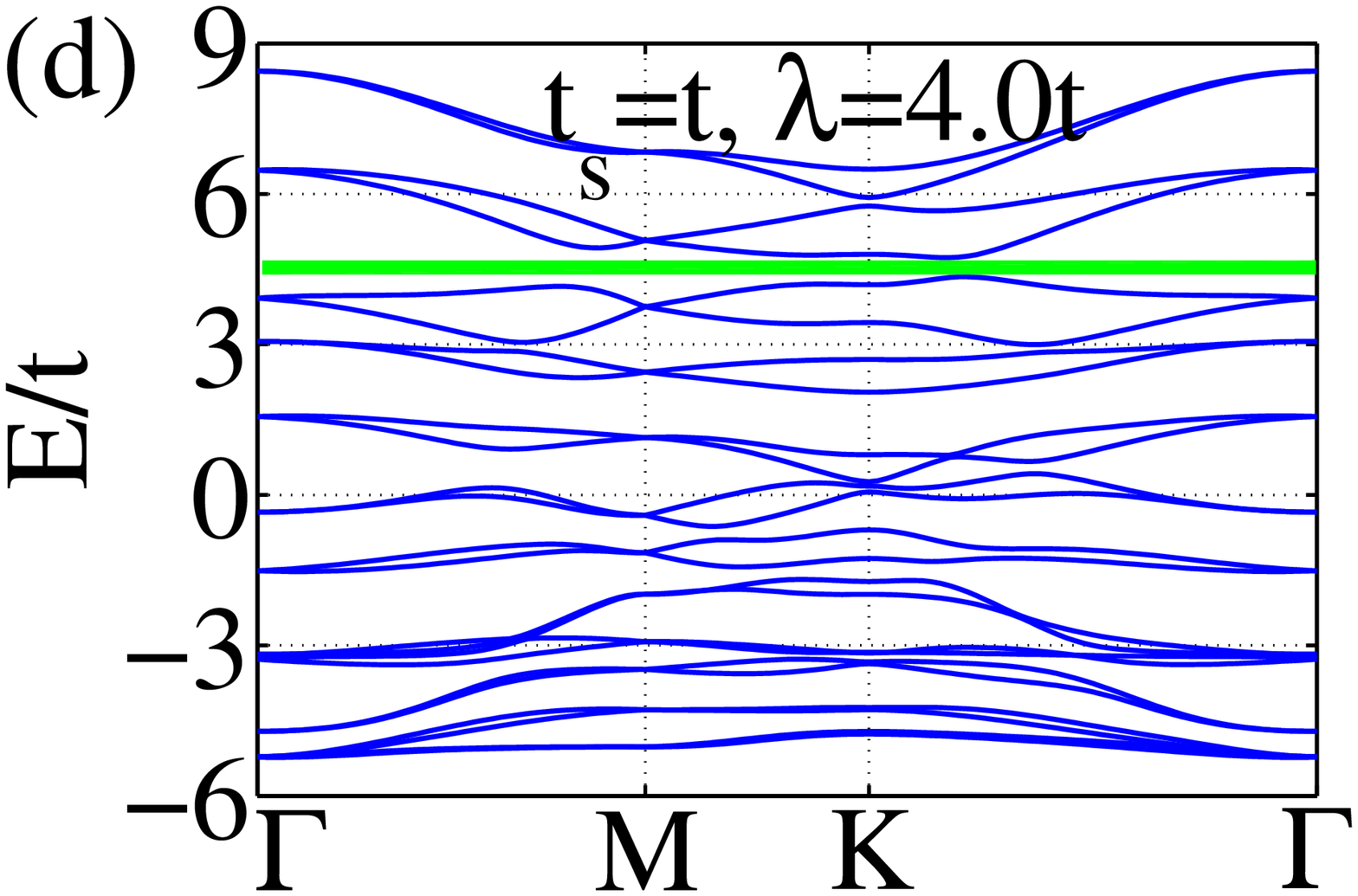}
\caption{(color online) The band structure of Eq.~\eqref{eq:H0} along the high symmetry directions [see Fig.~\ref{fig1b}] of a bilayer with $t_p=-2t_s/3$, $t_s=-t,t$ and $\lambda=2t, 4t$ as shown within the figure.  Green (light gray) lines indicate filling fractions of of a $Z_2$ TI.  The thickness of the line indicates the size of the gap. In the cases shown $Z_2$ TI occur at $t_{2g}$ filling fractions:
(a) $\frac 3 4$; (b) $\frac 3 4$;
(d) $\frac 5 6$.}
\label{fig:bilayer}
\end{figure}

\begin{figure}[htb]
\includegraphics[width=0.49\linewidth]{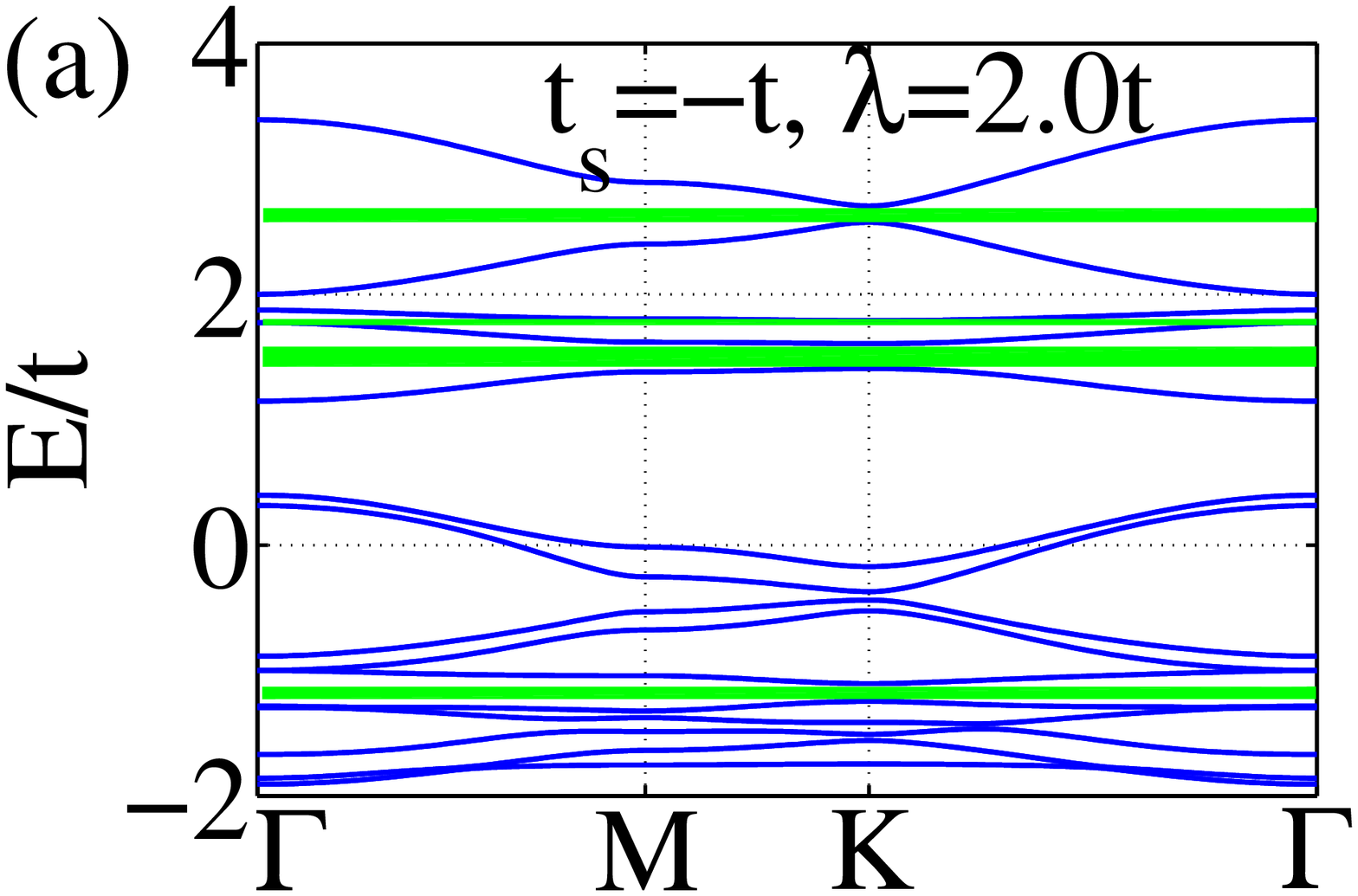}
\includegraphics[width=0.49\linewidth]{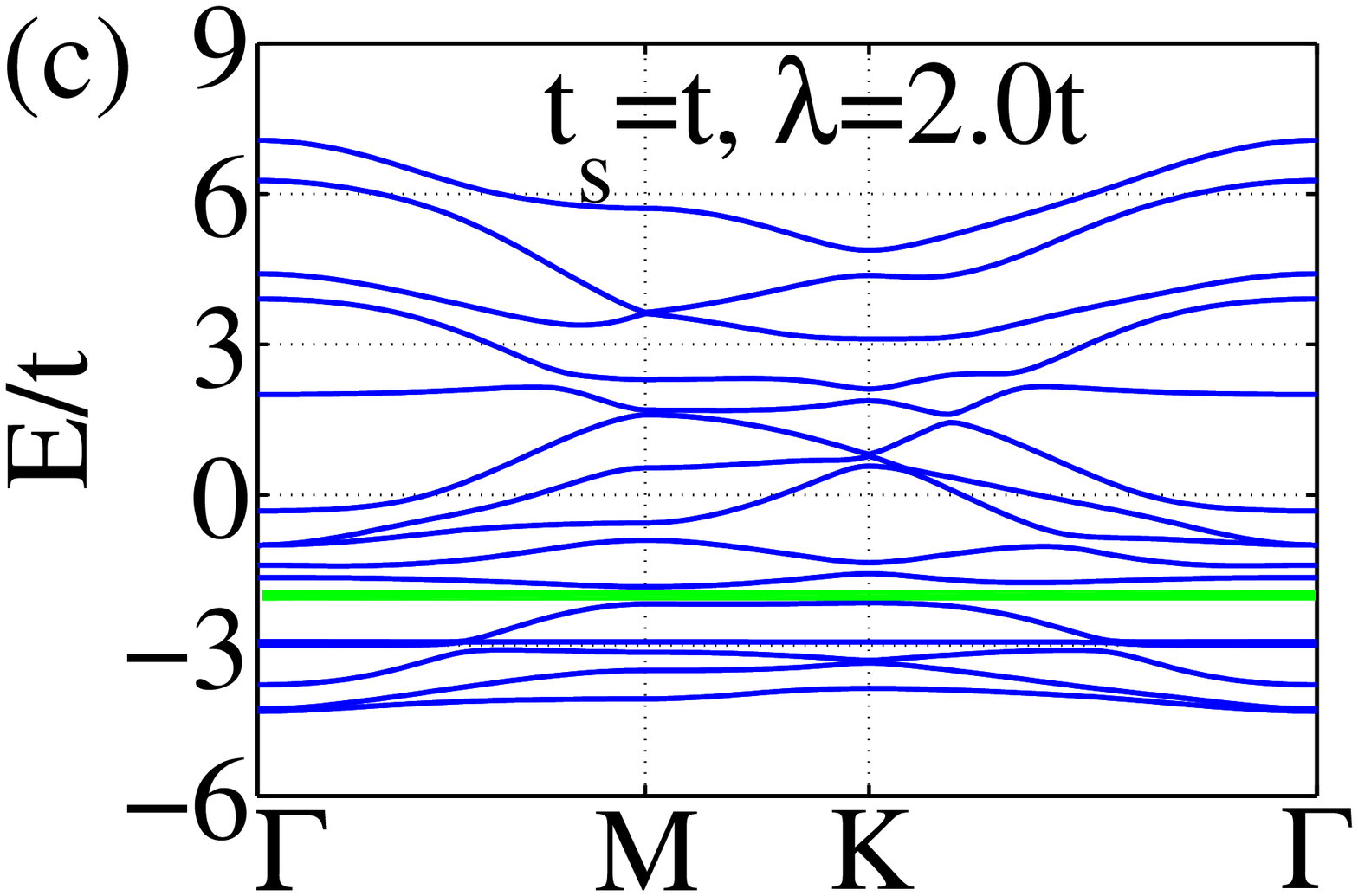}
\includegraphics[width=0.49\linewidth]{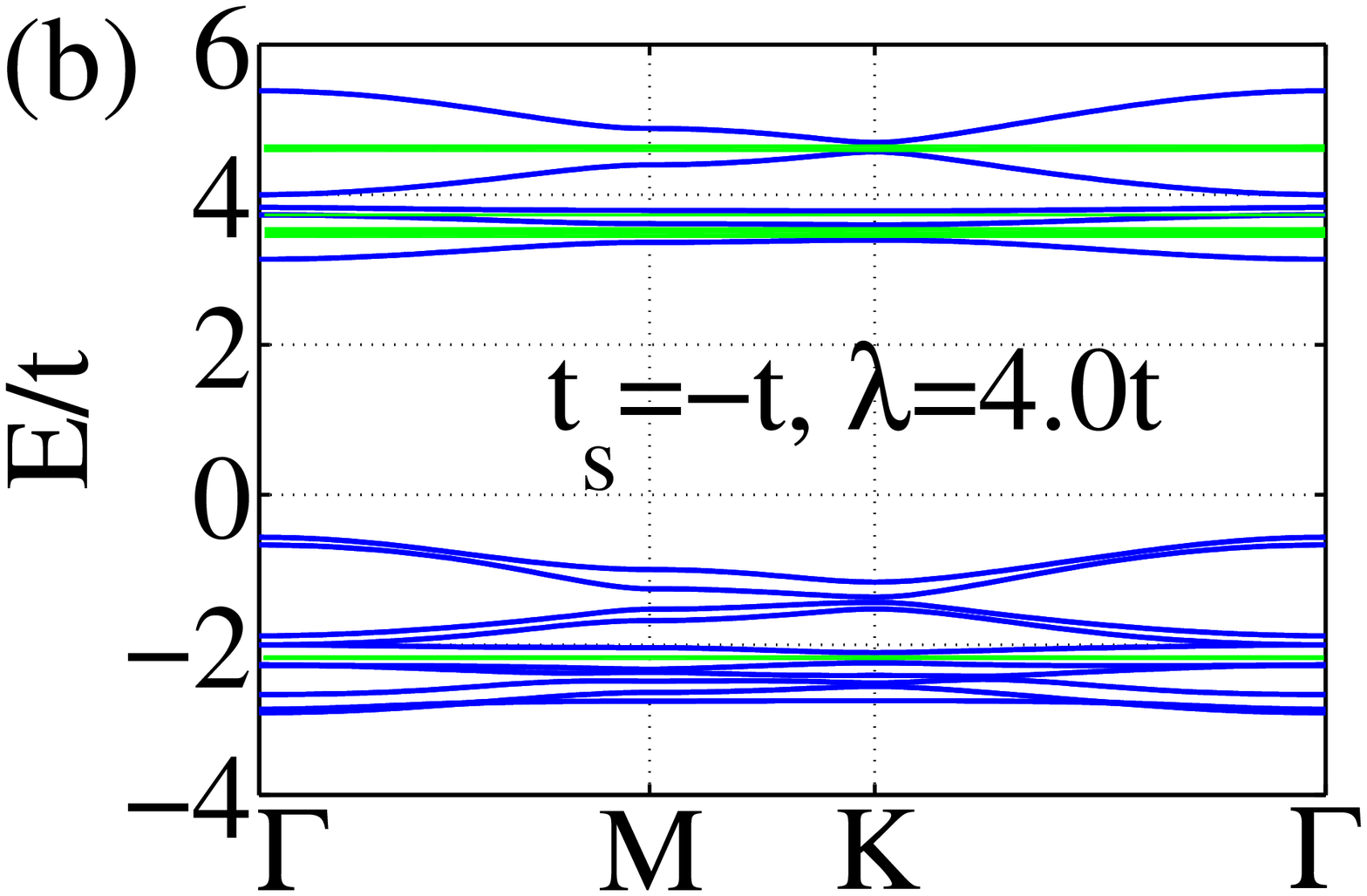}
\includegraphics[width=0.49\linewidth]{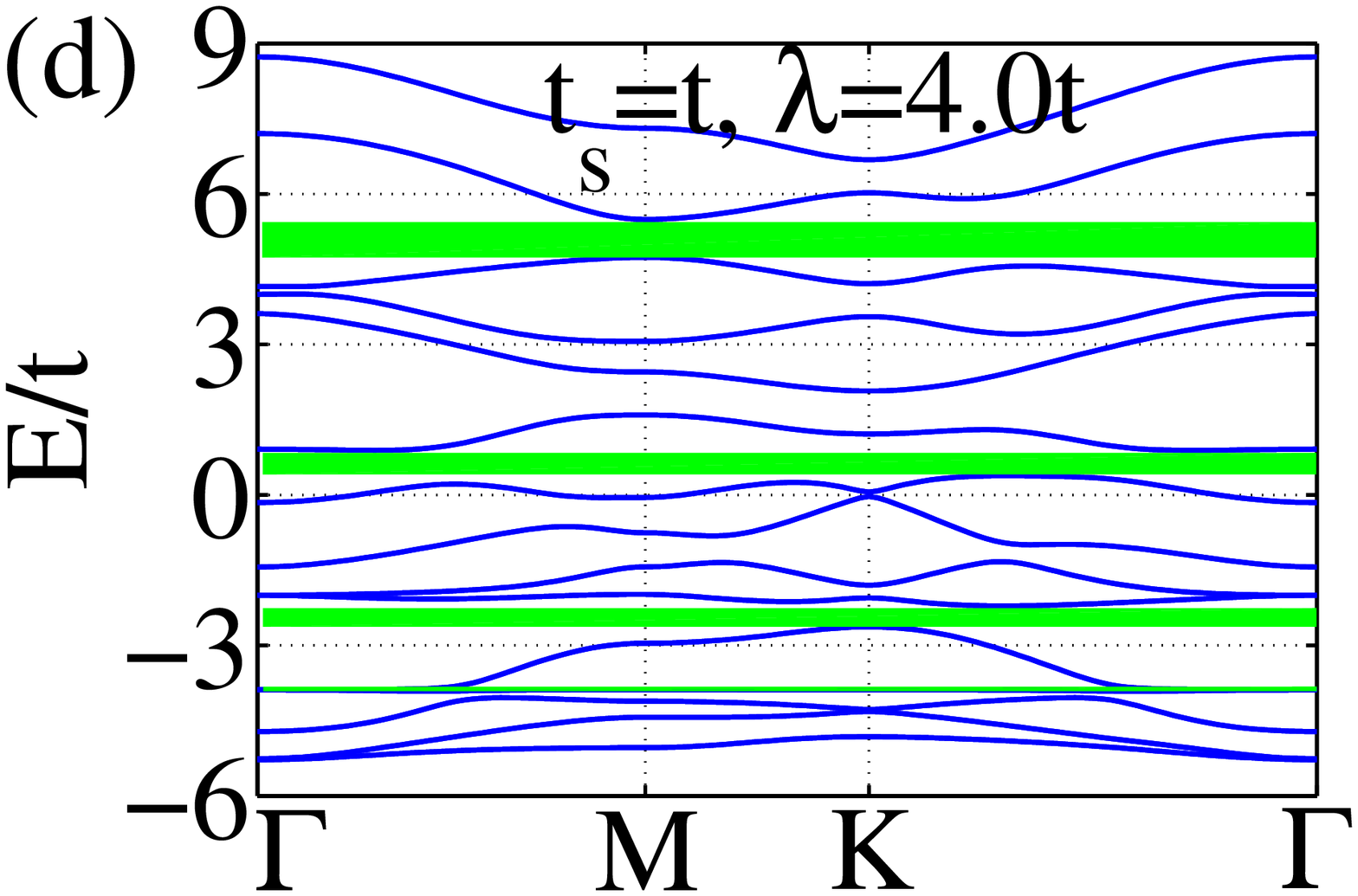}
\caption{(color online) The band structure of Eq.\eqref{eq:H0} along the high symmetry directions [see Fig.\ref{fig1b}] of a triangular-kagome-triangular layer with $t_p=-2t_s/3$, $t_s=-t,t$ and $\lambda=2t, 4t$ as shown within the figure.  Green (light gray) lines indicate filling fractions of of a $Z_2$ TI.  The thickness of the line indicates the size of the gap. In the cases shown $Z_2$ TI occur at $t_{2g}$ filling fractions:
(a) $\frac 1 3$, $\frac {11} {15}$, $\frac {4} {5}$, 
$\frac {14} {15}$; 
(b) $\frac 1 3$, $\frac {11} {15}$, $\frac {4} {5}$, 
$\frac {14} {15}$; 
(c) $\frac 1 3$; 
(d) $\frac 4 {15}$, $\frac 1 3$, $\frac 3 5$, $\frac {13} {15}$.}
\label{fig:TKT}
\end{figure}

\begin{figure}[htb]
\includegraphics[width=0.49\linewidth]{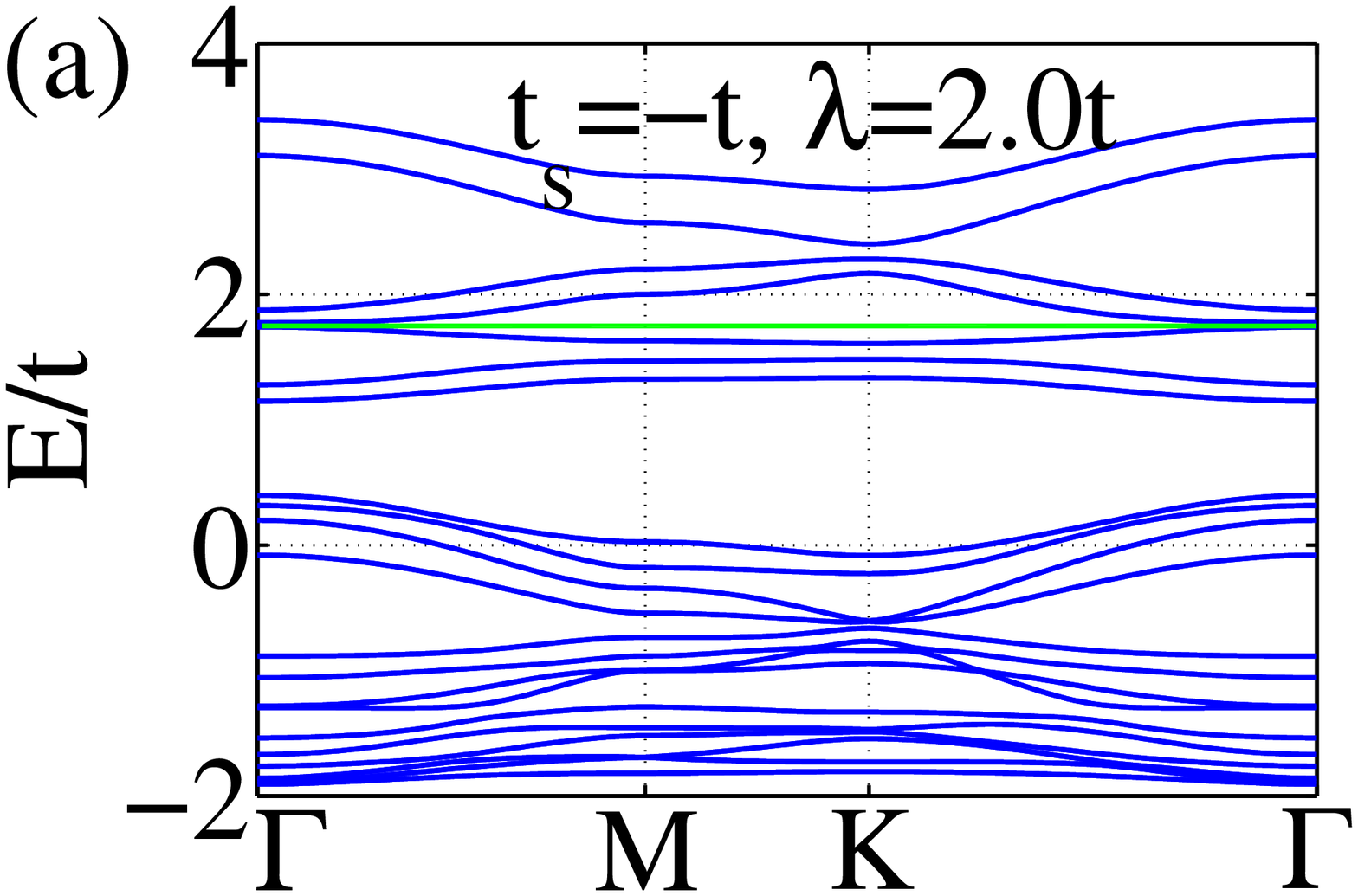}
\includegraphics[width=0.49\linewidth]{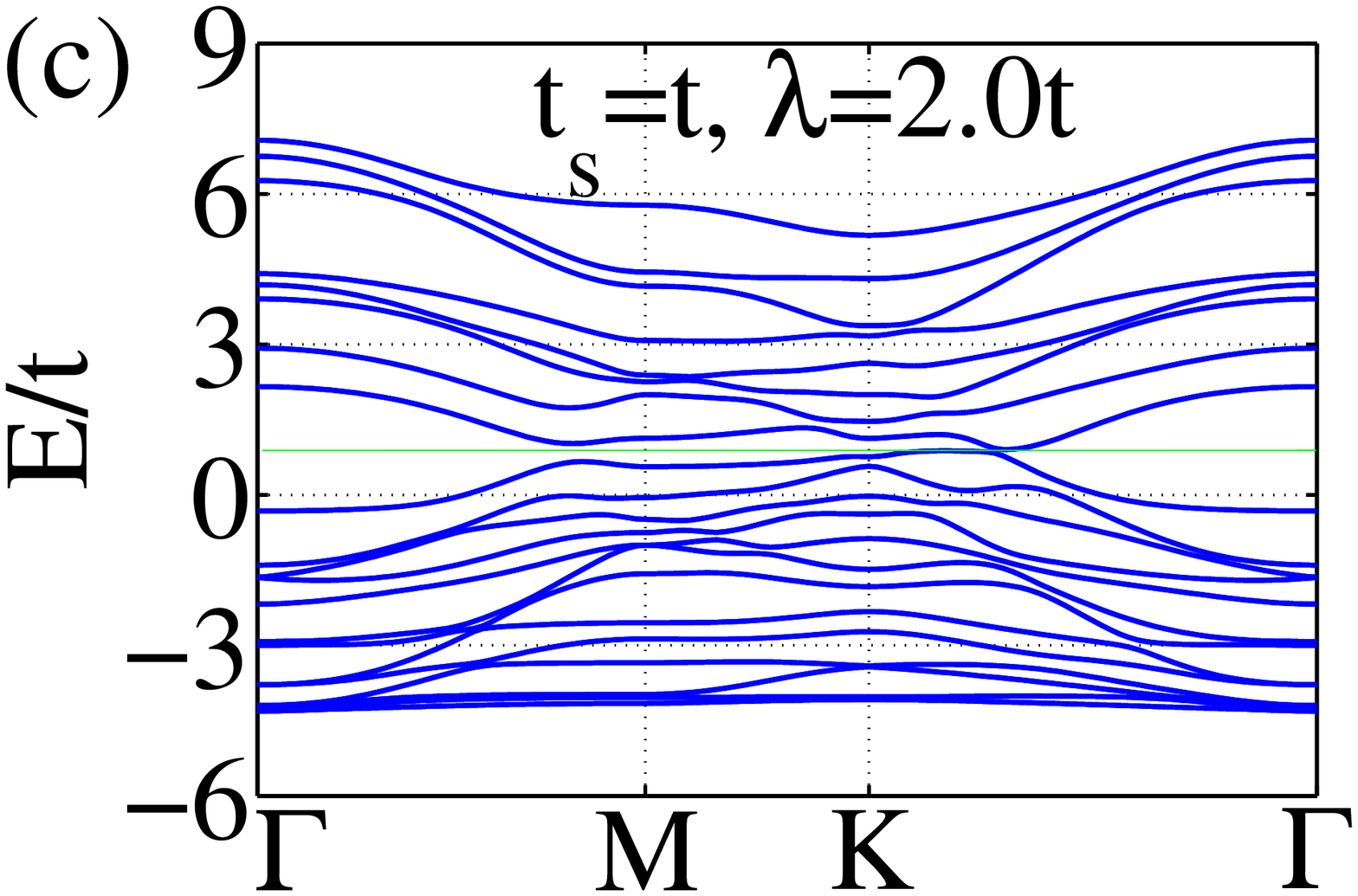}
\includegraphics[width=0.49\linewidth]{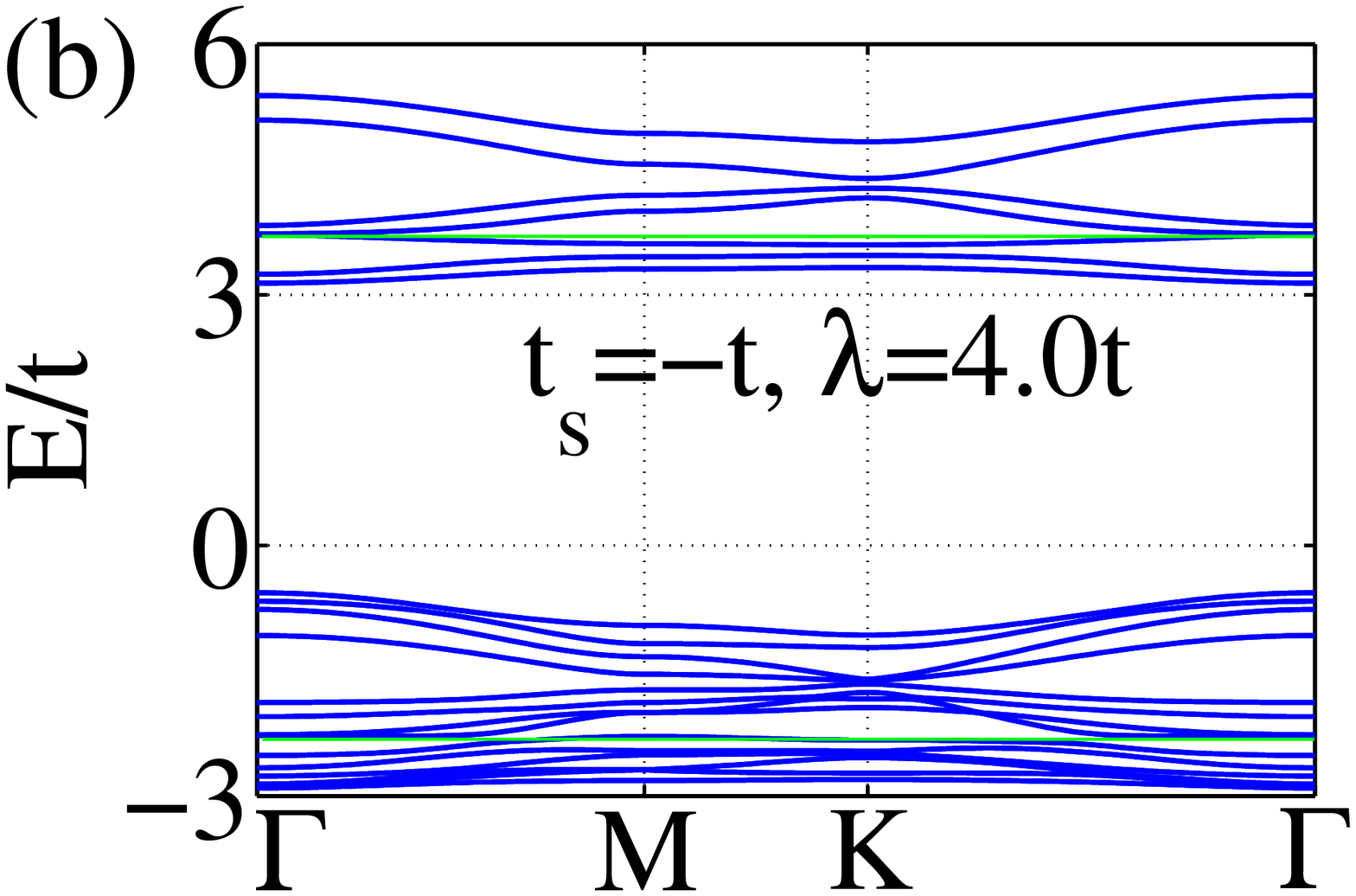}
\includegraphics[width=0.49\linewidth]{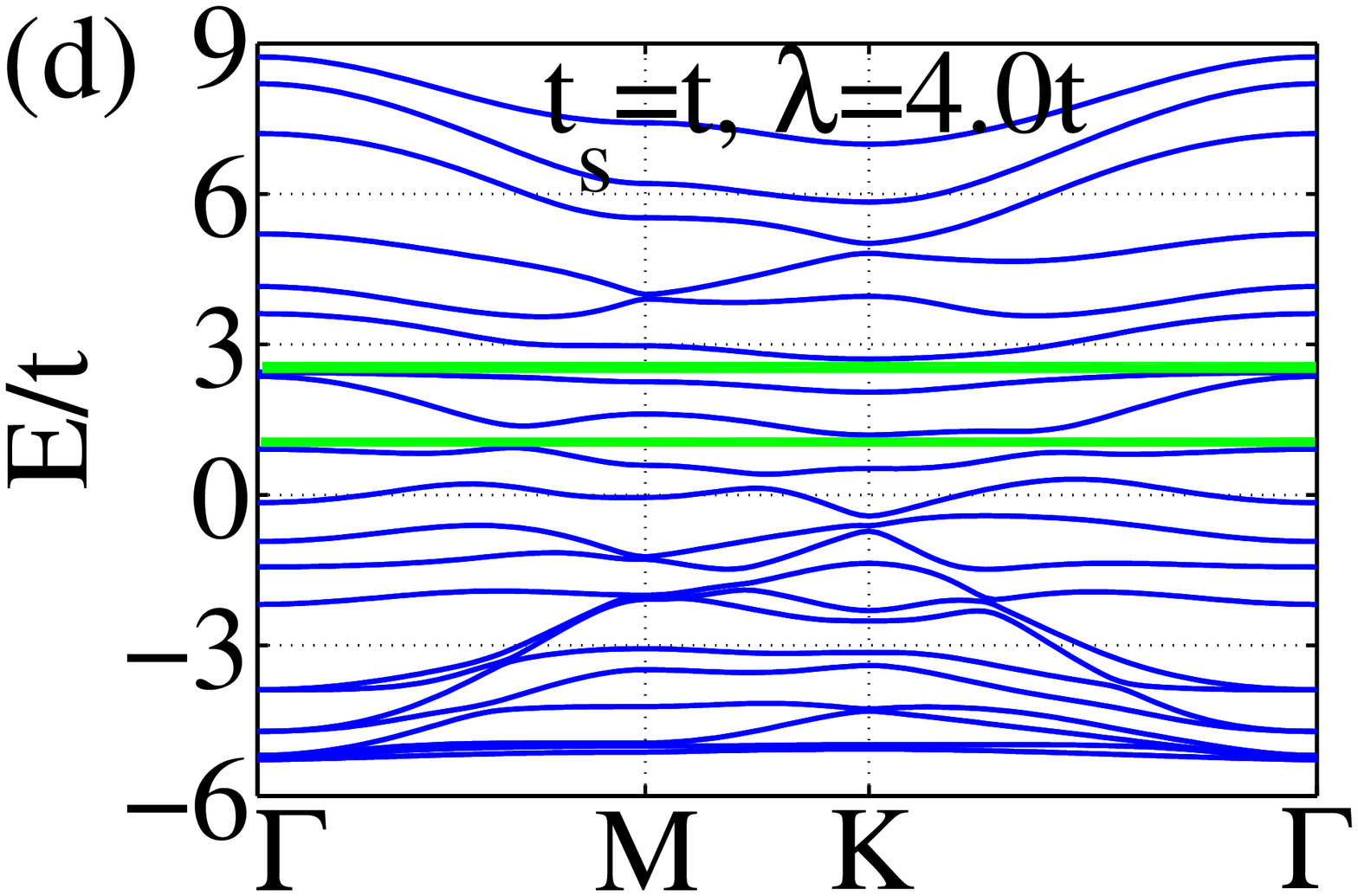}
\caption{(color online) The band structure of Eq.\eqref{eq:H0} along the high symmetry directions [see Fig.\ref{fig1b}] of a kagome-triangular-kagome layer with $t_p=-2t_s/3$, $t_s=-t,t$ and $\lambda=2t, 4t$ as shown within the figure.  Green (light gray) lines indicate filling fractions of of a $Z_2$ TI.  The thickness of the line indicates the size of the gap. In the cases shown $Z_2$ TI occur at $t_{2g}$ filling fractions:
(a) $\frac {17}{21}$;
(b) $\frac 2 7$, $\frac {17}{21}$;
(c) $\frac {13}{21}$;
(d) $\frac {13}{21}$, $\frac {15} {21}$.}
\label{fig:KTK}
\end{figure}

In this work, we will focus our attention primarily on a transition metal ion $d$-shelling filling of 6 electrons or less, which means we will focus on the $t_{2g}$ manifold. In the captions of Figs.~\ref{fig:singlelayer}-\ref{fig:KTK} we have used $t_{2g}$ filling fractions for convenience of counting the bands. This must be converted to a total $d$-shelling filling when considering actual materials by simply writing the $t_{2g}$ filling in the form $y/6$ and then dividing $y$ by 10.  For example, a $t_{2g}$ filling of 5/6 is a total $d$-shell filling of 5/10=1/2, and a $t_{2g}$ filling of 7/9=(2*7/3)/6 is a total $d$-shell filling of (2*7/3)/10=7/15.  We note that a half-filled $d$-shell corresponds to a half-filled $j=1/2$ shell when spin-orbit coupling is present.  

Throughout this work, we will focus on 4$d$ and 5$d$ transition metal ions. For these ions, the spin orbit-coupling is large--up to 2-4 times the value of the nearest neighbor hopping in the structures we consider.\cite{Yang_Kim:prb10,Xiao:nc11} The starting point of our study is a tight binding model of the $d$ electrons with an atomic spin-orbit coupling. Although the 4$d$ and $5d$ orbitals are rather localized compared to $s$ and $p$-type orbitals, they are still rather delocalized compared to 3$d$ orbitals and their direct overlap may be appreciable for the systems of interest here. We therefore include it\cite{Go:prl12,Witczak:prb12} in addition to the indirect hopping via the oxygen orbitals.\cite{Pesin:np10,Yang_Kim:prb10,Kargarian:prb11}  

The non-interacting Hamiltonian we consider is 
\begin{equation}
\label{eq:H0}
H_0=\sum_{\langle i,j \rangle,\alpha,\beta}t_{i\alpha, j\beta}c_{i\alpha}^{\dagger}c_{j\beta}-
\lambda\sum_i {\bf l}_i \cdot {\bf s}_i,
\end{equation}
where the $d$-orbital hopping takes the form\cite{Go:prl12,Witczak:prb12} 
\begin{equation}
\label{eq:t}
t_{i\alpha, j\beta}=t_{i\alpha, j\beta}^{in}+t_{i\alpha, j\beta}^{dir}.
\end{equation}
The hopping parameter Eq.~\eqref{eq:t} contains both an indirect and a direct hopping term between the $d$-orbitals. The details of the indirect hopping on the pyrochlores A$_2$B$_2$O$_7$, which is mediated by intermediate oxygen $p$-states, are given in Refs.~[\onlinecite{Pesin:np10,Yang_Kim:prb10,Kargarian:prb11}] and the details of the direct hopping are summarized in App.~\ref{app:tb}. The operator $c_{i\alpha}^{\dagger}$ creates a $d$-electron on site $i$ with spin and $t_{2g}$ orbital state $\alpha$, while the operator $c_{j\beta}$ annihilates a $d$-orbital electron on site $j$ with spin and $t_{2g}$ orbital state $\beta$. The matrix elements Eq.~\eqref{eq:t} are diagonal in the spin degree of freedom. In the presence of spin-orbit coupling in Eq.~\eqref{eq:H0}, it is however convenient to rotate the spin quantization axis to the local frame, in which case the effective hopping parameters become spin-dependent. Here, $\lambda>0$ is the intrinsic spin-orbit coupling in the system which acts within the $t_{2g}$ manifold so $|{\bf l}|=1$, and ${\bf s}_i$ is the spin of the electron in a $t_{2g}$ $d$-orbital on site $i$.\cite{Pesin:np10,Yang_Kim:prb10,Kargarian:prb11} In the 5$d$ oxides, the strength of the spin-orbit coupling is estimated to be 0.2-0.7~eV and the hopping strength is on the order of 0.4-0.6~eV.\cite{Shitade:prl09,Kargarian:prb11} In order to cover the physical ranges and emphasize the physics associated with strong spin-orbit coupling, we will consider $\lambda/t$ in the range of 2-4, with $t$ the strength of the hopping between $d$-orbitals mediated by the oxygen (indirect hopping).\cite{Pesin:np10,Yang_Kim:prb10,Kargarian:prb11} The hopping $t$ will set our reference energy scale in this paper.

\begin{figure}
\includegraphics[width=0.8\linewidth]{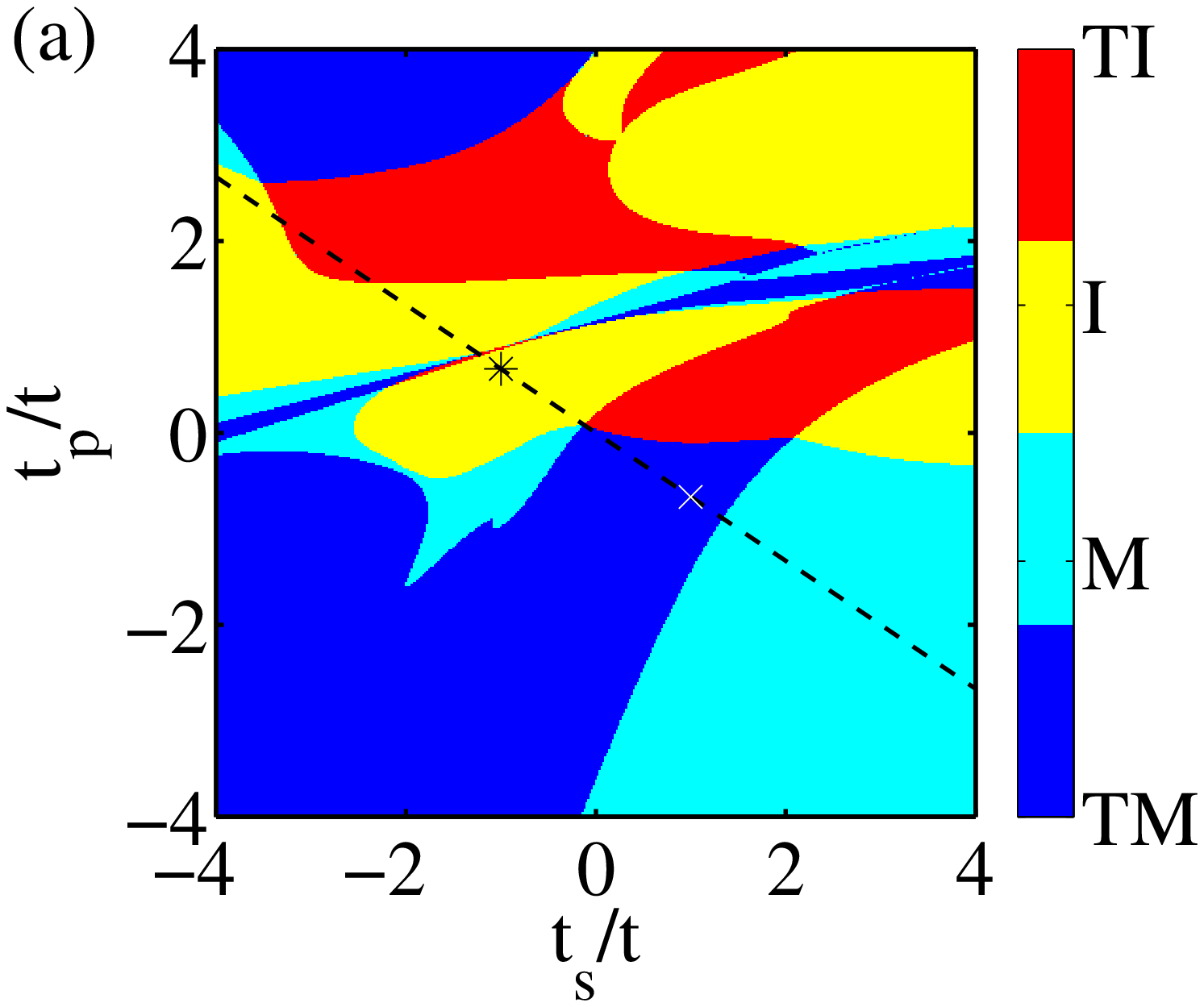}
\includegraphics[width=0.8\linewidth]{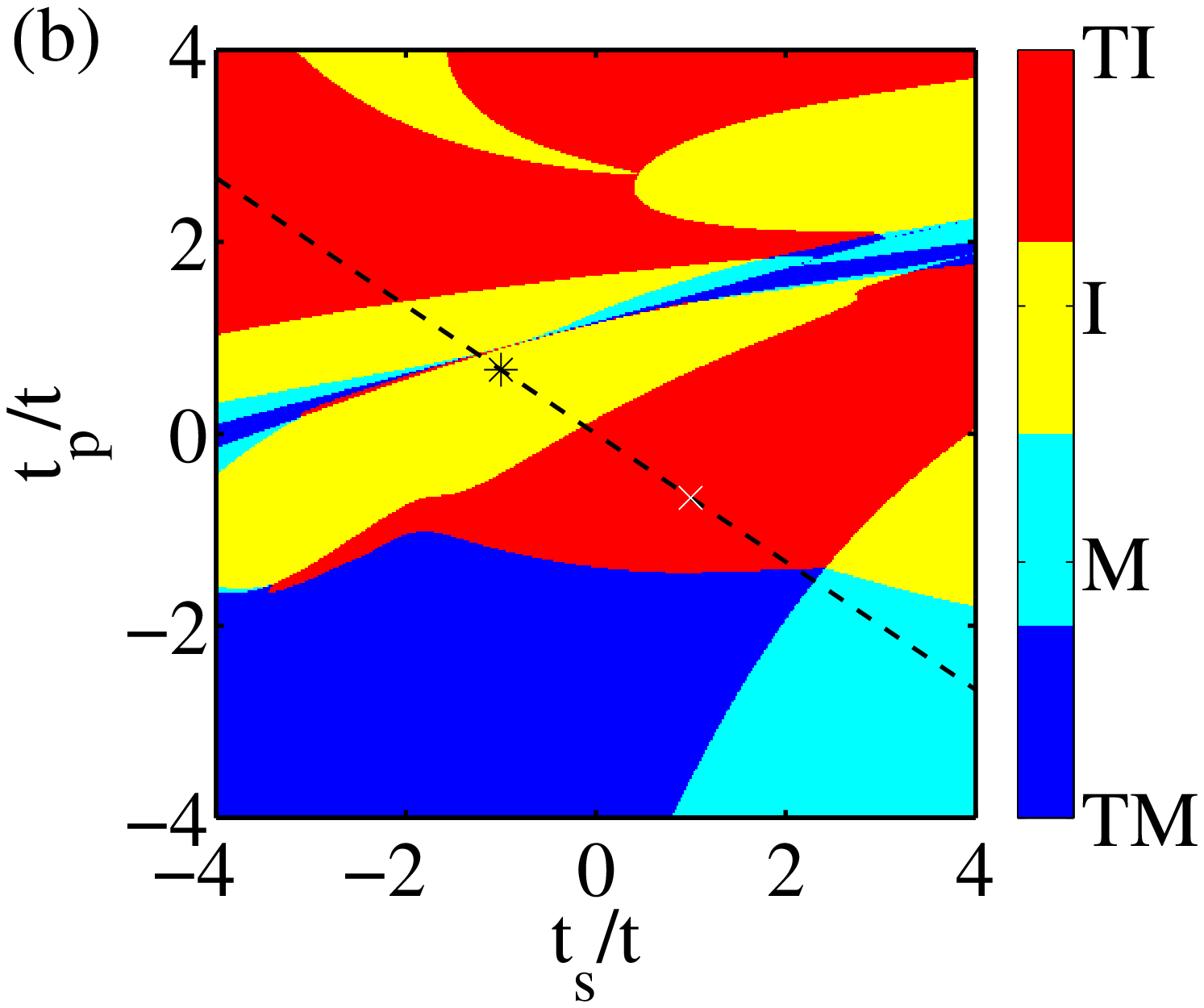}
\caption{(color online) Phase diagrams for the bilayer case in Fig.~\ref{fig2} at a $t_{2g}$ filling of 5/6 and a total $d$-shell filling of 1/2 for (a) $\lambda=2t$ and (b) $\lambda=4t$. The different colors represent different
phases: TM is a topological metal (finite direct band gap, negative indirect gap, with non-trivial $Z_2$ index for {\em fully} occupied bands), M is a metal, I is a trivial insulator, and TI is a topological 
insulator. The black line is $t_p=-2t_s/3$. The $\ast$ represent (-1, 2/3) and the $\times$ represent
(1, -2/3), corresponding to the two values of $t_s=\pm 1$ we considered in Figs.~\ref{fig:singlelayer}-\ref{fig:KTK}.}
\label{fig:phase_U_0}
\end{figure}

The hopping amplitude in Eq.~\eqref{eq:t} contains a direct $d$-$d$ hopping, $t_{i\alpha, j\beta}^{dir}$, in addition to the indirect hopping via the oxygen orbitals.\cite{Go:prl12,Witczak:prb12} The direct hopping is parameterized by the strength of the $\sigma$-bonds, $t_s$, and the $\pi$-bonds, $t_p$ (see App.~\ref{app:tb}). Following Refs.~[\onlinecite{Go:prl12,Witczak:prb12}], we consider a set of representative ratios to explore a realistic parameter space: We set 
 $t_p=-2t_s/3$ and consider the cases of $t_s=-t$ and $t_s=t$. For the bilayer system shown in Fig.~\ref{fig2}, the phase diagrams in Fig.~\ref{fig:phase_U_0} illustrate the evolution of the ground state of the system as a function of $t_p$ and $t_s$ for two representative values of spin-orbit coupling at a total $d$-shell filling of 5 electrons.

Our main results for the band structure of Eq.~\eqref{eq:H0} are shown in Figs.~\ref{fig:singlelayer}-\ref{fig:KTK}. Light green (gray) lines indicate filling fractions for which the system is a $Z_2$ TI, and the width of the line is proportional to the size of the gap at that filling. We note that the bilayer system is not inversion symmetric so one must use a more general formulation\cite{Fukui:jpsj07} to compute the $Z_2$ invariant than the parity eigenvalues at time-reversal invariant momenta.\cite{Fu:prb07} The numerical values of the $t_{2g}$ filling are given in the figure captions, and can be converted to the total $d$-shell filling as described a few paragraphs above. This information can be used by experimental groups to help identify candidate materials for the   A$_2$B'$_2$O$_7$/A$_2$B$_2$O$_7$/A$_2$B'$_2$O$_7$ sandwich structure.  We note that in order to obtain some of the desired filling fractions for topological phases, it may be necessary to use two different A-site elements to create structures of the form A$_2$B'$_2$O$_7$/A$_{2(1-x)}$A'$_{2x}$B$_2$O$_7$/A$_2$B'$_2$O$_7$ and tune $x$ appropriately. This is necessary in order to obtain a {\em naive} ``fractional" filling of the $t_{2g}$-orbitals of the type 7/9, 8/9, 3/4, etc. in which the fraction is not of the form $n$/6 for some integer $n$.

A few general patterns emerge from Figs.~\ref{fig:singlelayer}-\ref{fig:KTK}. First, while for the case of $t_s=-t$ the $j=1/2$ and $j=3/2$ manifolds are already clearly separated for $\lambda=2t$, the $\lambda\to\infty$ limit is well approximated by $\lambda=4t$ because the separation of the upper $j=1/2$ and lower $j=3/2$ manifold is large compared to the typical band widths already for $\lambda=4t$.  Second, for $t_s=t$ the $j=1/2$ and $j=3/2$ manifolds are not well separated even for $\lambda=4t$.  Third, it appears that both the cases $t_s=-t$ and $t_s=t$ are equally favorable for finding topological phases, both in terms of the number of filling fractions that admit a $Z_2$ TI and in terms of the sizes of the bulk gaps that the TI would possess. Fourth, we find that the simpler sandwich structure with only a single $A$-site element will only realize a $Z_2$ TI in the bilayer at $t_{2g}$ filling fraction 5/6 (a half-filled $d$ shell, and a half-filled $j=1/2$ manifold in the $\lambda \to \infty$ limit). This is because the bilayer is the only system with an even number of lattice sites in the unit cell and hence, the band structure can be gapped for a half-filled $d$ shell. If one expands the class of systems considered to include the A$_2$B'$_2$O$_7$/A$_{2(1-x)}$A'$_{2x}$B$_2$O$_7$/A$_2$B'$_2$O$_7$-type structure, our results suggest that one might be able to find $Z_2$ TI with a larger bulk gap.  We caution, however, that a more accurate estimate of the bandstructure should be obtained within density functional theory.

As the bilayer shown in Fig.~\ref{fig2} is the only system that exhibits a $Z_2$ TI for for a sandwich structure of the form A$_2$B'$_2$O$_7$/A$_2$B$_2$O$_7$/A$_2$B'$_2$O$_7$ in the parameter ranges of the Hamiltonian \eqref{eq:H0} we considered, we have presented more complete phase diagrams at $t_{2g}$ filling 5/6 (total $d$ shell {\em and} $j=1/2$ manifold half-filled) in Fig.~\ref{fig:phase_U_0}. Overall, the phase structure as a function of the parameters $t_s$ and $t_p$ is complicated, as might have been anticipated from earlier work.\cite{Hu:prb11} However, it is clear that a larger spin-orbit coupling does favor expanded regions of a TI phase.

Thus far, we have only considered the non-interacting Hamiltonian \eqref{eq:H0}. We will now add interactions to the Hamiltonian and see how this changes the overall picture presented above. Our main result is that interactions will drive magnetic order in the system which can give rise to a zero magnetic field quantum Hall state known as a Chern insulator or quantum anomalous Hall state.

\section{The interacting cases}
\label{sec:interacting}

\begin{figure}
\includegraphics[width=\linewidth]{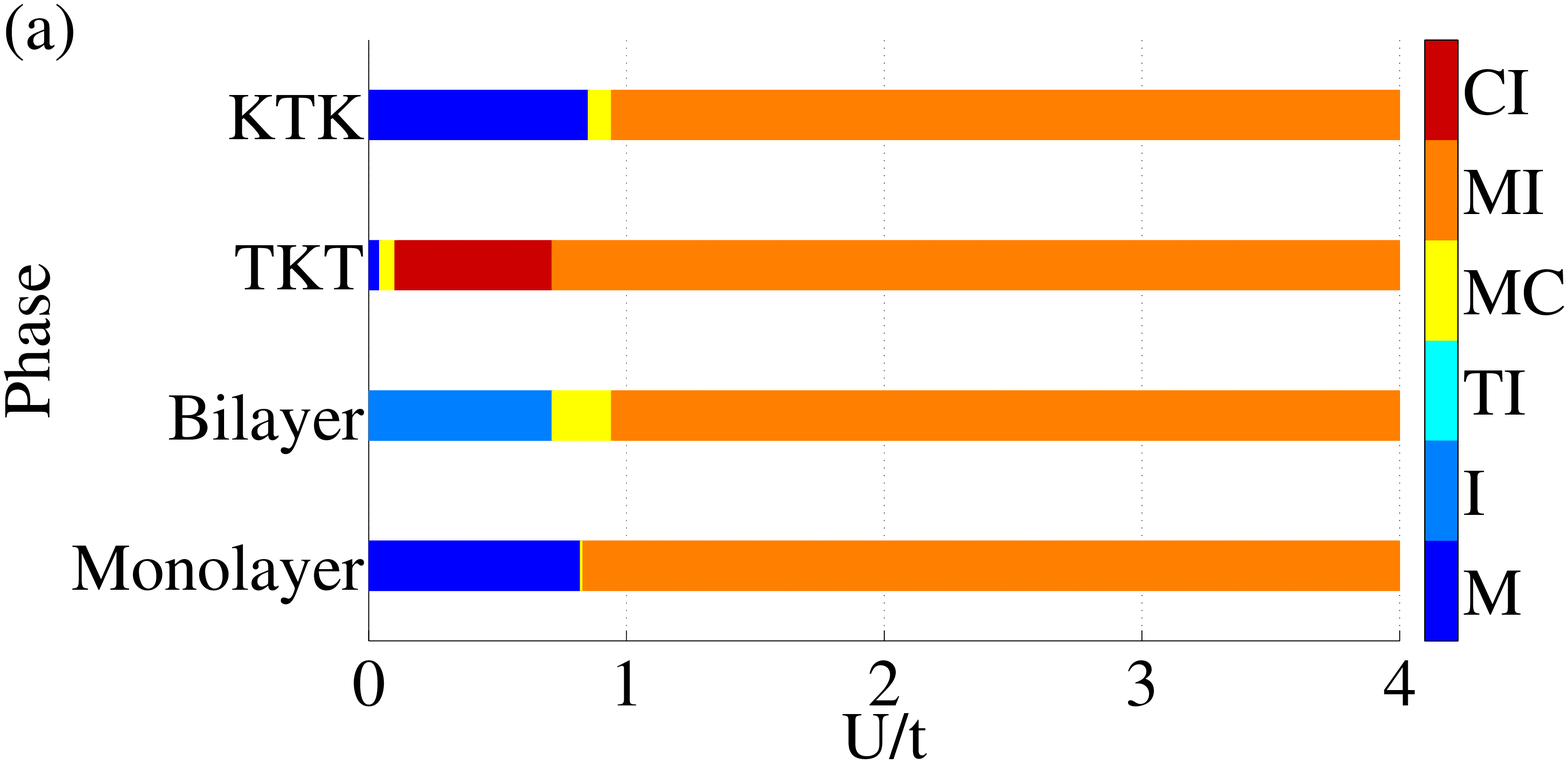}
\includegraphics[width=\linewidth]{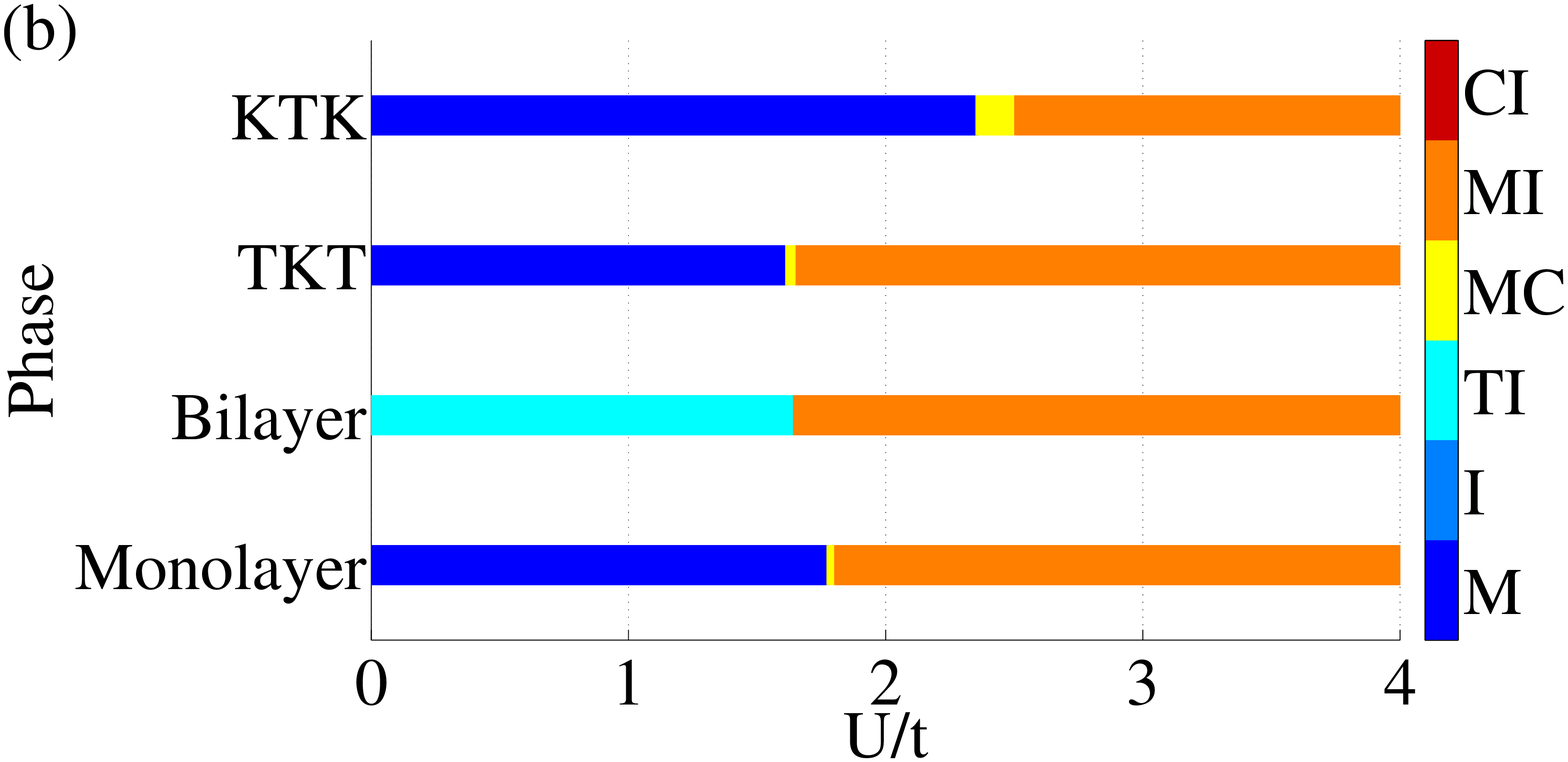}
\includegraphics[width=\linewidth]{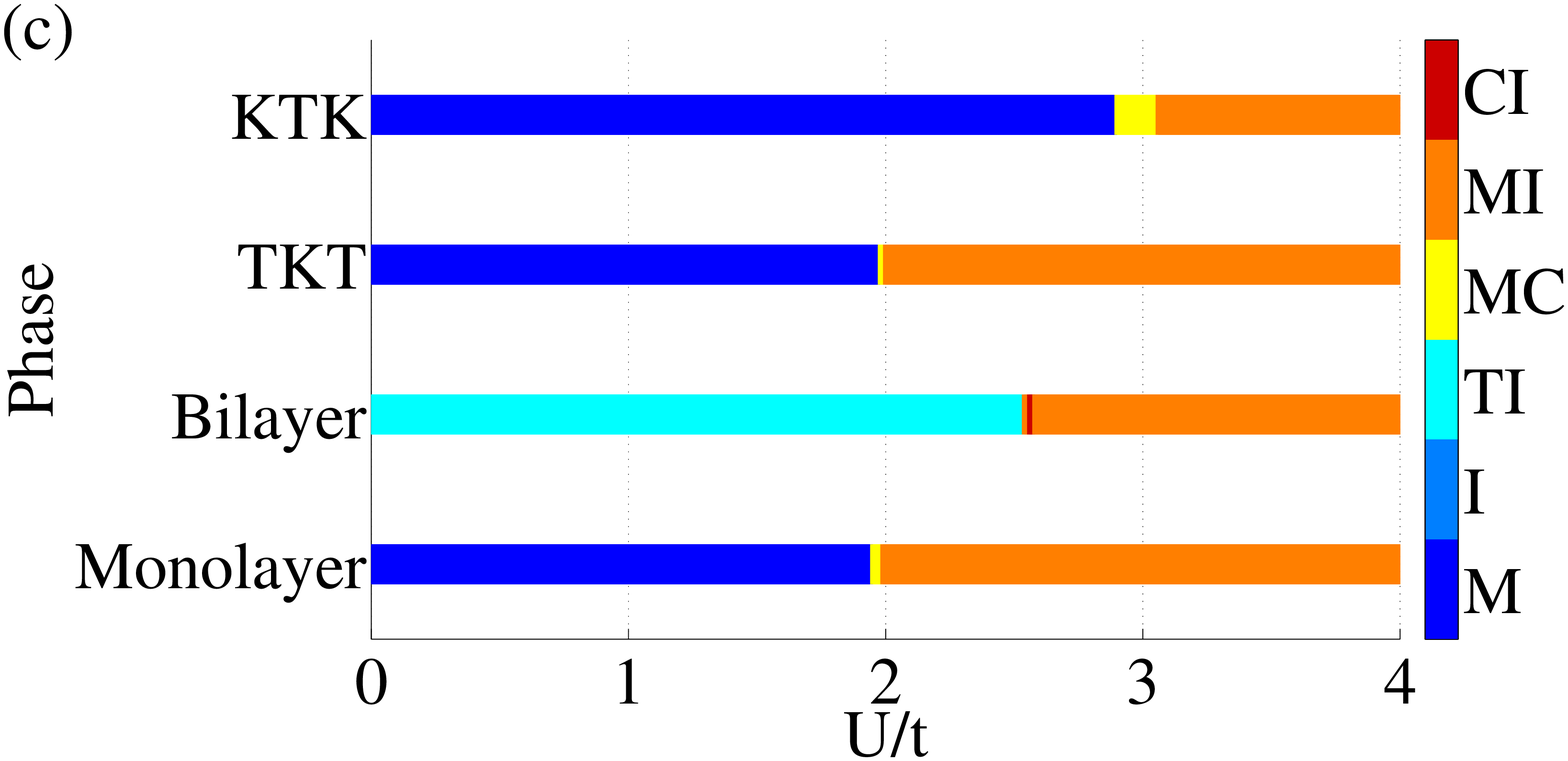}
\caption{(color online) The phases of the interacting model \eqref{eq:H0} and \eqref{eq:H_U} with (a) $t_s=-t$, (b) $t_s=0.25t$, and (c) $t_s=t$ for a $t_{2g}$ filling of 5/6 (a total $d$-shell and $j=1/2$ filling of 1/2) in the limit of strong spin-orbit coupling. Here M is a metal, I is an insulator, TI is a topological 
insulator,  MC is a  magnetic metal,  MI is a  magnetic insulator, and
CI is a Chern insulator (zero magnetic field quantum Hall state). The most robust Chern insulator is in the TKT structure for $t_s=-t$.  Note that there is a small region of the Chern insulator in the bilayer system for $t_s=t$ around $U/t\approx 2.5$ near the end of the TI phase.}
\label{fig:U_phase}
\end{figure}

Many of the bulk pyrochlore transition metal oxides exhibit magnetic phases\cite{Gardner:rmp10} and these phases are ultimately driven by electronic interactions.  It is thus important to consider the effect of interactions on the non-interacting results we presented in Sec.~\ref{sec:non-interacting}. To do so, we will add a Hubbard $U$ term to Eq.~\eqref{eq:H0} and consider only the $j=1/2$ manifold by assuming the $j=3/2$ manifold is well separated in energy ({\it i.e.}, spacing is large compared to $U$),
\begin{equation}
\label{eq:H_U}
H_U=U\sum n_{i\uparrow}n_{i\downarrow}.
\end{equation}
where $\uparrow,\downarrow$ refer the $j^z$ projections in the $j=1/2$ manifold. In our calculations, we further assume a half-filled $d$-shell (equivalent to a half-filled $j=1/2$ manifold). We study the problem within the Hartree-Fock approximation at zero temperature. In this approximation, $H_U\to -U \sum_i (2\langle {\bf j}_i\rangle \cdot  {\bf j}_i-\langle {\bf j}_i\rangle^2)$,
where ${\bf j}_i=\frac{1}{2}\sum_{\alpha\beta= \uparrow,\downarrow} c_{i\alpha}^\dagger {\bf \sigma}_{\alpha\beta}c_{i\beta}$.  The ${\bf j}_i$ is proportional to the spontaneous local moment of the $d$-orbitals.\cite{Witczak:prb12}  We mesh the two-dimensional Brillouin zone with a 150 by 150 grid of points and compute the band structure and the unrestricted local magnetic moments $\langle {\bf j}\rangle$ self-consistently. Details of the calculation are given in App.~\ref{app:HF}. The resulting phase diagrams for a half-filled $d$-shell are shown in Fig.~\ref{fig:U_phase}. For gapped phases with zero magnetic moments (where time-reversal symmetry is preserved) we compute the $Z_2$ invariant from the formulation of Ref.~[\onlinecite{Fukui:jpsj07}]. For the gapped phases that break time-reversal symmetry, we compute the Chern number.\cite{Fukui:jpsj05}  When the Chern number is nonzero (due to the non-trivial topological feature of the bulk bands) the system is guaranteed to possess gapless edge modes which support a quantum Hall effect.

\begin{figure}
\includegraphics[width=\linewidth]{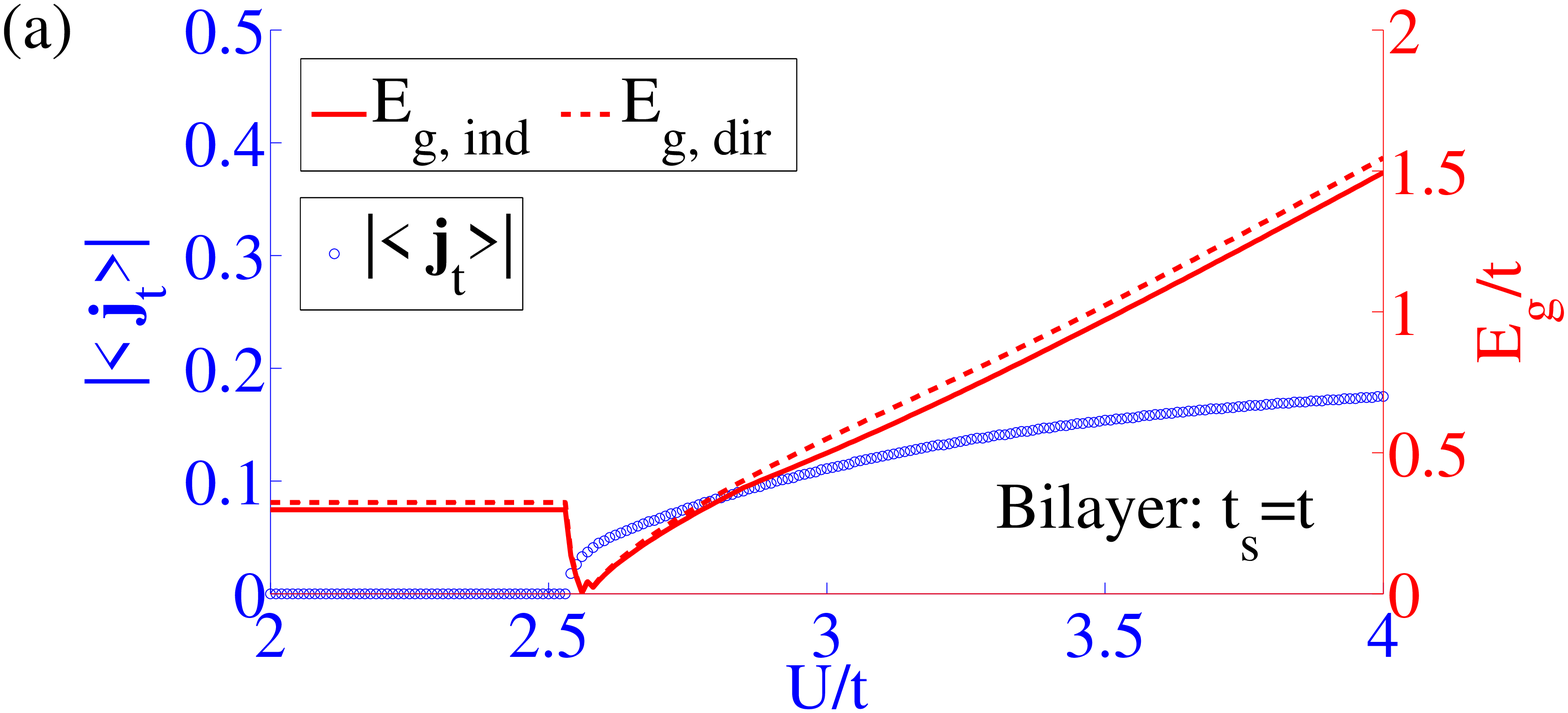}
\includegraphics[width=\linewidth]{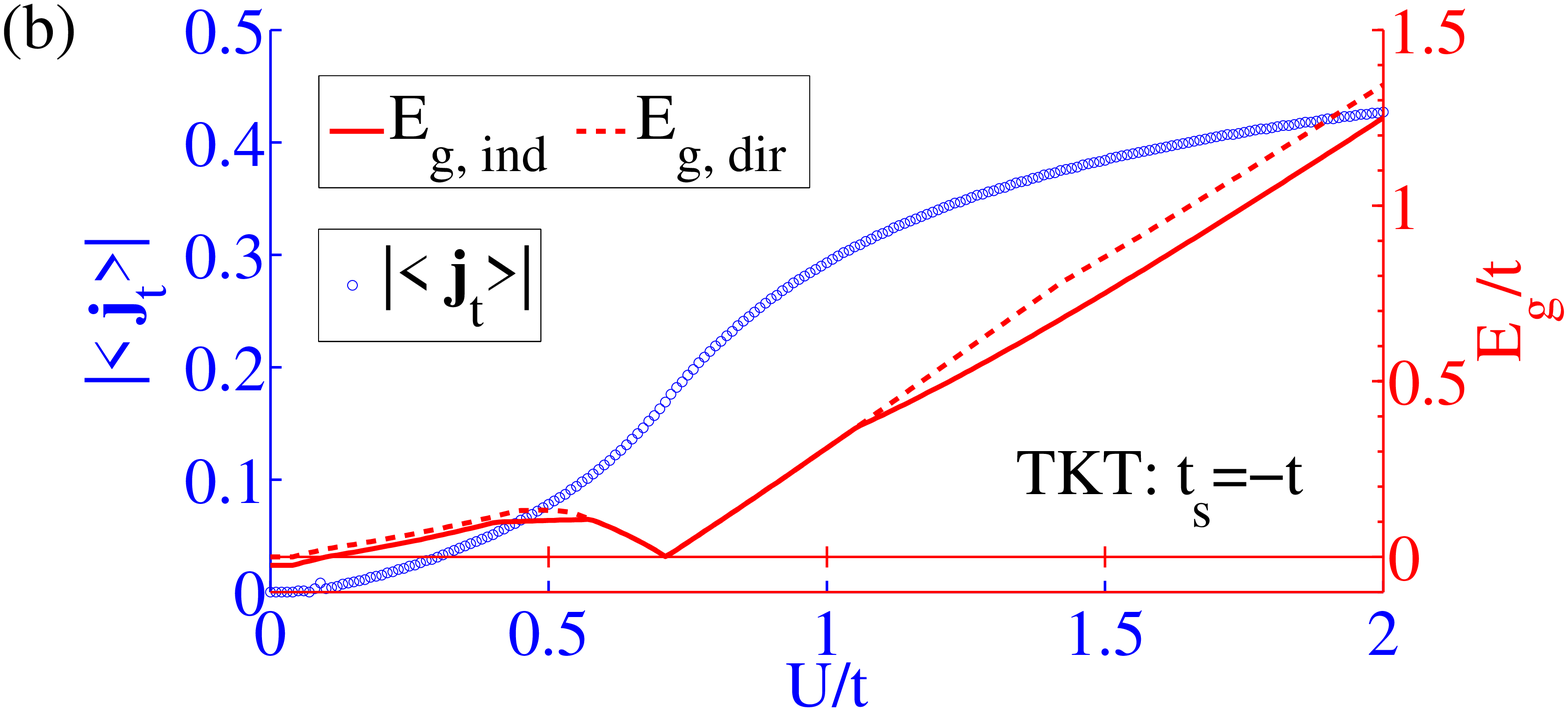}
\caption{(color online) Top: The magnitude of the net magnetic moment, $|\langle {\bf j}_t\rangle|$, and the energy gap, $E_g/t$ ($E_{g,ind}$ denotes the indirect gap and $E_{g,dir}$ denotes the direct gap), as a function of $U/t$ for the bilayer system with parameters corresponding to Fig.~\ref{fig:U_phase}. Bottom:  The same for the TKT system.}
\label{fig:magap}
\end{figure}

\begin{figure}
\includegraphics[width=0.49\linewidth]{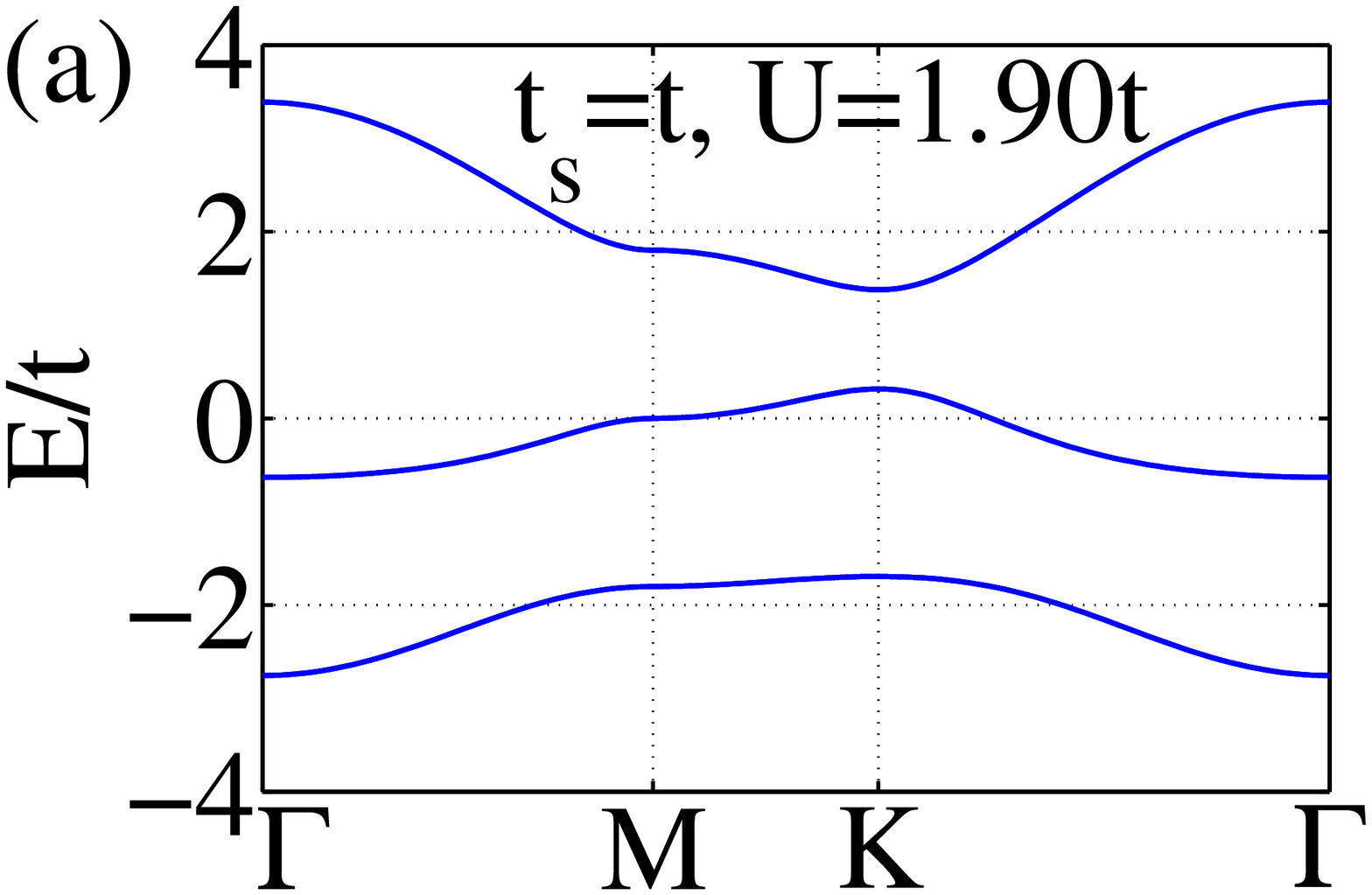}
\includegraphics[width=0.49\linewidth]{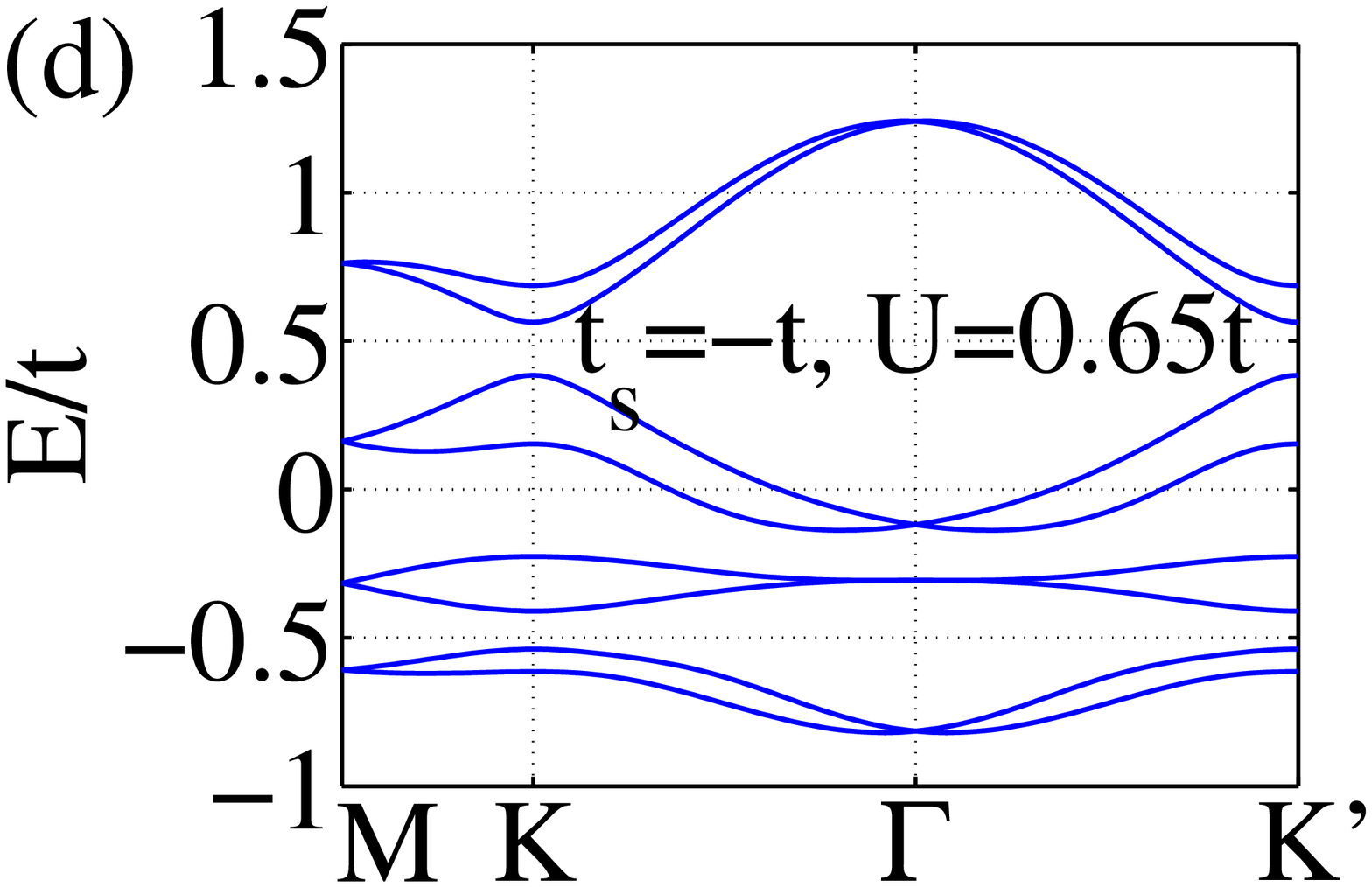}
\includegraphics[width=0.49\linewidth]{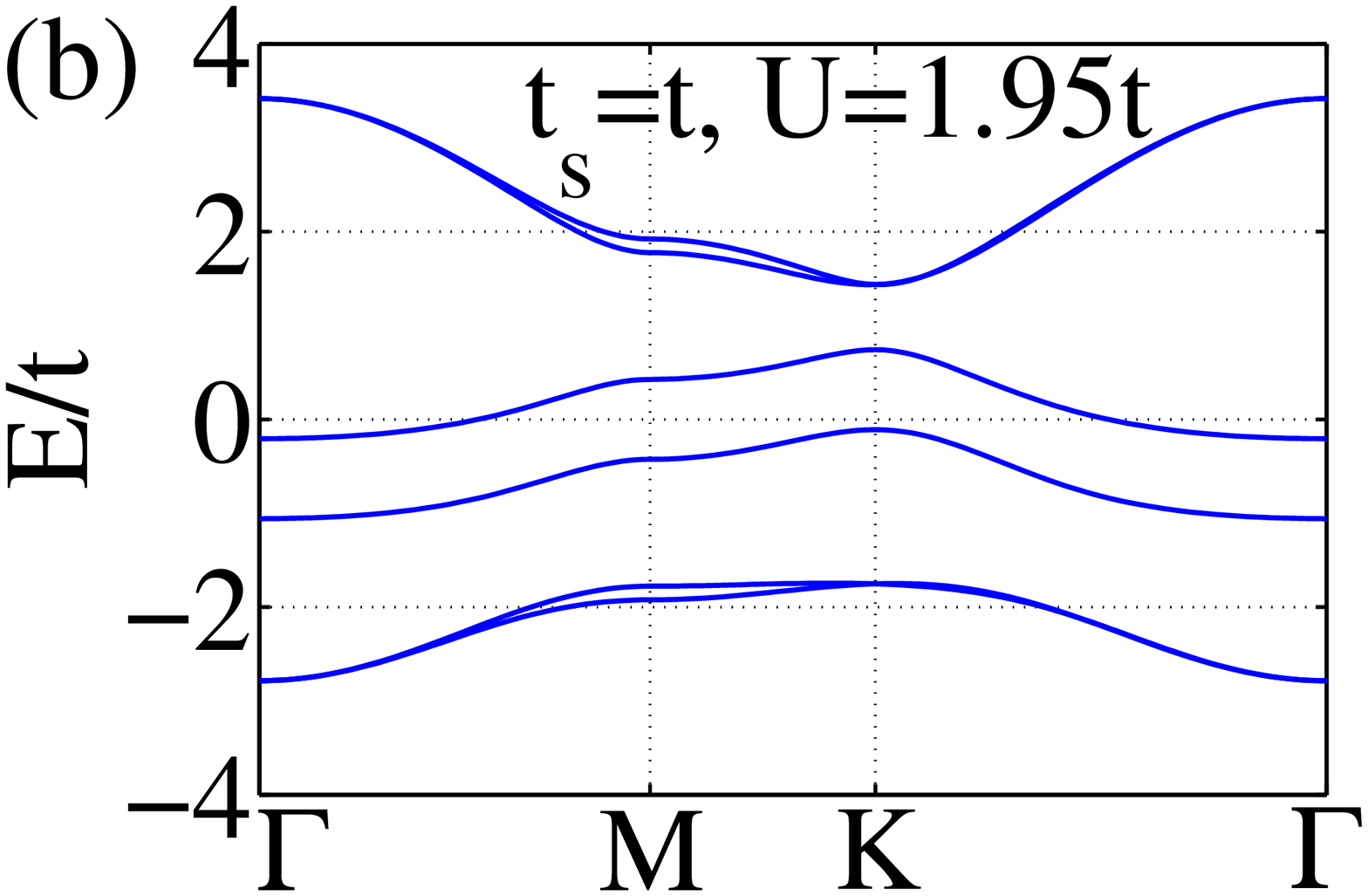}
\includegraphics[width=0.49\linewidth]{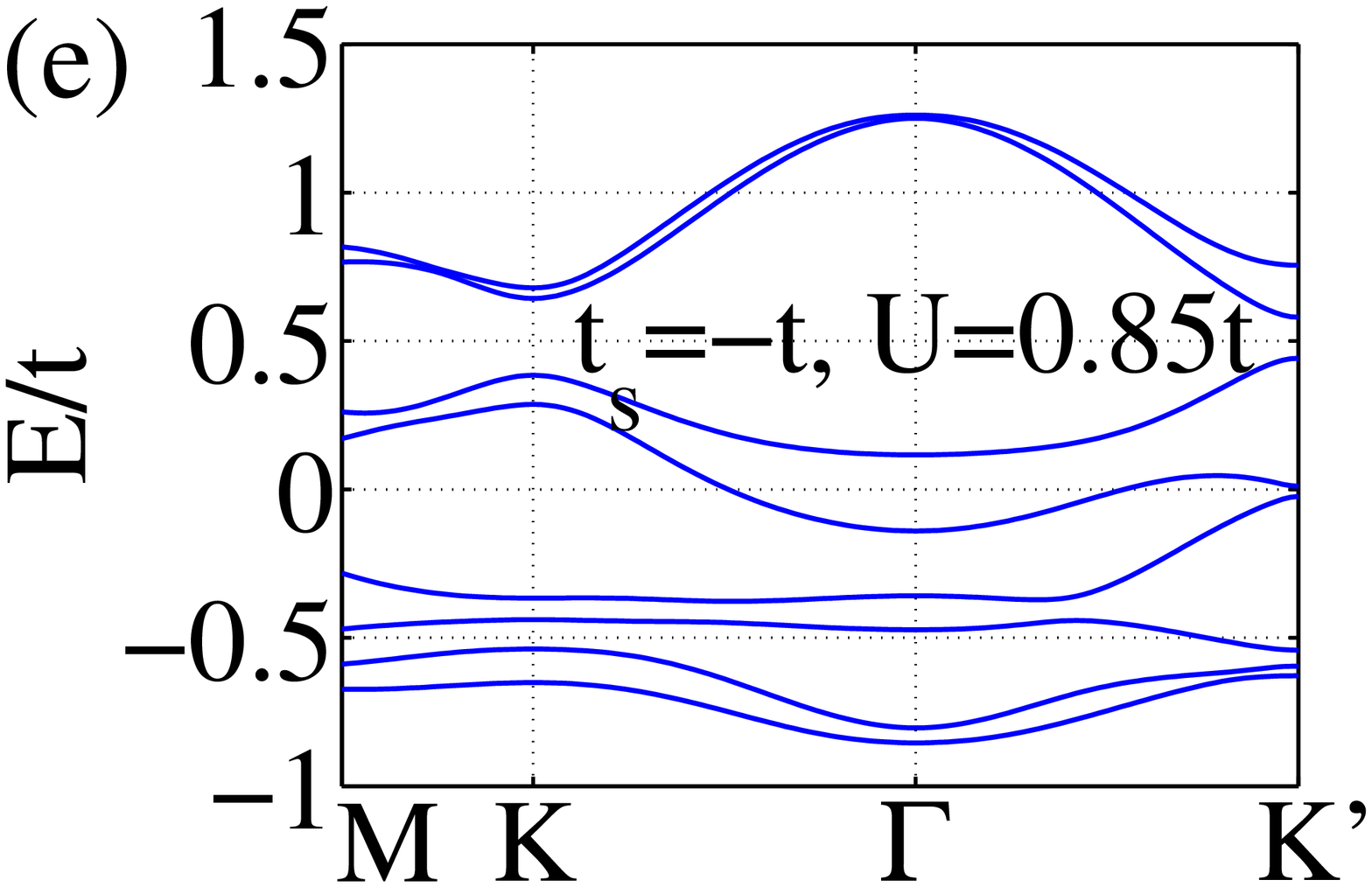}
\includegraphics[width=0.49\linewidth]{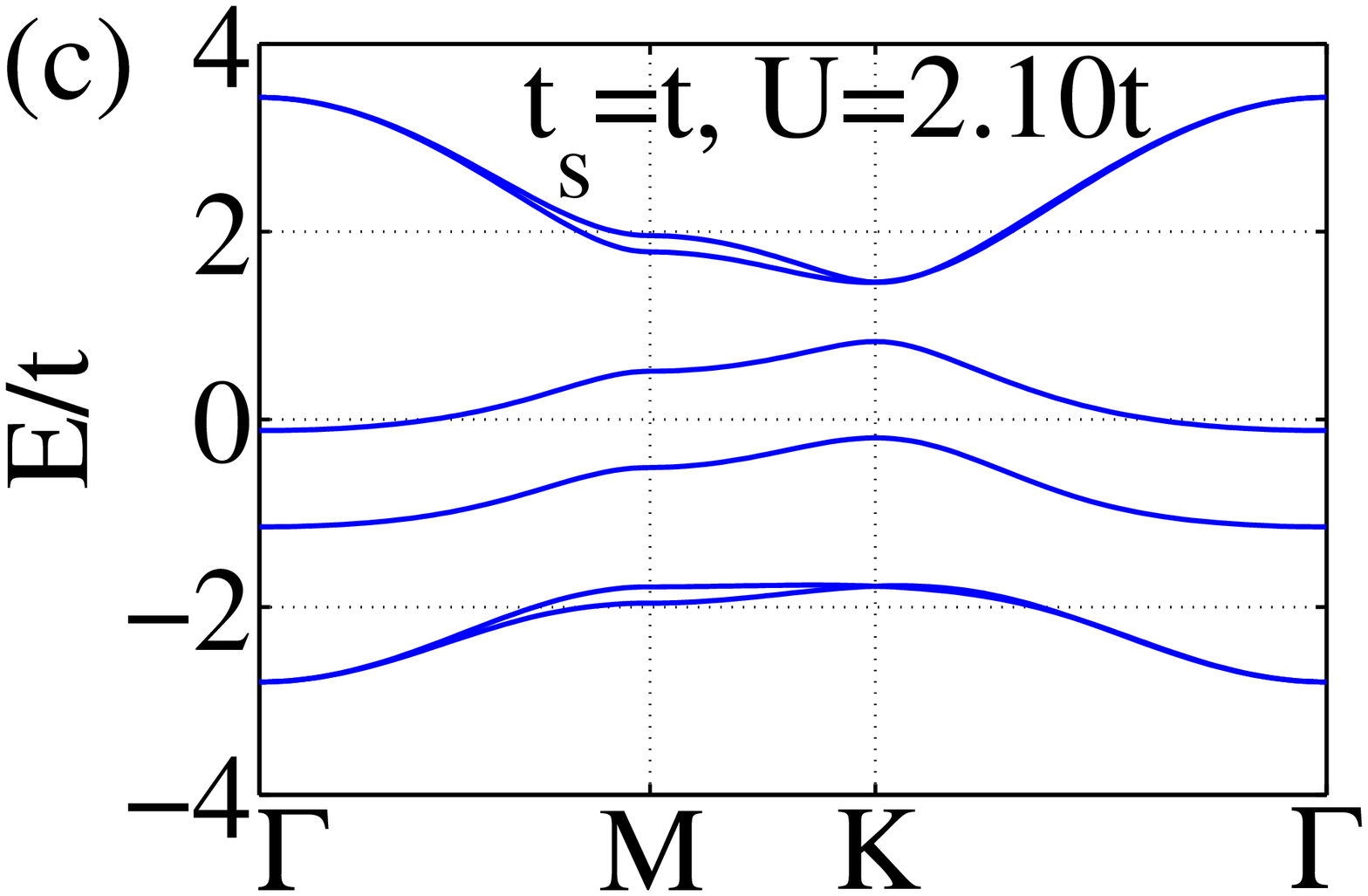}
\includegraphics[width=0.49\linewidth]{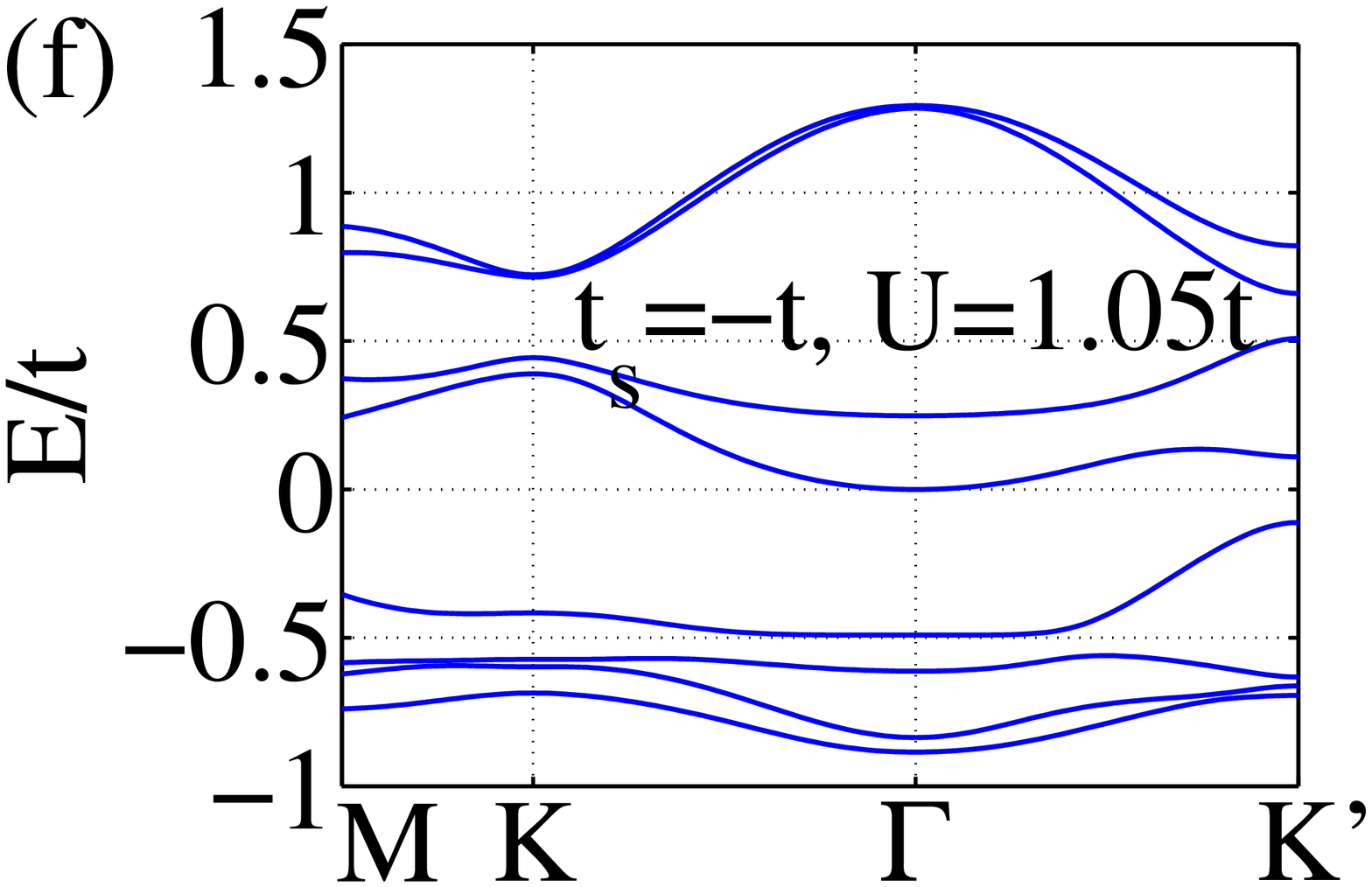}
\caption{(color online) The evolution of bands in a single kagome layer in the Hartree-Fock approximation when $t_s=t$ for (a) $U=1.90t$, (b) $U=1.95t$, (c) $U=2.10t$ and for a bilayer when $t_s=-t$ with (d) $U=0.65t$, (e) $U=0.85t$, (f) $U=1.05t$.  For the single kagome layer this corresponds to the M$\to$MC$\to$MI transition in Fig.~\ref{fig:U_phase} and for the bilayer this corresponds to the I$\to$MC$\to$MI transition. Note the Kramer's degeneracy at the time-reversal invariant momenta M and $\Gamma$ in (d) and the difference between $\Gamma$K and $\Gamma$K' in (e) and (f).}
\label{fig:kag_m}
\end{figure}

A few features of Fig.~\ref{fig:U_phase} are worth calling attention to. First, for all systems and tight-binding parameter regimes considered, there is finite range of $U$ for which the system remains time-reversal invariant, while for $U\gtrsim 3.0t$ all systems have become magnetic, as expected for sufficiently large $U$. The particular value of $U$ for which this happens depends on the density of states at the Fermi level if the system is metallic (if larger, then critical $U$ is smaller), or the magnitude of the gap if the system is gapped (a larger gap will have a larger critical $U$). These features can be seen by comparing the critical $U$ values in Fig.~\ref{fig:U_phase} with the band features in Figs.~\ref{fig:singlelayer}-\ref{fig:KTK}.  As illustrated in Fig.\ref{fig:magap}, we observe a second-order (or weakly first-order) transition between the non-magnetic and magnetic phases in the systems that possess a Chern insulator phase.  For the other systems (and parameter regimes), the magnetic transitions (not shown) most often appear first order.

Second, the Hubbard interaction $U$ never has the effect of shifting the bands around in such a way to obtain a $Z_2$ TI; if $U=0$ is not already a $Z_2$ TI one is not later obtained by increasing $U$. Third, the bilayer is the only system for a half-filled $d$-shell that possess both a TI and a Chern insulator. {\em However, the TKT system for $t_s=-t$ is an excellent candidate for realizing the much sought after zero magnetic field quantum Hall state, the Chern insulator (also known as quantum anomalous Hall state).}

\begin{figure}
\includegraphics[width=0.49\linewidth]{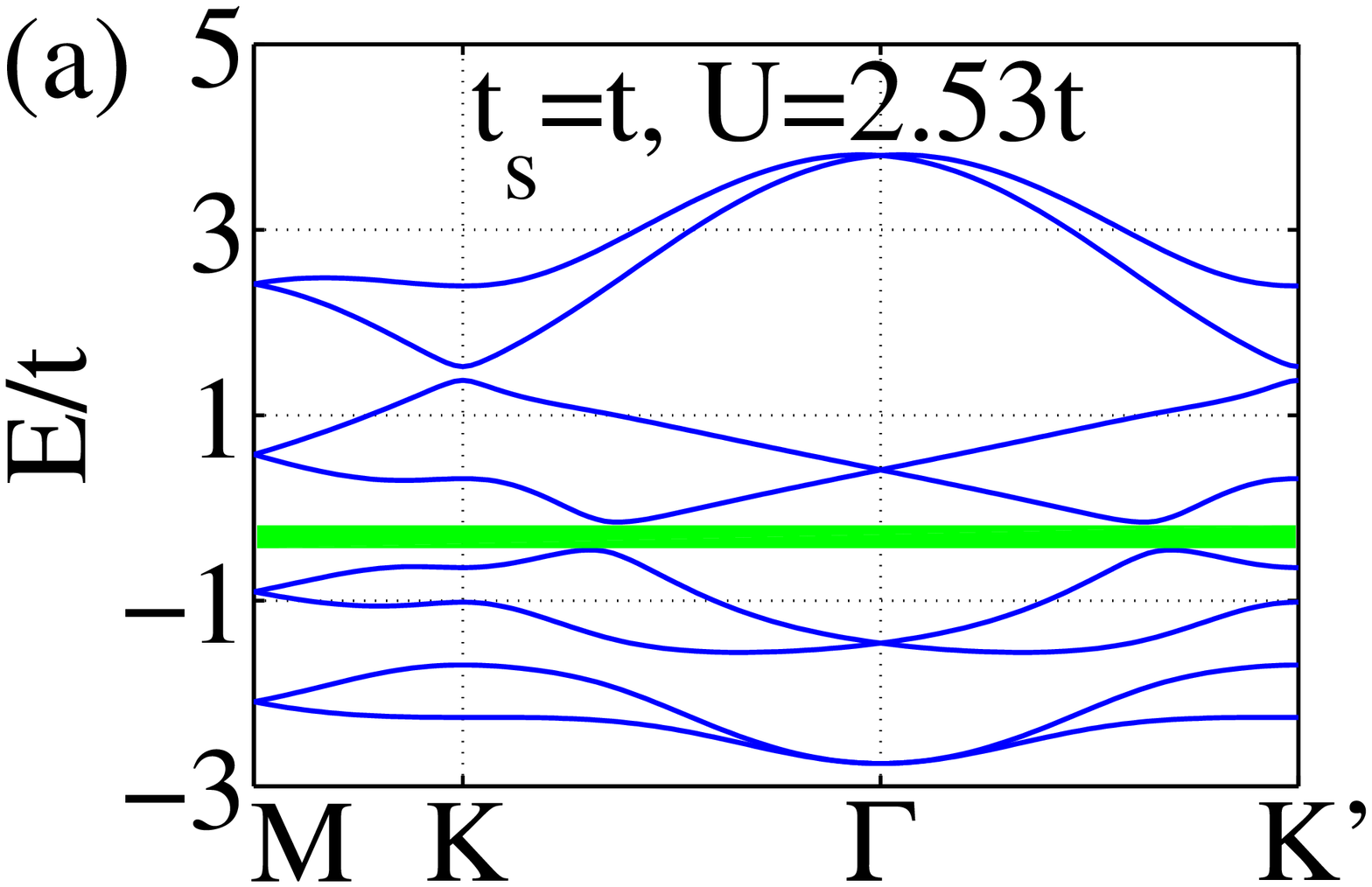}
\includegraphics[width=0.49\linewidth]{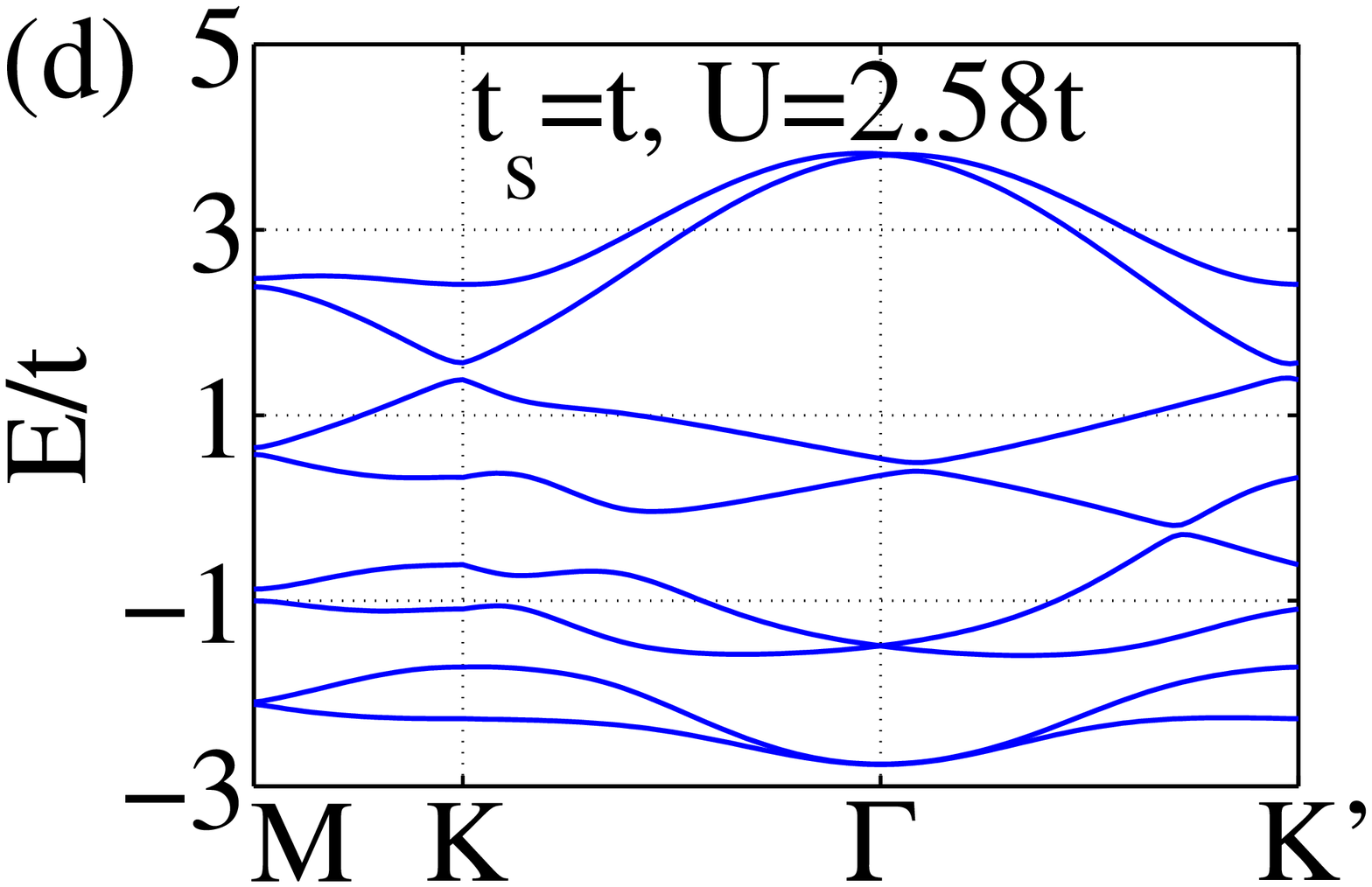}
\includegraphics[width=0.49\linewidth]{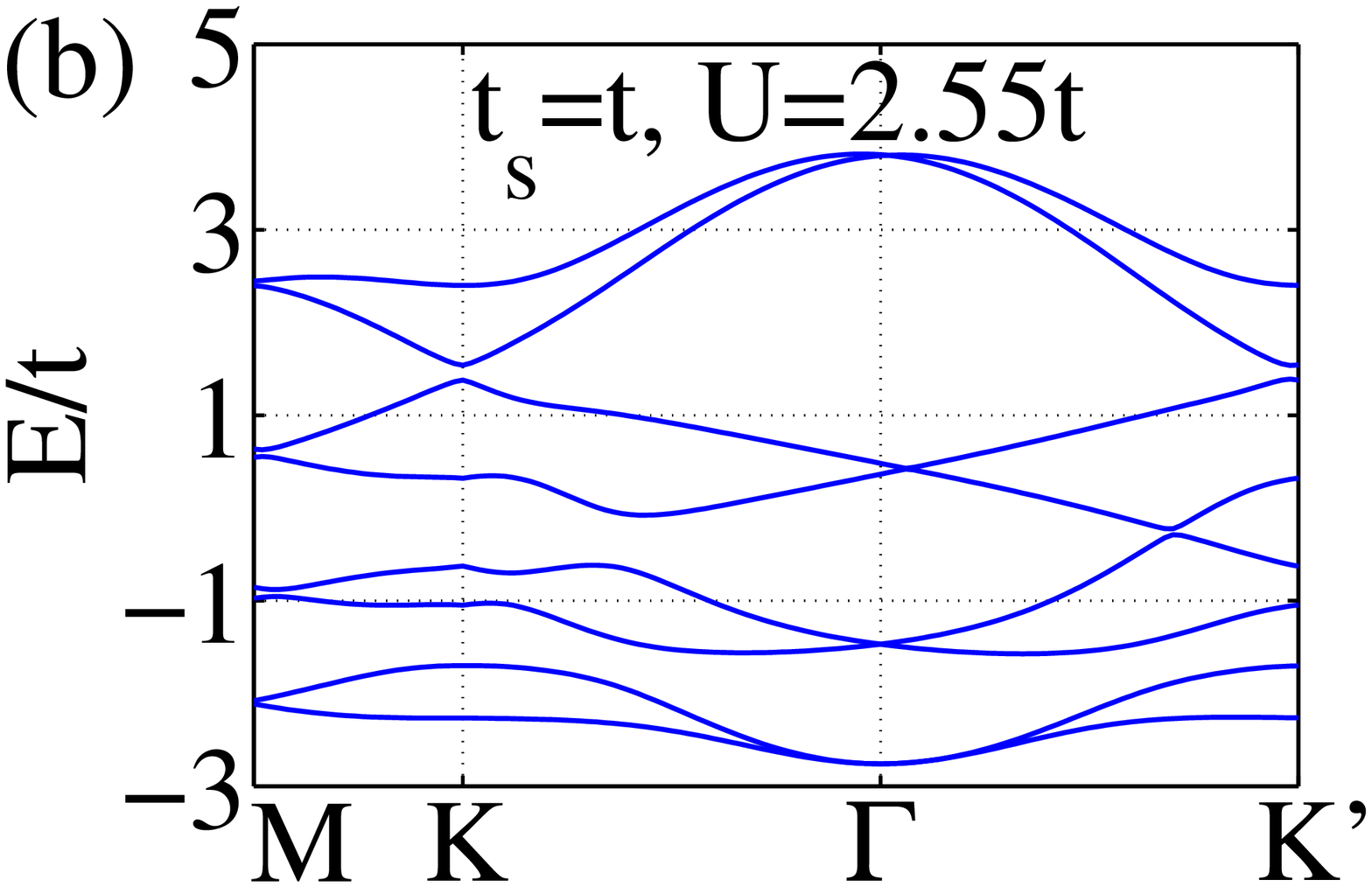}
\includegraphics[width=0.49\linewidth]{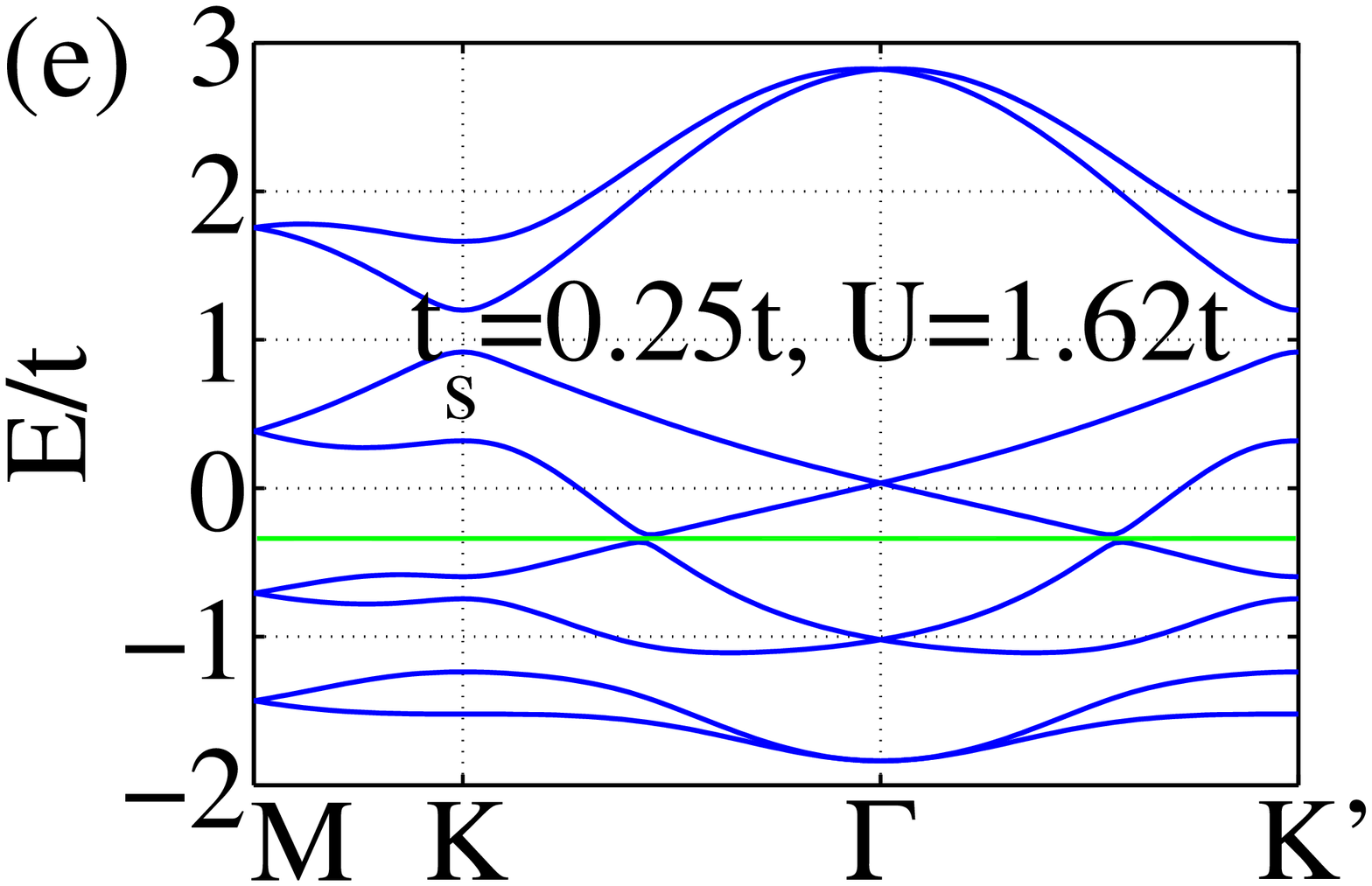}
\includegraphics[width=0.49\linewidth]{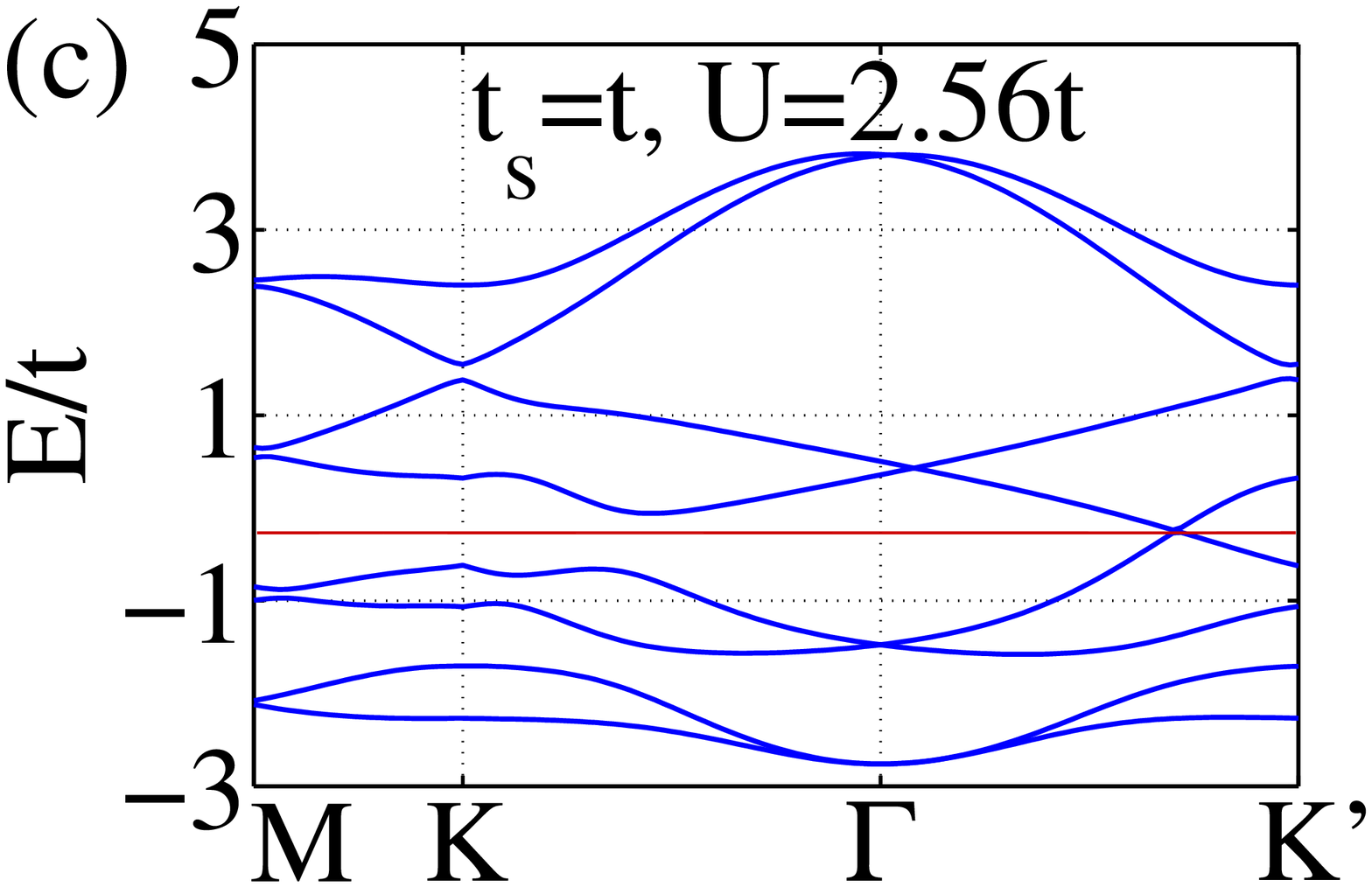}
\includegraphics[width=0.49\linewidth]{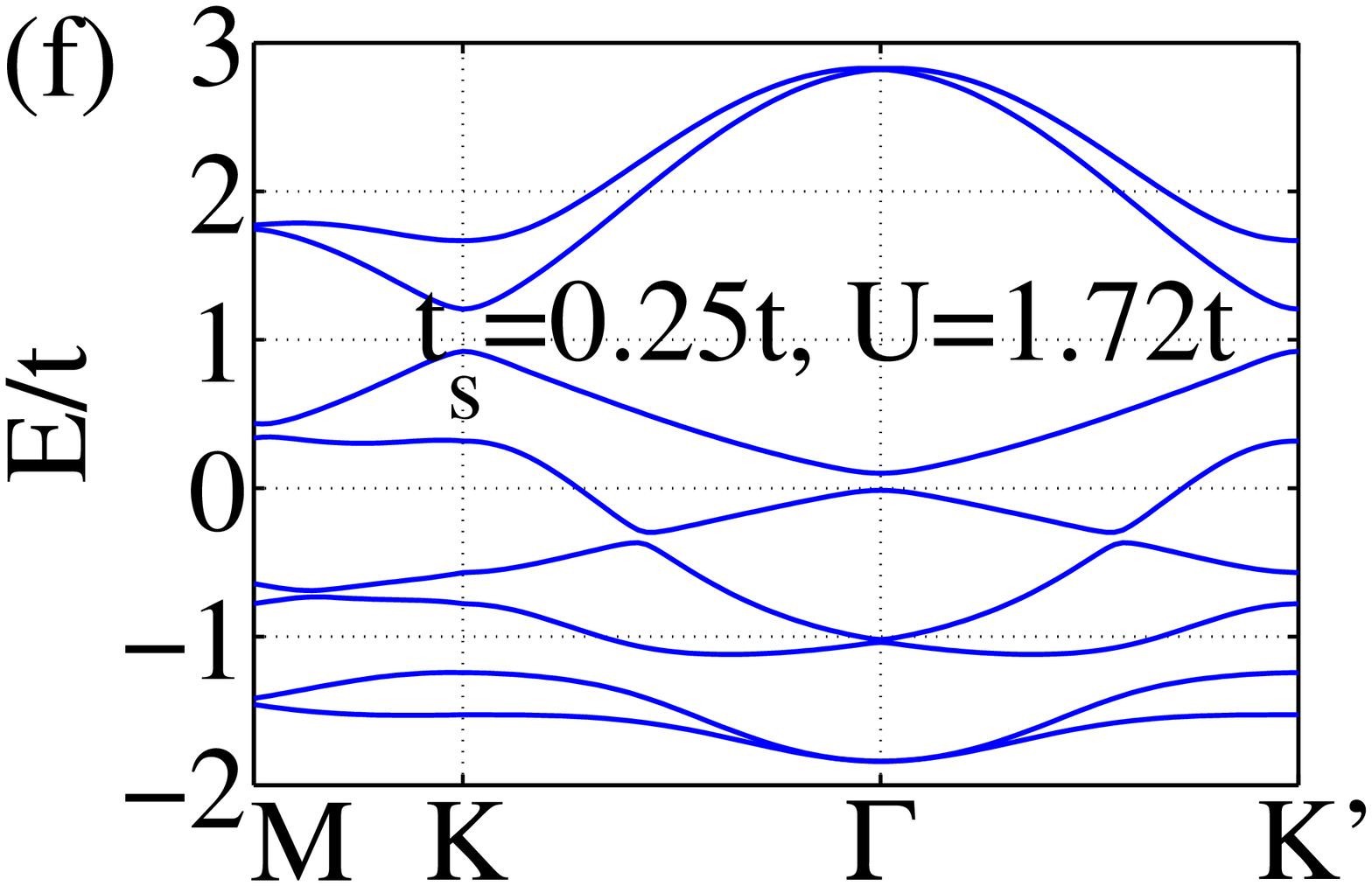}
\caption{(color online) The evolution of the bands when $t_s=t$ for a bilayer with (a) $U=2.53t$, (b) $U=2.55t$, (c) $U=2.56t$, (d) $U=2.58t$, and for a bilayer when $t_s=0.25t$ with (e) $U=1.62t$, and (f) $U=1.72t$. The figures clearly show that for $t_s=t$, the phase changes from (a) a TI to (b) a magnetic insulator (gap is too small to be visible)  to (c) a Chern insulator and back to (d) a magnetic insulator. The red (gray) horizontal line in (c) reveals the gap of the Chern insulator. For $t_s=0.25t$, the (e) TI changes to 
a magnetic insulator (f) directly. Note the Kramer's degeneracy at the time-reversal invariant momenta M and $\Gamma$ in the TI in (a) and (e) and the difference between $\Gamma$K and $\Gamma$K'.}
\label{fig:bi_m}
\end{figure}

\begin{figure}
\includegraphics[width=0.49\linewidth]{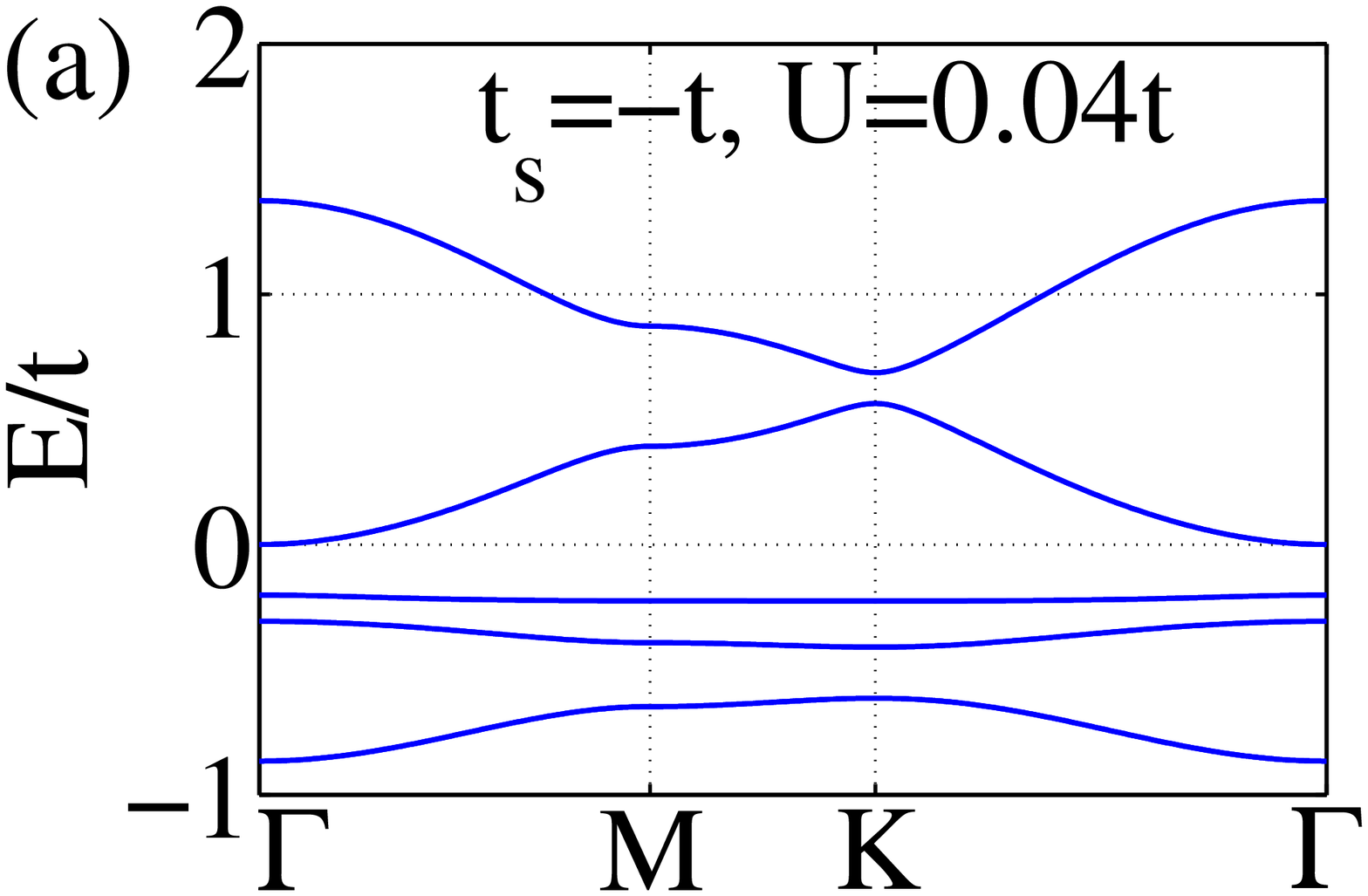}
\includegraphics[width=0.49\linewidth]{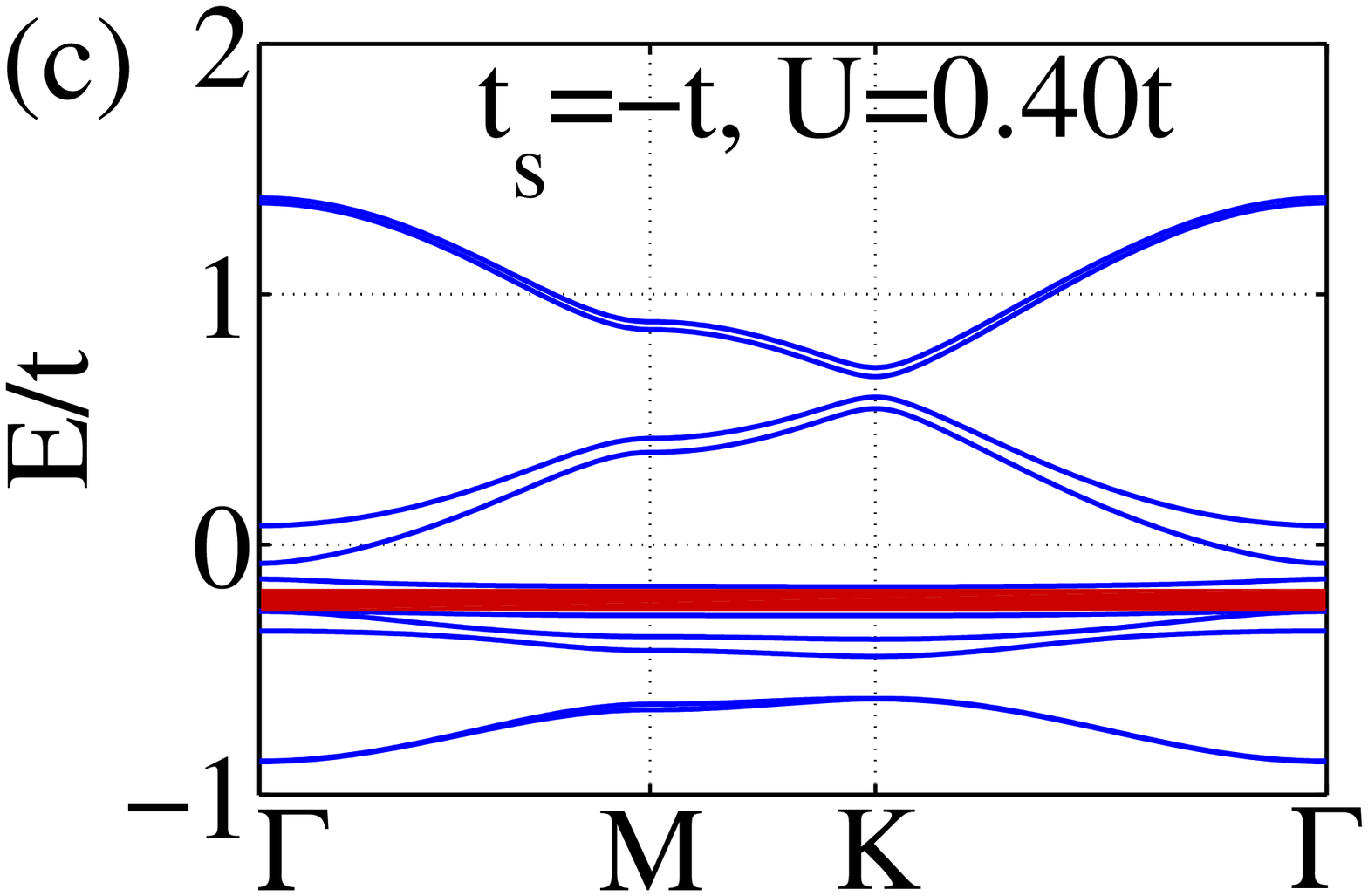}
\includegraphics[width=0.49\linewidth]{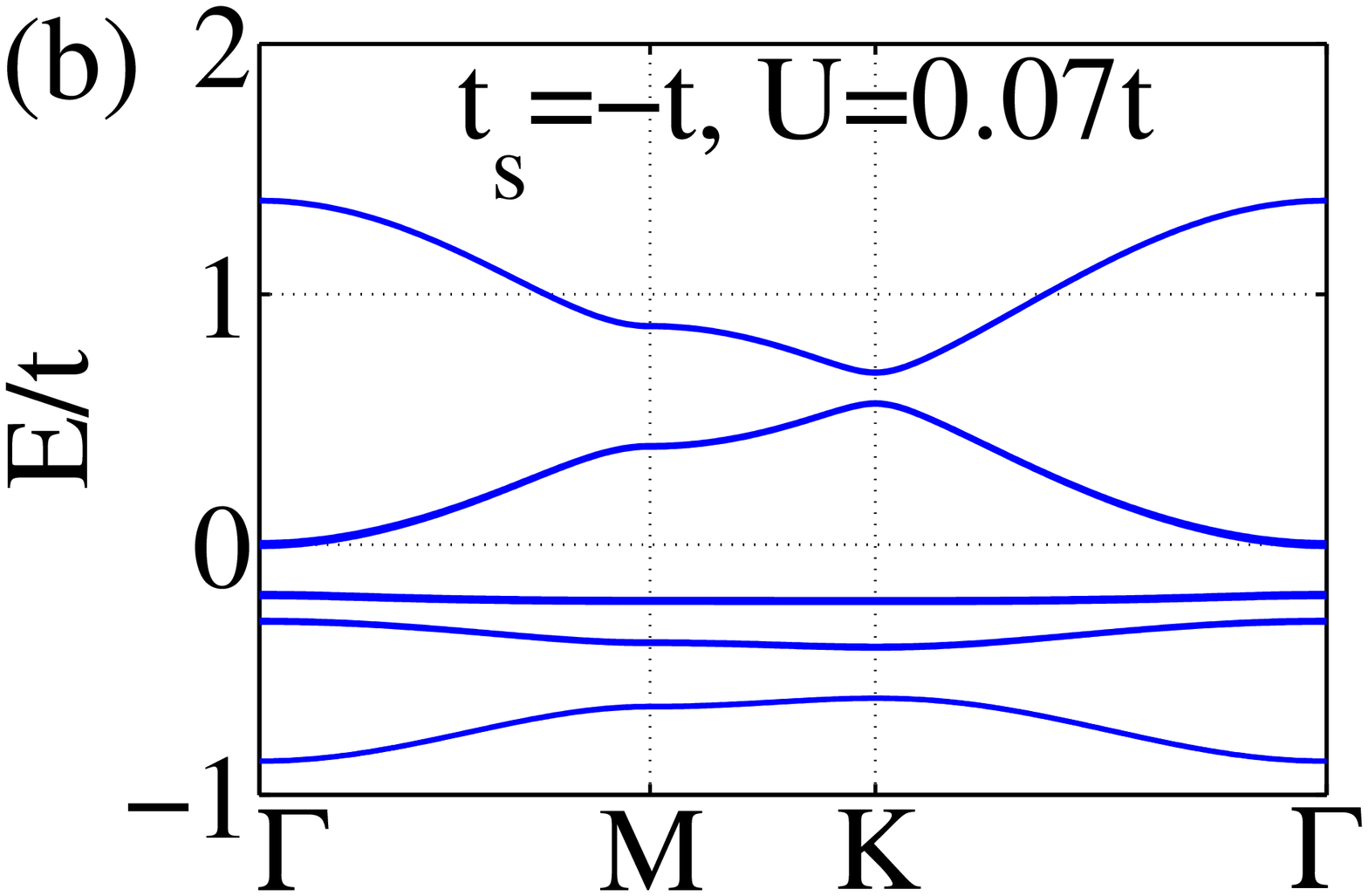}
\includegraphics[width=0.49\linewidth]{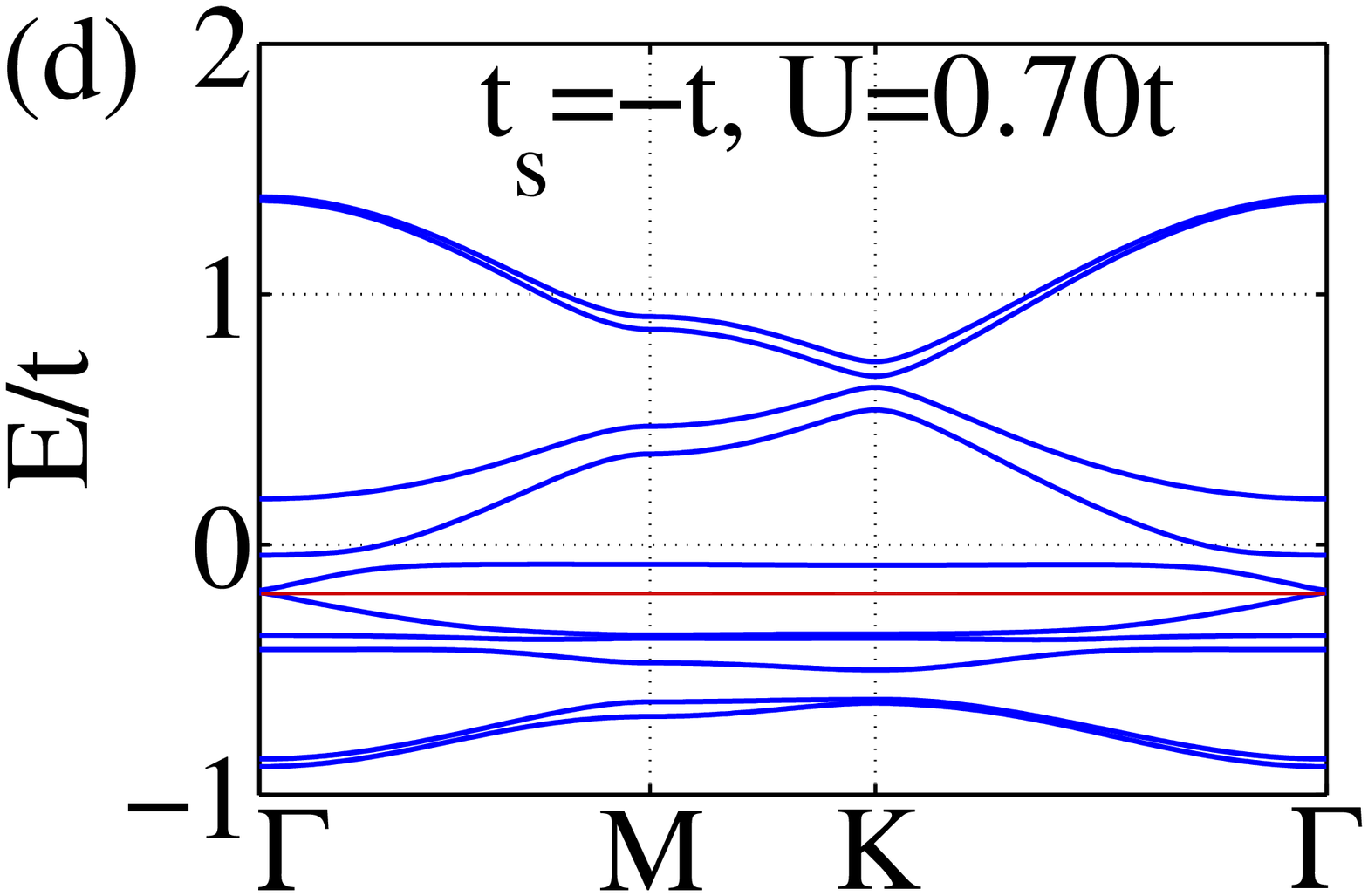}
\caption{(color online) The evolution of the bands when $t_s=-t$ for a TKT layer with (a) Metal $U=0.04t$, (b) Magnetic conductor $U=0.07t$, (c) Chern insulator $U=0.40t$, (d) Chern insulator $U=0.70t$. From (a) to (b),(c) the gap opens gradually, but later the gap closes again, and the tendency reveals near the $\Gamma$-point in (d). The pattern of transitions is M$\to$MC$\to$CI.}
\label{fig:tri_m}
\end{figure}

In Figs.~\ref{fig:kag_m}-\ref{fig:tri_m} we detail the band structure as a function of $U$ for different systems and tight-binding parameter values to show which band features correspond to various transitions in the phase diagrams in Fig.~\ref{fig:U_phase}. We note that the band dispersions are generally quite different from their non-interacting counterparts in Figs.~\ref{fig:singlelayer}-\ref{fig:KTK}. Figure~\ref{fig:kag_m} shows the evolution of the bands around $U\approx 2t$ where the M$\to$MC$\to$MI transition occurs in the kagome layer for $t_s=t$, and around $U\approx0.9t$ where the I$\to$MC$\to$MI transition occurs in the bilayer for $t_s=-t$. Band features in the bilayer $t_s=t$ transitions TI$\to$MI$\to$CI$\to$MI and in the bilayer $t_s=0.25t$ transition TI$\to$MI are shown in Fig.~\ref{fig:bi_m}. In the case of the bilayer, inversion symmetry is broken resulting in a dispersion from $\Gamma\to K$ being generally different from the dispersion from $\Gamma \to K'$, so we have plotted both. Finally, in Fig.\ref{fig:tri_m} we show the band evolution for the M$\to$MC$\to$CI transition in the $t_s=-t$ for a TKT layer.

\begin{figure}[ht]
\centering
\includegraphics[width=\linewidth]{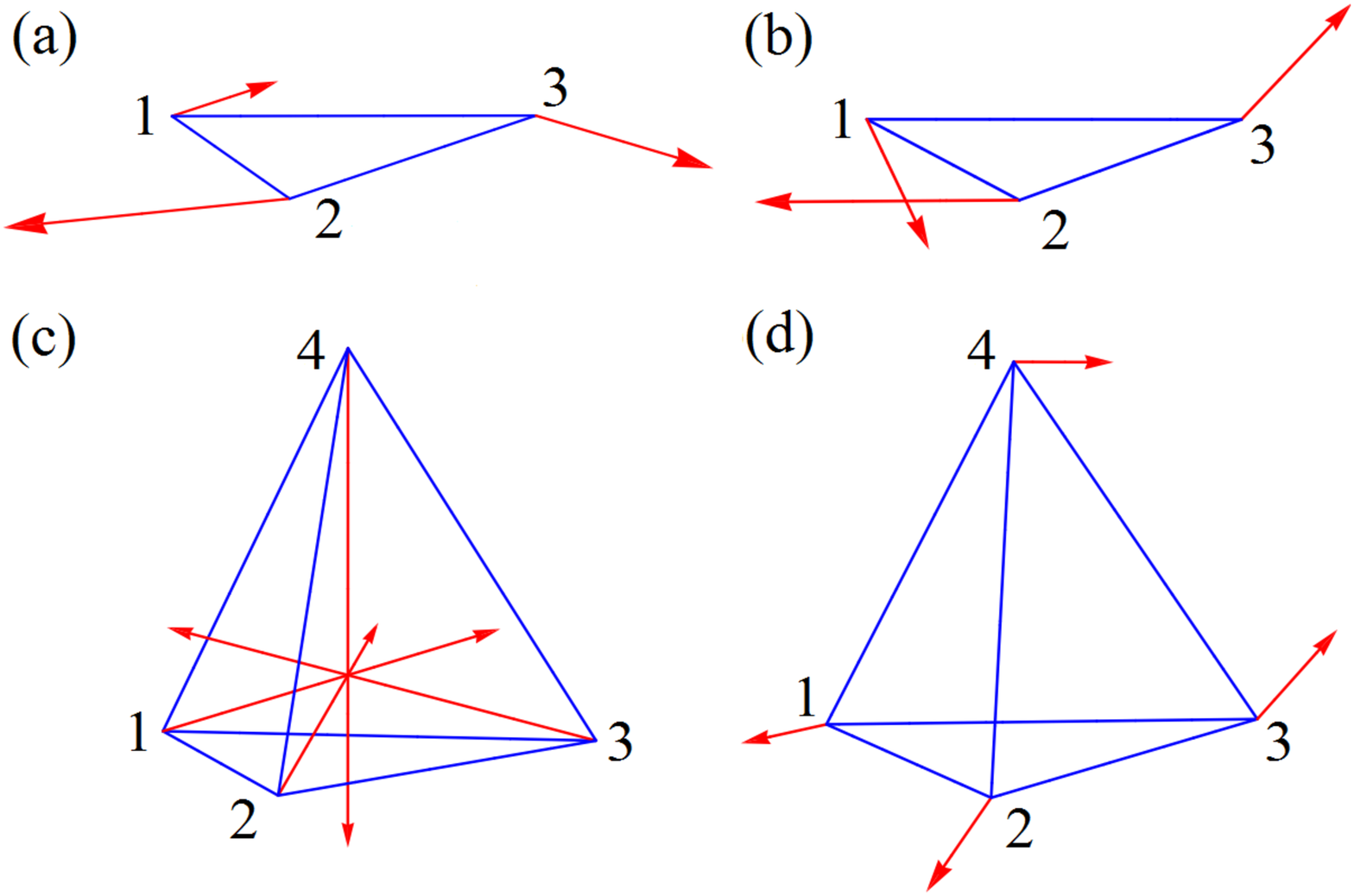}
\caption{(color online) Two kinds of magnetic configurations obtained within the Hartree-Fock approximation for a
single kagome layer with (a) $t_s=-t, U=2.5t$, (b) $t_s=t, U=3.0t$ and a bilayer with (c) $t_s=-t, U=3.0t$, (d) $t_s=t, U=3.0t$. For the kagome layer, two configurations exist. One (a) is ``rotating" and keeps the $C_3$ symmetry, while (b)
is an antiferromagnetic phase without this property. For the bilayer, the $C_3$ symmetry
is also retained for $t_s=-t$. In the other phase shown in (d), the magnetic moment 4 is antiparallel to 
moment 1, and moment 3 lies in the plane 413 while moment 2 lies in plane 412.}
\label{fig:spin_kb}
\end{figure}

\begin{figure}[ht]
\centering
\includegraphics[width=\linewidth]{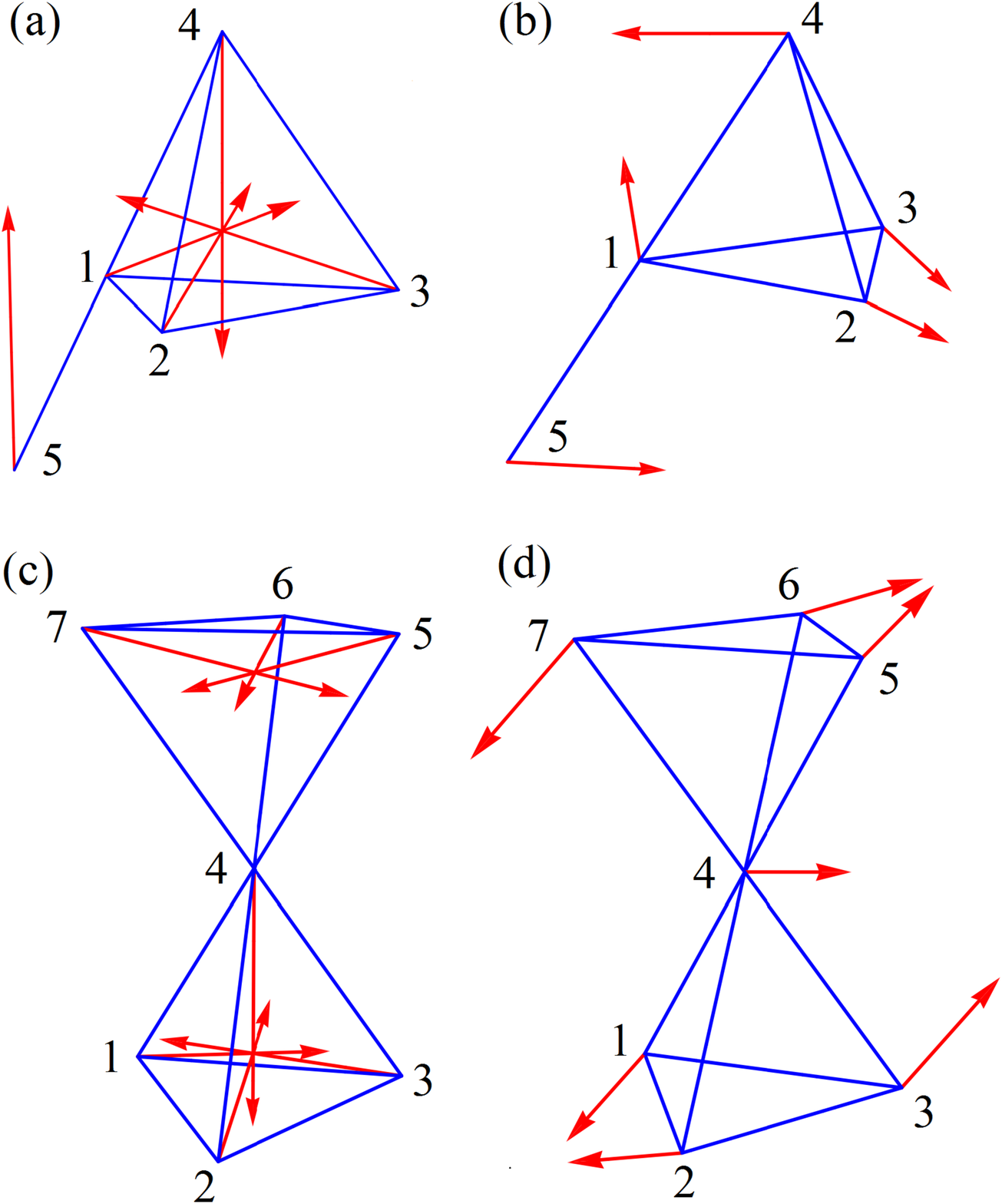}
\caption{(color online) The two kinds of magnetic configurations
for a TKT layer: (a) $t_s=-t, U=3.0t$, (b) $t_s=t, U=3.0t$ and for a KTK layer: (c) $t_s=-t, U=2.5t$, (d) $t_s=t, U=3.0t$. In (a), 
the moments 4,5 are antiparallel to each other, and perpendicular
to the 123 plane. In (b), the moments 4,5 are antiparallel to each other, and parallel
to the 123 plane. One moment in 123 will lie in the same 
plane as moment 4,5, and the other two moments will point in an 
almost opposite direction. In (d) the moment 4 is parallel to the 123 plane.}
\label{fig:spin_tri}
\end{figure}

It is interesting to supplement the information about the changes in the band structure as a function of increasing $U$ with figures illustrating the magnetic configurations of the local moments on the $d$-orbitals when a magnetic transition occurs.  The main results are shown in Figs.~\ref{fig:spin_kb}-\ref{fig:spin_tri}. One striking feature is that the sign of $t_s$ affects the magnetic order. This is evident as well in closely related three-dimensional studies.\cite{Witczak:prb12} The magnetic moments point in different directions on different sites of the unit cell which reduces the overall magnetization. Yet,  we observe in general a net magnetic moment, see Fig.~\ref{fig:magap}.

A final point worth emphasizing before moving to the conclusions is that all the magnetic orders shown in Figs.~\ref{fig:spin_kb}-\ref{fig:spin_tri} are non-coplanar. This condition can be favorable for obtaining a Chern insulator, but is not strictly required in a mult-band system.\cite{Ruegg11_2,Ruegg:prb12}  In fact, the physics of the Chern insulator we obtain in our Hartree-Fock approximation is rather similar to the large exchange limit studied in Ref.~[\onlinecite{Ohgushi:prb00}].  In both cases, the ``spin" of the electron is forced to adiabatically follow the local moments residing on the lattice.

\section{Conclusions}
\label{sec:conclusions}

In this paper, we extended previous work of perovskite heterostructures\cite{Xiao:nc11,Ruegg11_2,Ruegg:prb12,Yang:prb11a,Wang:prb11} grown along the [111] direction to pyrochlore heterostructures grown along [111].  Because (111) and equivalent planes are natural cleavage planes of the pyrochlore systems, it is expected that [111] is a good growth direction. We focused on sandwich structures where the electronically active region is confined to a quasi-two dimensional few-layer system as shown in Fig.~\ref{fig2}. The analysis of the relevant tight-binding models for the $d$-shell electrons in the presence of spin-orbit coupling revealed that the bilayer system is the most promising candidate to realize a TI phase for a half-filled $d$-shell (as, e.g., realized in pyrochlore iridates). For the other systems, TI phases occur at ``fractional" fillings which would require one to adjust the $d$-shell electron occupation by considering systems of the form A$_{2(1-x)}$A'$_{2x}$B$_2$O$_7$ for the electronically active region. 

The effect of the electron-electron interaction was studied within the Hartree-Fock mean-field theory. We found that the electron-electron interaction in general stabilizes magnetic order with non-planar local moments.\cite{Witczak:prb12} Interestingly, we have also identified parameters at intermediate interaction strength where a Chern insulator coexists with magnetic order.  Within our self-consistent calculation we find that the TKT structure is the most likely system (among those we considered) to realize such a zero-magnetic field quantum Hall state.

While our work based on semi-realistic lattice models revealed several scenarios to obtain topological phases in layered pyrochlore oxide thin films grown along the [111] direction, it also suggests that the expected energy scales (i.e.~the gap) of these phases are in general small. Moreover, both the weakly interacting phases as well as the type of magnetic order at larger interactions crucially depend on the details of the hopping parameters. In the future, it is therefore desirable to incorporate band structure calculations based on the density-functional theory to have better estimates of the hopping parameters and the strength of the interaction (relative to the band width) needed to obtain a Chern insulator phase.

\acknowledgements
We thank Mehdi Kargarian, Jun Wen, Qi Chen and Zhenhua Qiao for helpful discussions. 
We gratefully acknowledge funding from ARO grants W911NF-09-1-0527, W911NF-12-1-0573,
 and NSF Grant DMR-0955778. GAF acknowledges the hospitality of the Aspen Center for Physics under NSF Grant PHY-1066293 where part of this work was done. AR acknowledges partial support through the Swiss National Science Foundation. The authors acknowledge the Texas Advanced Computing Center (TACC) at The University of Texas at Austin for providing the necessary computing resources. URL: \url{http://www.tacc.utexas.edu}. Figure~\ref{fig1c} and Fig.~\ref{fig2} are based on the free crystal structure drawing software of Ref.~\onlinecite{Ozawa:2009}.
\begin{appendix}
\section{Tight-binding parameters for the direct overlap of $d$-orbitals in Eq.\eqref{eq:H0}}
\label{app:tb}

The tight-binding hopping amplitudes used in the present work were derived from the nearest-neighbor hopping amplitudes on the pyrochlore lattice, including both oxygen-mediated (indirect) as well as direct hopping processes:
\begin{equation}
t_{i\alpha,j\beta}=t_{i\alpha,j\beta}^{in}+t_{i\alpha,j\beta}^{dir}.
\end{equation}
The procedure to obtain the indirect hopping $t_{i\alpha,j\beta}^{in}$ between the $t_{2g}$ electrons on the pyrochlore lattice mediated by the intermediate oxygens was discussed in Ref.~[\onlinecite{Pesin:np10}]. Due to the extended nature of the 5$d$ orbitals, we also consider the direct overlap between the $d$ orbitals which yields an additional contribution to the hopping matrix $t_{i\alpha,j\beta}^{dir}$.\cite{Witczak:prb12} In the following, we discuss the derivation of $t_{i\alpha,j\beta}^{dir}$.

Note that the oxygen octahedra enclosing the transition metal ions are rotated with respect to each other at different sites of the unit cell. The splitting of the vacuum $d$ levels into the $t_{2g}$ and $e_g$ manifolds results from the local octahedral crystal field. Hence, the $t_{2g}$ orbitals are defined with respect to the local frame. Following the convention in Ref.~[\onlinecite{Pesin:np10}], the transformation to the local coordinates are
\begin{equation}
(x_i,y_i,z_i)=(x,y,z)R_i,
\end{equation}
where $(x,y,z)$ are the coordinates in the global reference frame and $(x_i,y_i,z_i)$ are the coordinates in the local frame at site $i$. The index $i=1,2,3,4$ labels the sites in the unit cell of the pyrochlore lattice.\cite{Pesin:np10,Yang_Kim:prb10,Kargarian:prb11} Explicitly, the rotation matrices are given by
\begin{equation}
R_1=\begin{pmatrix}
 2/3& -1/3& -2/3\\
 -1/3&2/3&-2/3\\
 2/3& 2/3& 1/3
\end{pmatrix},
\end{equation}
\begin{equation}
R_2=\begin{pmatrix}
 2/3& 2/3& 1/3\\
 -2/3&1/3&2/3\\
 1/3& -2/3& 2/3
\end{pmatrix},
\end{equation}
\begin{equation}
R_3=\begin{pmatrix}
 1/3& -2/3& 2/3\\
 2/3&2/3&1/3\\
 -2/3& 1/3& 2/3
\end{pmatrix},
\end{equation}
\begin{equation}
R_4=\begin{pmatrix}
 1/3& -2/3& 2/3\\
 -2/3&-2/3&-1/3\\
 2/3& -1/3& -2/3
\end{pmatrix}.
\end{equation}
For later use, we also give the corresponding Euler angles in the convention of Ref.~[\onlinecite{Su:1994}]:
\begin{eqnarray}
\alpha_1&=&\frac{5\pi}{4},\nonumber\\
\beta_1&=&\arccos(1/3),\nonumber\\
\gamma_1&=&3\pi/4;\label{eq:a1}
\end{eqnarray}
\begin{eqnarray}
\alpha_2&=&\arccos(2/3),\nonumber\\
\beta_2&=&\arccos(1/\sqrt{5}),\nonumber\\
\gamma_2&=&2\pi-\arccos(-1/\sqrt{5});\label{eq:a2}
\end{eqnarray}
\begin{eqnarray}
\alpha_3&=&\arccos(2/3),\nonumber\\
\beta_3&=&\arccos(2/\sqrt{5}),\nonumber\\
\gamma_3&=&\arccos(2/\sqrt{5});\label{eq:a3}
\end{eqnarray}
\begin{eqnarray}
\alpha_4&=&\arccos(-2/3),\nonumber\\
\beta_4&=&2\pi-\arccos(2/\sqrt{5}),\nonumber\\
\gamma_4&=&2\pi-\arccos(-2/\sqrt{5})\label{eq:a4}.
\end{eqnarray}
The general procedure to obtain the direct overlap is to expand the local $t_{2g}$ orbitals in the $d$-orbitals of the global frame from which the hopping amplitudes are obtained through the use of the Slater-Koster integrals.\cite{Slater:1954} 

To expand the local $t_{2g}$ orbitals in the global frame, we make use of the transformation law of spherical harmonics under a rotation with Euler angles $(\alpha,\beta,\gamma)$:\cite{Su:1994}
\begin{equation}
Y_l^m(\theta_i,\phi_i)=\sum_{m'=-l}^lY_l^{m'}(\theta,\phi)D_{m'm}^l(\alpha_i,\beta_i,\gamma_i).
\label{eq:D}
\end{equation}
Here, $D_{m'm}^l(\alpha,\beta,\gamma)=e^{-im'\alpha}d_{m'm}^l(\beta)e^{-im\gamma}$ is a $(2l+1)\times(2l+1)$ dimensional rotation matrix and the elements $d_{m'm}^l(\beta)$ are given in Ref.~[\onlinecite{Su:1994}] with $l=2$. To proceed, we introduce the standard real orbitals
\begin{eqnarray}
d_{yz}&=&\frac{i}{\sqrt{2}}\left(Y_{2}^{-1}+Y_{2}^{1}\right),\\
d_{zx}&=&\frac{1}{\sqrt{2}}\left(Y_{2}^{-1}-Y_{2}^{1}\right),\\
d_{xy}&=&\frac{i}{\sqrt{2}}\left(Y_{2}^{-2}-Y_{2}^{2}\right),\\
d_{3z^2-r^2}&=&Y_{2}^{0},\\
d_{x^2-y^2}&=&\frac{1}{\sqrt{2}}\left(Y_{2}^{-2}+Y_{2}^{1}\right).
\end{eqnarray}
Using the Euler angles in Eqs.~\eqref{eq:a1}-\eqref{eq:a4} together with Eq.~\eqref{eq:D} and the definition of the real orbitals, we find that the local $t_{2g}$ orbitals are expanded in the following way
\begin{equation}
\begin{pmatrix}
|y_iz_i\rangle\\
|z_ix_i\rangle\\
|x_iy_i\rangle
\end{pmatrix}=M_i^T
\begin{pmatrix}
|yz\rangle\\
|zx\rangle\\ 
|xy\rangle\\ 
|3z^2-r^2\rangle\\
|x^2-y^2\rangle
\end{pmatrix}.
\end{equation}
The rotation matrices $M_i$ are $5\times 3$ matrices given by
\begin{equation}
M_1=
\begin{pmatrix}
-2/9 & -5/9 & 2/9\\
-5/9 & -2/9 & 2/9\\
-2/9 & -2/9 & 5/9\\
2/(3\sqrt{3}) & 2/(3\sqrt{3}) & 4/(3\sqrt{3})\\
2/3 & -2/3 & 0
\end{pmatrix},
\end{equation}
\begin{equation}
M_2=
\begin{pmatrix}
-2/9 & -2/9 & 5/9\\
2/9 & 5/9 & -2/9\\
5/9 & 2/9 & -2/9\\
-4/(2\sqrt{3}) & 2/(3\sqrt{3}) & -2/(3\sqrt{3})\\
0 & 2/3 & 2/3
\end{pmatrix},
\end{equation}
\begin{equation}
M_3=
\begin{pmatrix}
5/9 & 2/9 & -2/9\\
-2/9 & -2/9 & 5/9\\
2/9 & 5/9 & -2/9\\
2/(3\sqrt{3}) & -4/(3\sqrt{3}) & -2/(3\sqrt{3})\\
-2/3 & 0 & -2/3
\end{pmatrix},
\end{equation}
\begin{equation}
M_4=
\begin{pmatrix}
5/9 & 2/9 & -2/9\\
2/9 & 2/9 & -5/9\\
-2/9 & -5/9 & 2/9\\
2/(3\sqrt{3}) & -4/(3\sqrt{3}) & -2/(3\sqrt{3})\\
-2/3 & 0 & -2/3
\end{pmatrix}.
\end{equation}
With the above matrices, the direct hopping between the $t_{2g}$ orbitals from site $j$ to site $i$ is written as
\begin{equation}
t_{i\alpha, j\beta}^{dir}=M_i^T S_{dd}(\hat{e}_{ij}) M_j\otimes \left(D_i^{\dag}D_j\right)_{\alpha\beta}.
\end{equation}
Here, the matrices $D_i$ account for the rotation to the local frame in spin space as described in Refs.~[\onlinecite{Pesin:np10,Yang_Kim:prb10,Kargarian:prb11}]. $S_{dd}(\hat{e}_{ij})$ is the $5\times5$-matrix containing the Slater-Koster integrals between $d$-orbitals along the unit direction $\hat{e}_{ij}$ between site $j$ and $i$ and is parameterized by $(dd\sigma)$, $(dd\pi$), and $(dd\delta)$. Because the $\delta$-bonds are typically small, we only keep the $\sigma$ and $\pi$-bonds and introduce the parameters $t_s\equiv t_{\sigma}=(dd\sigma)$  and $t_p \equiv t_\pi=(dd\pi)$ to quantify the direct hopping.

\section{Some details of the Hartree-Fock mean-field calculation}
\label{app:HF}

We studied the interacting half-filled d-shell within the unrestricted 
Hartree-Fock mean field theory. A typical calculation starts from twenty different randomly generated 
initial magnetic configurations $\langle{\bf j}\rangle^{in}$ which define the initial mean-field Hamiltonians via the mean-field decoupled version of the interaction Eq.~\eqref{eq:H_U} together with the non-interacting Bloch Hamiltonian Eq.~\eqref{eq:H0} (projected to the $j=1/2$ manifold).

For a given filling fraction, the occupied single-particle states at zero temperature are easily identified on our meshed Brillouin zone (150 by 150). We then obtain the new 
local magnetic moments by evaluating the expectation values $\langle{\bf j}\rangle^{out}$ on each site. This process is iterated until
the difference between $\langle{\bf j}\rangle^{in}$ and $\langle{\bf j}\rangle^{out}$
is less than a fixed tolerance, in our case
\begin{equation}
|\langle{\bf j}\rangle^{in}_{i,n}-\langle{\bf j}\rangle^{out}_{i,n}|<10^{-8},
\end{equation}
where $n=x,y,z$. We
compare the results from twenty groups of such calculations and take the one
with minimum free energy.

Generally, we find magnetic phases for $U>U_c$ and the magnitudes of the local moments $\langle{\bf j}\rangle$ increase with $U$. However, even for $U<U_c$, a finite magnetic moment may be obtained due to finite-size effects. The finite size effect has been identified by varying the {\bf k}-mesh of the Brillouin zone. We found $|\langle{\bf j}\rangle|<0.01$ to be a robust criteria for which the system can be considered non-magnetic. In these cases, we reset the magnetic moments to zero.

Figure~\ref{fig:magap} show the magnitudes of the net magnetic moments as function of $U/t$ for the bilayer and TKT systems, respectively,
\begin{equation}
\langle{\bf j}_t\rangle=\sum_i\langle {\bf j}_i\rangle
\end{equation}
where the sum runs over the sites of the unit cell. In addition, Fig.~\ref{fig:magap} show the direct and indirect gaps. This information was used to distinguish between insulating and metallic phases. The topological properties were addressed by computing the $Z_2$ invariant for the time-reversal symmetric phases and the Chern number for the magnetic phases.\cite{Fukui:jpsj05,Fukui:jpsj07}

\end{appendix}


\begin{thebibliography}{10}
\bibitem{Note_Strain}%
  \BibitemOpen
  \bibinfo {note} {Strain effects are interesting to explore as well, but they
  are beyond the scope of the present work.}%
  \bibAnnoteFile{Stop}{Note_Strain}%
\end{thebibliography}
%

\end{document}